\documentclass[%
reprint,
showpacs,
 amsmath,amssymb,
 aps,
 pre,
 showkeys,
floatfix,
longbibliography
]{revtex4-2}

\usepackage{graphicx} 
\usepackage{dcolumn} 
\usepackage{bm} 
\usepackage{booktabs}
\usepackage{array}
\usepackage[utf8]{inputenc}
\usepackage[T1]{fontenc}
\usepackage{times}
\usepackage{flafter} 
\usepackage{mathrsfs} 
\usepackage[hidelinks,urlcolor=blue, colorlinks=true,citecolor=blue, linkcolor=blue ]{hyperref}
\providecommand{\abs}[1]{\left\lvert#1\right\rvert}

\makeindex

\newcommand{\iek}[1]{\left(#1\right)}
\newcommand{\fiek}[1]{\left\{#1\right\}}

\providecommand{\vect}[1]{\pmb{#1}}

\begin{document}

\title{Energetically favorable configurations of hematite cube chains}

\author{M.\ Brics}
\email[Corresponding author's email: ]{martins.brics@lu.lv}
\author{V.\ Šints}
\author{G.\ Kitenbergs}
\author{A.\ C\={e}bers}
\affiliation{%
 MMML lab, Department of Physics,
University of Latvia, Jelgavas 3, Rīga, LV-1004, Latvia
}

\date{\today}

\begin{abstract}
Hematite at room temperature is a weak ferromagnetic material. Its permanent magnetization is three orders smaller than for magnetite. Thus, hematite colloids allow us to explore a different physical range of particle interaction parameters compared to ordinary ferromagnetic particle colloids. In this paper we investigate a colloid consisting of hematite particles with cubic shape. We search for energetically favorable structures in an external magnetic field with analytical and numerical methods and molecular dynamics simulations and analyze whether it is possible to observe them in experiments. We find that energetically favorable configurations are observable only for short chains. Longer chains usually contain kinks which are formed in the process of chain formation due to the interplay of energy and thermal fluctuations as an individual cube can be in one of two alignments with an equal probability. 
\end{abstract}


\maketitle


\section{Introduction}
\label{sec:intro}

Hematite colloids represent a unique system to study interactions between two particles. They can be synthesized in different shapes: cubes, disks, ellipsoids, peanuts, and others \cite{Rossi_phd,hematite_col_syntheis,hematite_col_syntheis2,Rossi2021}   and maintain a permanent dipole moment even at large sizes (up to 15 $\mu$m)\cite{Philipse2018,lowrie_2007}. Apart from a new physical regime where steric forces  compete with magnetic forces, compared to ordinary ferromagnetic colloids, they provide also an opportunity to directly observe the different structures in colloids with an optical microscope \cite{Rossi_phd, hem_ellipsoids, hem_ellipsoids3, Philipse2018}. As hematite is a weak ferromagnetic at room temperature, thermal fluctuations play an important role. Hematite particles can form chains and rings, in which fluctuations are clearly visible  \cite{hem_ellipsoids,hem_ellipsoids2,hem_ellipsoids3}.

In this work we concentrate our study to colloids made of hematite particles with cubic shape. In scientific literature several experiments can be found with such colloids. It was demonstrated in \cite{Soni} that a two-dimensional chiral fluid can be created using hematite colloids. The hematite cubes in rotating magnetic field behave like two-dimensional fluid showing characteristic instabilities. In article \cite{Petrichenko_2020} swarming of micron-sized hematite cubes in a rotating magnetic field was examined. It was shown that in an external  rotating magnetic field particles form swarms which start to rotate. Experimental results for rotational speed were in a good agreement  with the proposed theoretical model. 
Targeted assembly and synchronization of self-spinning microgears or rotors made of hematite cubes and chemically inert polymer beads were demonstrated in\cite{Aubret_2018}.
Micron sized polymer colloids with embedded hematite cubes were used to demonstrate unstable fronts and motile structures formed by microrollers in \cite{Driscoll_2016}. Potential application of hematite colloidal cubes for the enhanced degradation of
organic dyes was investigated in \cite{Castillo_2014}.
The sedimentation of hematite
cubes and their crystal structure were investigated in \cite {Meijer_2013, Rossi2, Meijer2015}. Formations of light activated two-dimensional “living crystals” were examined in \cite{Palacci_2013}.

\begin{figure}[htbp]
\includegraphics[width=0.99\columnwidth ]{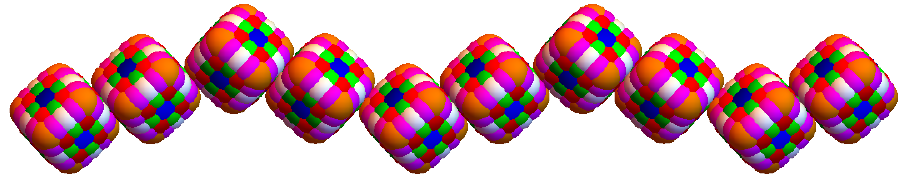}\\[2ex]
\includegraphics[width=0.99\columnwidth ]{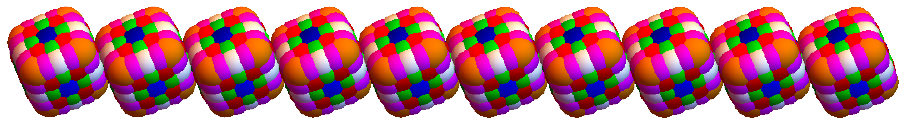}\\
\caption{Chain with kinks (top) vs straight chain (bottom).  
The top chain has four kinks. 
}
\label{fig:kinks1}
\end{figure}

\begin{figure}[h]
\includegraphics[width=0.48\columnwidth]{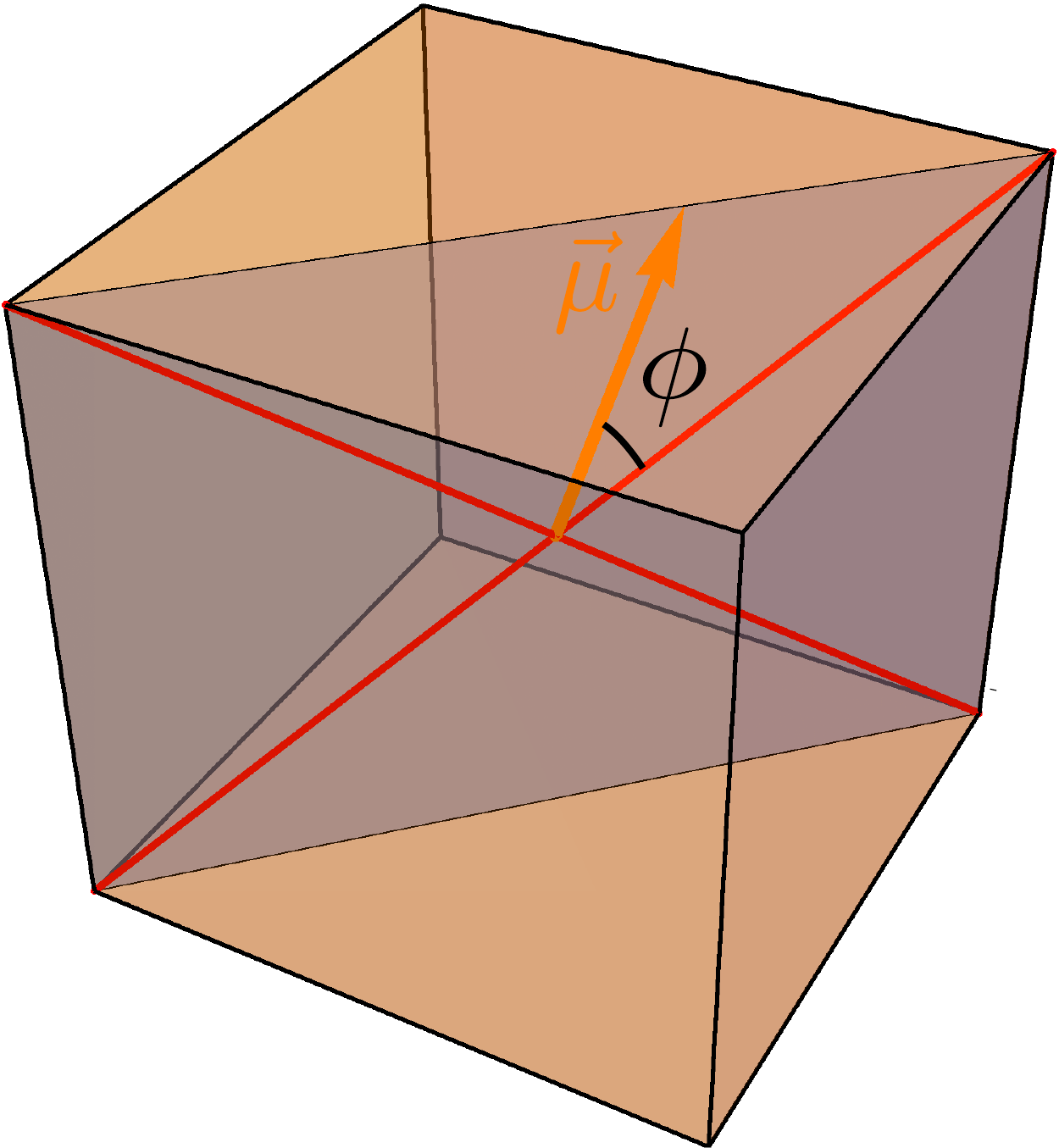}\hfill
\includegraphics[width=0.48\columnwidth]{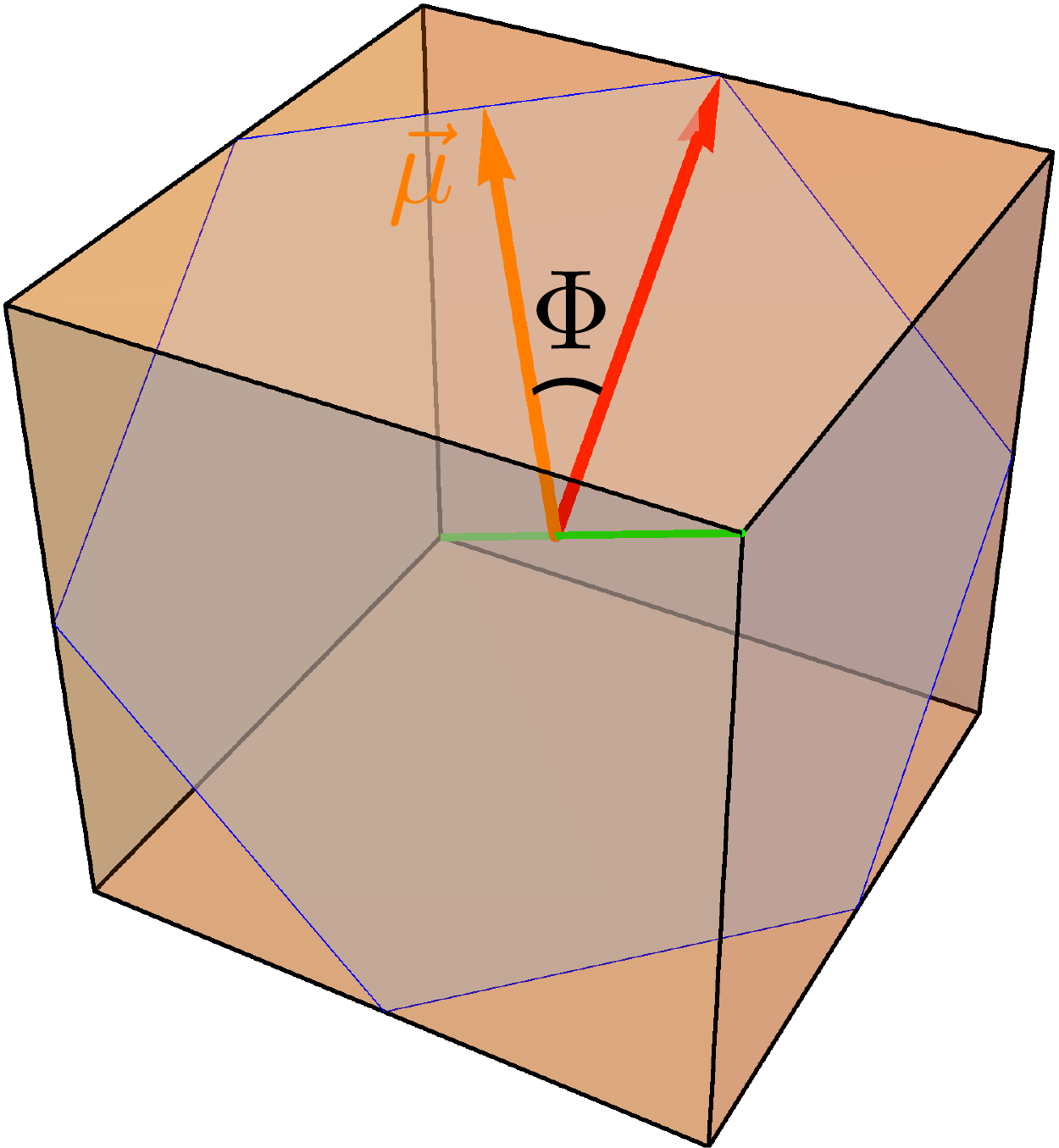}
\caption{Orientation of the magnetic moment $\vect \mu$. The left figure corresponds to the first case and the right to the second case of orientation of magnetic moment. In the first case we choose the angle $\phi$ to be positive if $\vect \mu$ points to the face and negative if it points to the edge.}
\label{fig:cube}
\end{figure}

Particularly interesting are experiments about structures of magnetic particles at low particle concentration in external magnetic field \cite{Rossi_phd, Philipse2018} as they provide building blocks for understanding behavior at higher concentrations\cite{Soni}. The magnetic particles in a colloid tend to align and form straight chains along the direction of the applied field. An increase in the strength of the applied magnetic field causes an additional rearrangement of the chains. Chains reorganize in the kinked structures (see Fig.~\ref{fig:kinks1} for illustration).  The Langevin molecular dynamics simulation was carried out in \cite{Philipse2018} for several orientation angles of the magnetic moment.  From the statistics of chains with kinks it is concluded that the amount of kinks is less pronounced at the magnetic moment orientation perpendicularly to the main diagonal in comparison with the case when the magnetic moment makes an empirical angle $12^{\circ}$ with the main diagonal. However, direct calculations of the energy of kinked configurations for these two orientations of magnetic moment were not carried out.  
In the literature, the orientation of the magnetic moment $\vect \mu$ in a hematite cube is under debate. Some say that there is an empirical angle $12^{\circ}$ with the diagonal in the plane defined by two diagonals \cite{Philipse2018} (Fig.~\ref{fig:cube} [left]). Others  claim that the magnetic moment is perpendicular to a diagonal of a cube stating that the moment is in the plane defined by the light blue hexagon in Fig.~\ref{fig:cube} [right] \cite {hematite1,hematite2,hematite3}. We try to solve this dispute and investigate the broader range of magnetic moment orientations to search for energetically favorable configurations. We find that there is a range of orientation angles where the kinked configurations are energetically favorable.

In this paper we are trying to understand how big of a role the energetically favorable structures play in the explanation of the observed kinked structures. We do this by searching and analyzing the energetically favorable configurations for different magnetic moment orientations. We speculate that the energetically favorable configurations should play a larger role in the explanation of kink formation in hematite chains.

For our investigation we choose to examine configurations for every orientation of magnetic moment mentioned in literature. As the first case we call a situation when the moment is in the plane defined by two diagonals and makes an angle $\phi$ with the diagonal (see Fig. \ref{fig:cube} [left]). Due to symmetries one actually has to examine a range $\phi\in[-\arctan(\frac{\sqrt{2}}{2}), 90^\circ-\arctan(\frac{\sqrt{2}}{2})]$. The angles $\phi=-\arctan(\frac{\sqrt{2}}{2})\approx -35^\circ$ and $\phi=90^\circ-\arctan(\frac{\sqrt{2}}{2})\approx 55^\circ$ correspond to the case when the magnetic moment is pointing to the midpoint of an edge or aligned perpendicular to the face of the cube respectively.  As the second case we call a situation where the magnetic moment is perpendicular to the cube's diagonal  and makes an angle $\Phi$ with a vector pointing from the center of the cube to the midpoint of an edge (see Fig. \ref{fig:cube} [right]). Due to symmetries it is only necessary to explore the region where $\Phi\in[0^\circ, 30^\circ]$. Note that in the given interval of angles there are exactly two situations ($\phi=\arcsin(\frac{1}{3})\approx19^\circ$ or $\Phi=30^\circ$ and $\phi=-\arctan(\frac{\sqrt{2}}{2})$ or $\Phi=0^\circ$) when the orientations of the magnetic moment are the same.

The content of this paper is divided into four sections. The Sec.~\ref{sec:intro} is introduction followed by a Sec.~\ref{sec:theor} where theoretical methods are described. The results are given in Sec.\ref{sec:res} and conclusions and discussion in Sec.~\ref{sec:concl}. This article  contains also two appendices. The Appendix~\ref{sec:app_A} demonstrates approximations used to construct superball out of spherical particles. In the Appendix~\ref{sec:app_B} it is shown how analytically energy is minimized  in the case of two cube chains and in the Appendix~\ref{sec:app_C}  error introduced by the dipole approximation is estimated.

\section{Theoretical methods}
\label{sec:theor}

To find energetically favorable structures we use two different methods. One is direct analytical/numerical energy minimization and the other is molecular dynamics (MD) simulations with simulated annealing.

To find energetically favorable structures  of hematite cubes in an external magnetic field using energy minimization one has to minimize the total energy of the system with constraints that cubes do not overlap. Hematite has a larger density $\rho_h=5.25\,\mathrm{ g/cm^3}$  than the solvent (water in \cite{Philipse2018} and our experiments, $\rho_s=1.00\,\mathrm{ g/cm^3}$), thus gravity effect is larger than buoyancy. Therefore, hematite particles may sediment. Gravitational length has to be calculated to check whether this happens  \cite{grav_len}
\begin{equation}
 l_g=\frac{k_B T}{(\rho_h-\rho_s)g a^3}, 
\end{equation}
where $k_B$ is Boltzmann constant, $T$ is temperature, $g$ is gravitational acceleration and $a$ is the edge length of a hematite cube. For the case of $a>1\,\mathrm{\mu m}$ and $T=300\,\mathrm{K}$ one obtains $l_g<0.1 a$. This means that all particles sediment and form 2D structures unless they are formed during sedimentation process and do not disassemble during collision with the bottom surface of a capillary. 

The total energy $E_{tot}$ of the hematite cube system is the sum of magnetic energy $E_{mag}$ and gravitational energy $E_g$.
If we assume that hematite cubes have a permanent magnetic dipole moment $\vect{\mu}_i$ and cubes are in a homogeneous external magnetic field $\vect{B}$ then magnetic energy of this system can be written as  
\begin{equation}
E_{mag}=\sum_i \sum_{j>i}W_{ij}-\sum_i \vect{\mu}_i\cdot\vect{B}, 
\end{equation}
where $W_{ij}$ is the magnetic interaction energy between two hematite cubes. For interaction energy we use the dipole approximation as it significantly accelerates computations, make it feasible to do analytical calculations, and typically introduces an error which  qualitatively  does not change any conclusions:
\begin{equation}
W_{ij}=\frac{\mu_0}{4\pi}\iek{\frac{\vect \mu_i\cdot\vect\mu_j}{r_{ij}^3}-\frac{3(\vect\mu_i\cdot\vect r_{ij})(\vect\mu_j\cdot\vect r_{ij})}{r_{ij}^5}}, 
\label{eq:dip_e}
\end{equation} 
where $\vect r_{ij}$ is the radius vector between centers of $i$-th and $j$-th cubes. Details of error estimation introduced by the dipole approximation are given in Appendix \ref{sec:app_C}.

Gravitational potential energy depends on the orientation of cubes. Hematite has a permanent magnetization $M=2.2\times10^3\,\mathrm{A/m}$ ($\mu=Ma^3$), thus for a two cube chain $E_g\leq(\rho_h-\rho_s)g a^4$, it is an order smaller quantity as compared to the dipole magnetic energy \eqref{eq:dip_e}. Therefore, gravitation is not explicitly included in analytical calculations of energy. We assume that all cubes have sedimented, calculate energetically favorable structures and reorient the calculated configuration so that the gravitational potential energy is minimal. We assume that every hematite cube touches the bottom surface of a capillary. If all cubes in the chain touch with faces, this is equivalent to a statement that all centers of cubes in the chain are at the same height above the bottom surface of a capillary. This statement is checked by comparing results of MD simulations with gravity effects included and quasi two-dimensional MD simulations with only magnetic and steric forces enabled. In quasi two-dimensional simulation all cubes are allowed to move only in the plane (can move in $x$, $y$ directions, but $z$ coordinate is fixed) but all rotations are allowed. For comparison we include also the three-dimensional MD calculations without explicit gravity treatment. Here  only steric and magnetic forces are explicitly taken into account, but the final configuration is rotated around axis parallel to the magnetic field in a way that the potential gravitational energy is minimal. In MD calculations this is equivalent to a case where particles can move and rotate in all directions (no constraints apart from overlapping).

\subsubsection{Energy minimization}

In our investigations we assume that all cubes are identical. All of them have the same edge length $a$ and the same orientation and magnitude of the magnetic moment $\mu_i$. Thus, it is beneficial to introduce dimensionless quantities: $\tilde{ \vect{r}}_{ij}=\frac{\vect r_{ij}}{a}$, $\tilde{\vect{\mu}}_{i}=\frac{\vect \mu_{i}}{\mu_m}$, $\tilde E_{tot}=\frac{4\pi E_{tot} a^3}{\mu_0 \mu_m^2}$, $\tilde{\vect{B}}=\frac{4\pi B a^3}{\mu_0 \mu_m}$.

The equation for the total energy, which has to minimized, in dimensionless units reads:
\begin{equation}
\begin{split}
\tilde E_{tot}=&\sum_i \sum_{j>i}\iek{\frac{\tilde{\vect \mu}_i\cdot\tilde {\vect  \mu}_j}{\tilde r_{ij}^3}-\frac{3(\tilde{\vect  \mu}_i\cdot\tilde{\vect  r}_{ij})(\tilde{\vect\mu}_j\cdot\tilde{\vect r}_{ij})}{\tilde r_{ij}^5}}+\\
&-\sum_i \tilde{\vect{\mu}}_i\cdot\tilde{\vect{B}}. 
\end{split}
\label{eq:eos}
\end{equation}

The direct approach to minimize energy \eqref{eq:eos} for a given $\phi$ or $\Phi$ and $\tilde{\vect{B}}$ with a condition that no particles overlap requires a nonlinear optimization with nonlinear constrains. This is computationally too demanding for systems with more than five particles. Fortunately, the computational costs can be reduced significantly by a simple consideration. If two cubes touch, then they touch with faces (touching faces are parallel). This fact is confirmed by MD simulations described in the next subsection. Such constraints lead to aggregates arranged in a simple cubic lattice or formation of chains (can be with kinks). However, this is not the case for other structures which are not energetically favorable. Further in this paper we give examples were cubes touch with edges or with a face and a vertex (see e.g. top figure of Fig.~\ref{fig:kinks1}).  

\begin{figure}[h]
\includegraphics[width=0.48\columnwidth]{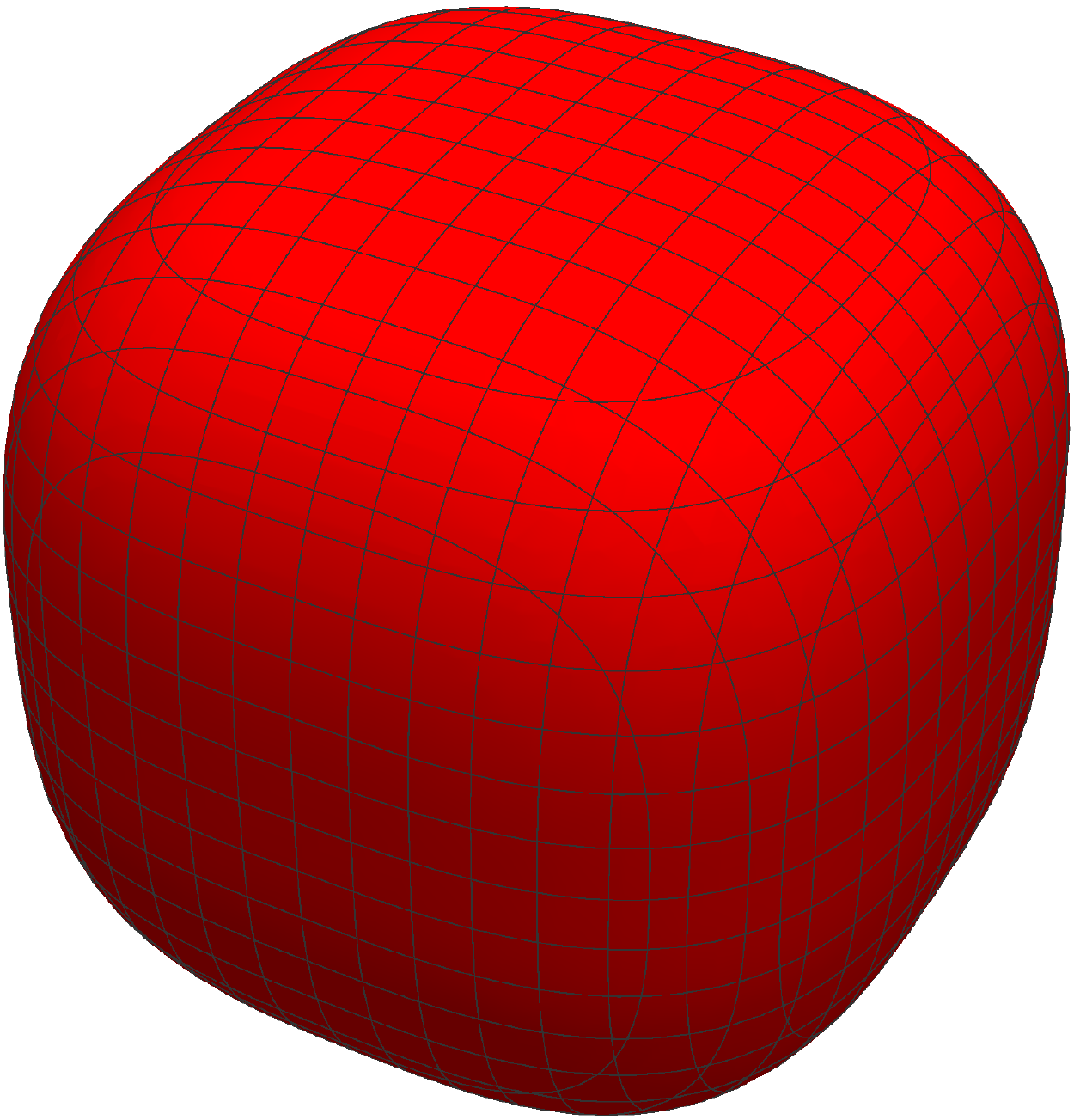}\hfill
\includegraphics[width=0.48\columnwidth]{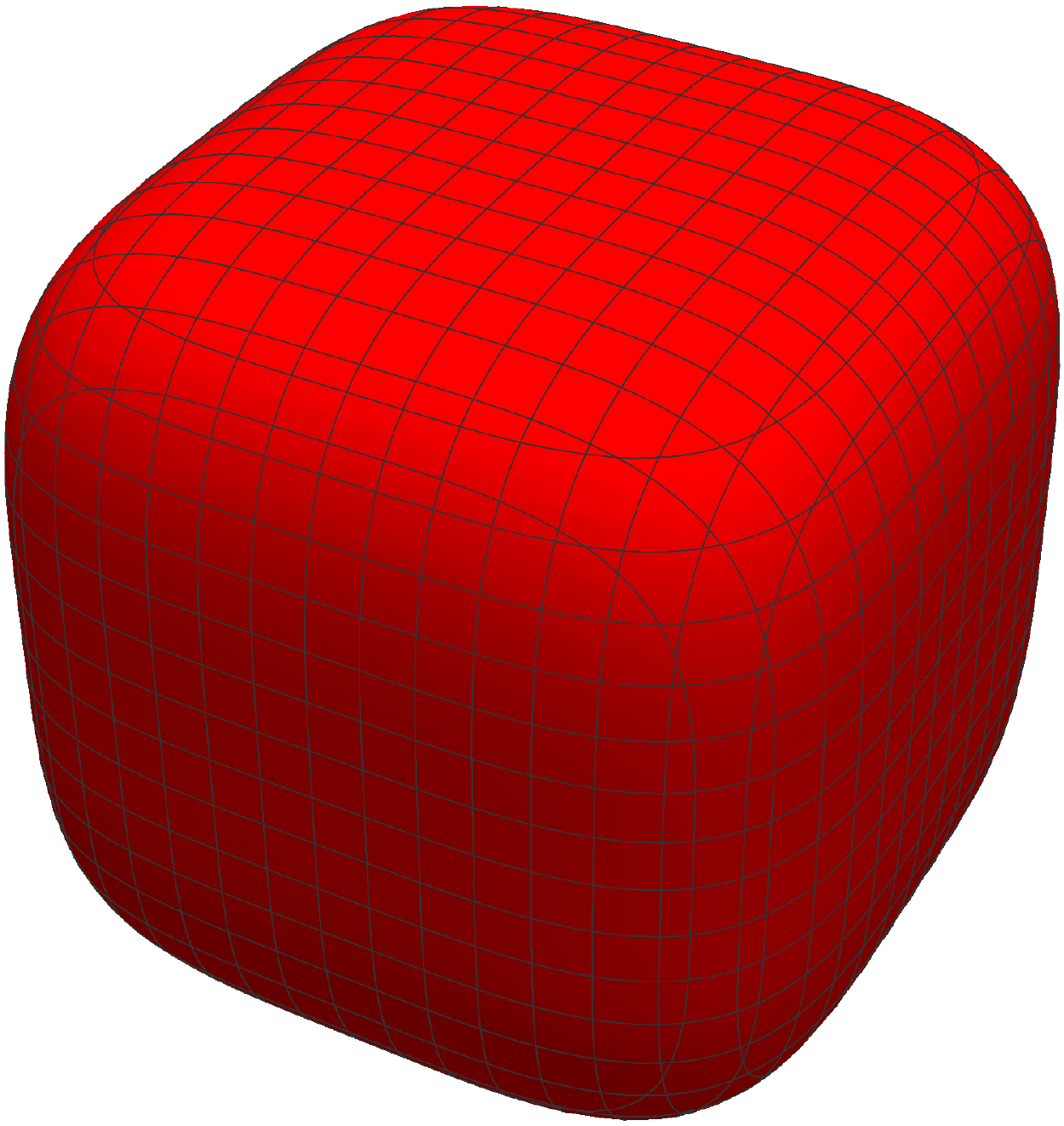}
\caption{Visualization of superballs with different shape parameters~$q$. On the left $q=1.5$ and on the right $q=2$.}
\label{fig:sb}
\end{figure}

To compare results with experiments it has to be taken into account that in experiment the shape of cubic particles in colloid is not a perfect cube, but rather a cube with rounded corners also know as a superball (see Fig. \ref{fig:sb}). Shape parameter $q$ describes how much the corners are rounded. For hematite cubes it typically is
 $q\in[1.5,2]$ \cite{Rossi2,Meijer2015}. Note that in the literature often a shape parameter $m=2q$ is used to describe a superball instead of the shape parameter $q$. The superball with an edge length $a$ in a particle centered coordinate frame is defined by:
\begin{equation}
 \abs{\frac{2x}{a}}^{2q}+\abs{\frac{2y}{a}}^{2q}+\abs{\frac{2z}{a}}^{2q}\leq1\,.
\end{equation}
The value of a shape  parameter $q=1$ corresponds to a sphere and the limit $q\rightarrow\infty$ corresponds to a cube, thus values in between are cubes with more or less rounded corners as can be seen from Fig.~\ref{fig:sb}. Visually looking at a superball with $q=2$ we see that it is already a good approximation to a cube and vice versa. However, it is not clear how large differences this approximation introduces e.g. for the energetically favorable configurations. Thus, parallel to an energy minimization of cubic particles also calculations with superball shaped particles with different  parameter $q$ are performed.

\subsubsection{Molecular dynamics simulations}
\label{Sec:md}

\begin{figure}[h]
\includegraphics[width=0.48\columnwidth]{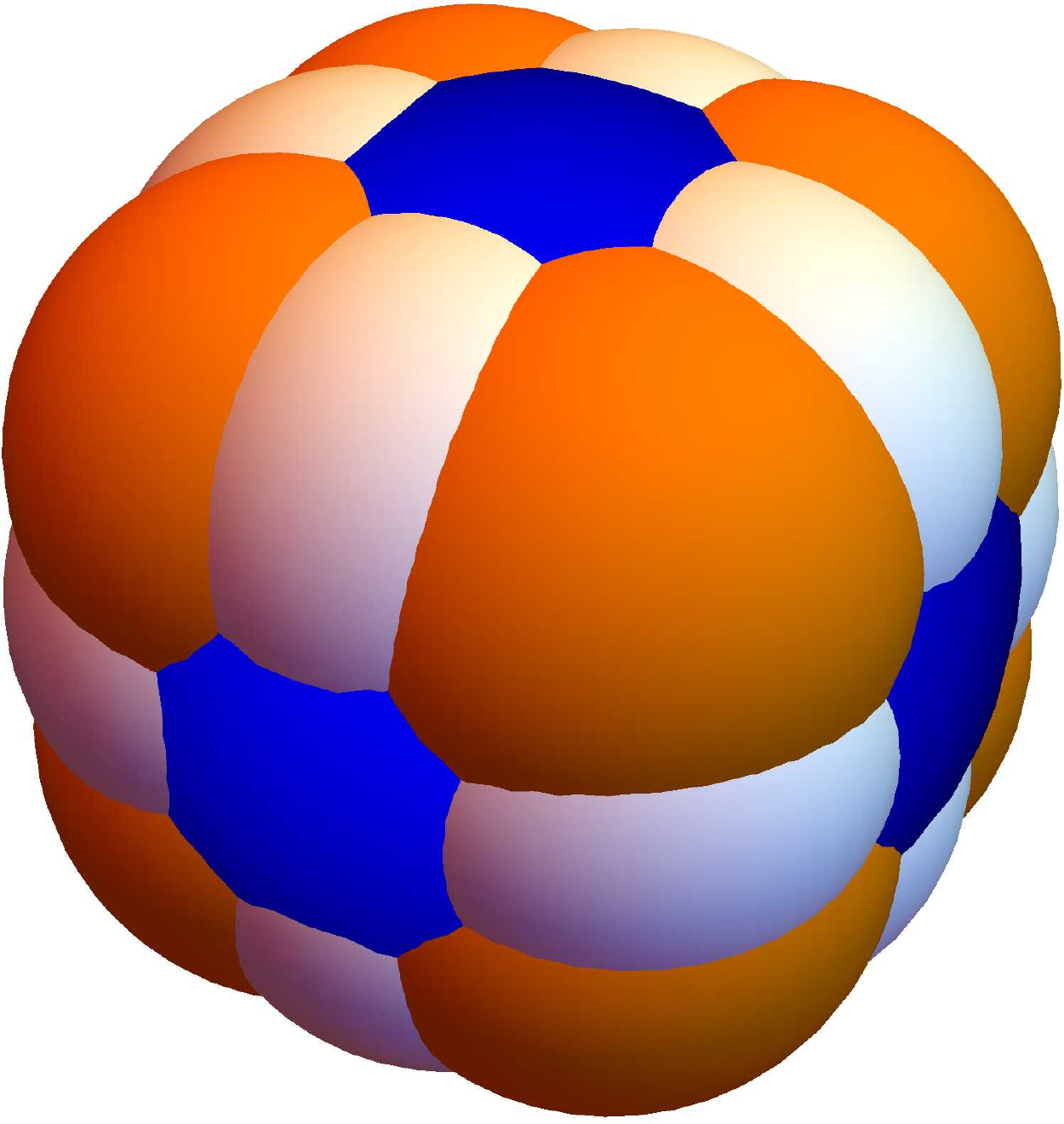}\hfill
\includegraphics[width=0.48\columnwidth]{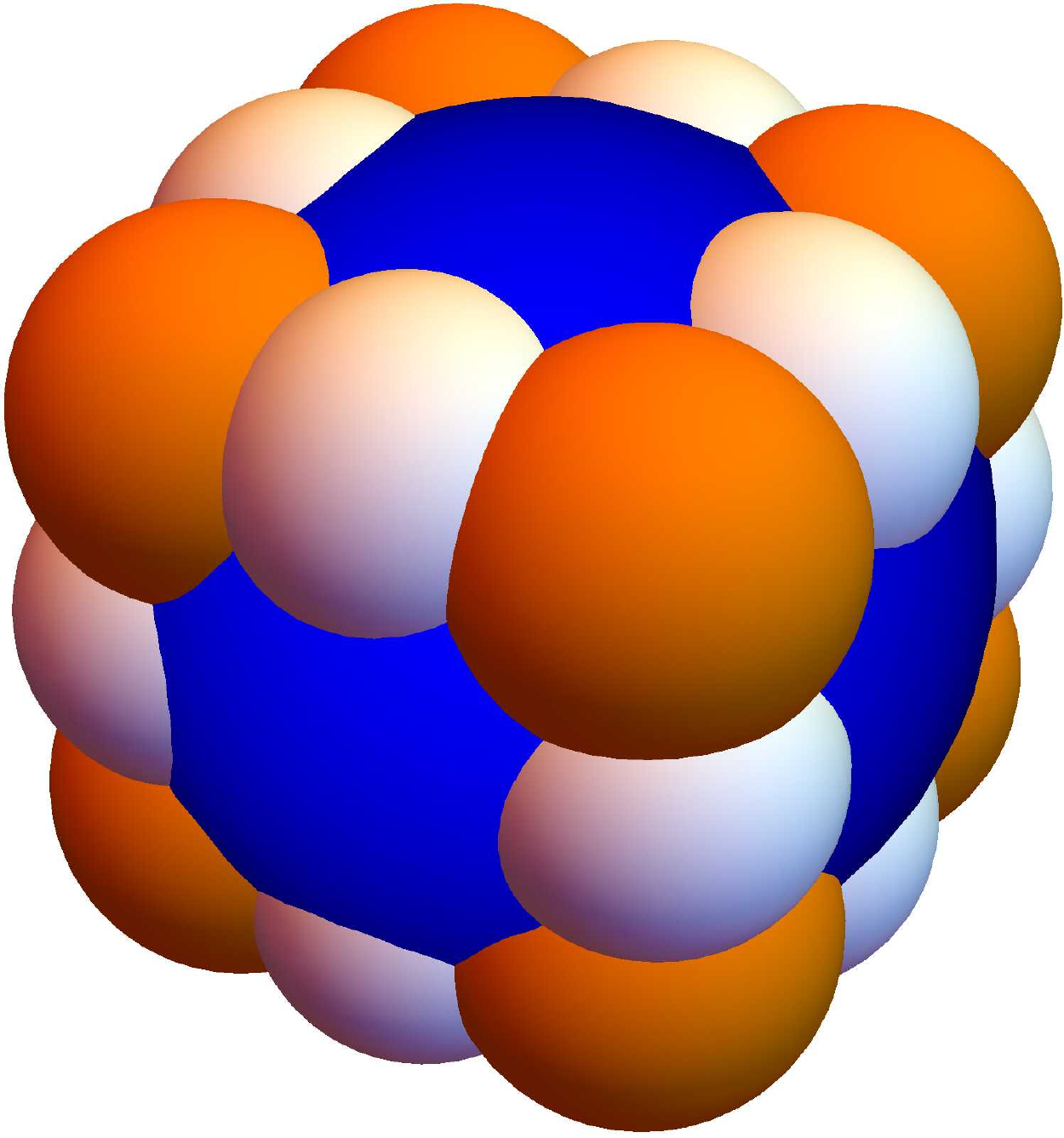}
\caption{Superball with $q=1.5$ (left) and $q=2.0$ (right) approximated with 21 spheres, as proposed in \cite{Kantorovich}.}
\label{fig:sball}
\end{figure}
Every physical model has to be benchmarked, however, in real experiments the orientation of the dipole moment  is unknown and it is thus hard to benchmark the results. Numerical computer experiments on the other hand do not have this ambiguity and thus are preferred over real experiments. Probably the most suitable method for this task is molecular dynamics (MD) simulations. For this purpose we chose to use the computer program ESPResSo 4.1.2   \cite{Espresso, Espresso2, Espresso3} as it is well optimized, calculations are easily parallelizable and  magnetic interactions are included. The only difficulty is that in ESPResSo 4.1.2 the steric repulsions are implemented only for spherical or elliptical particles but not for cubes and superballs with $q\in[1.5,2]$. To overcome this limitation those shapes  have to be constructed out of spherical particles. To construct a cube out  of spheres one can use an approach proposed in \cite{Kantorovich2}, however, for a decent approximation one needs about 100 spheres. It is easier to construct cubes with rounded corners - superballs, especially with small values of shape parameter $q$. Particularly elegant solution for this problem is demonstrated in \cite{Kantorovich} where for $q\leq1.5$ it was proposed to do this with only 21 spheres (one real particle and 20 virtual particles). Virtual particles in this case have no mass, charge and magnetic moment, but interact with all other particles with streric interaction (do not allow overlapping). To calculate the motion of real particles in ESPResSo, the total force and torque acting on the real and all virtual particles belonging to that real particle are calculated and summed and then trajectory is determined. Virtual particles on the other hand are moved according to the motion of the corresponding real particle.

 As it can be seen by comparing Fig.~\ref{fig:sb} and Fig.~\ref{fig:sball}, this approximation is rather good for $q=1.5$, but becomes worse with increasing $q$. To overcome this,  extra spheres have to be added in the approximation as the curvatures at vertices and edge midpoints become larger and radii of spheres thus smaller. To simulate superballs with $q=2.0$ we proceeded the idea proposed in \cite{Philipse2018} and added two extra spheres per edge. This then gives 45 spheres in total (for details see appendix\ref{sec:app_A}). To check whether 45 sphere approximation is sufficient we compare it to an approximation with 93 spheres.

To find the energetically favorable configurations with ESPResSo 4.1.2 we are using simulated annealing. In other words this means that the real particles are propagated by solving Langevin equations and temperature is slowly decreased. Note that for a full model the gravity and the flat bottom surface of a capillary are explicitly taken into account. The Langevin equation for transnational components in our case reads:
\begin{equation}
 m\frac{\mathrm{d} \vect{v}}{\mathrm{d} t}=-\vect{\nabla}(E_{tot}+V_{LJ}) -\gamma \vect{v}+\sqrt{2\gamma k_B T}\vect{\xi}(t),
\end{equation}
where $m$ is mass, $\gamma$ is viscosity, $k_B$ is is Boltzmann's constant, $T$ is the temperature, $\xi(t)$ is a white noise, and $V_{LJ}$ is a repelling potential making sure that no two cubes overlap. For steric repulsion we are using Weeks-Chandler-Anderson potential \cite{WCA}.

\section{Results}
\label{sec:res}

We used  two methods (energy minimization and MD simulations) to determine the energetically favorable configurations. To compare these two methods we performed experiments with our experimental system, similar to \cite{Philipse2018}, where the external magnetic filed is applied in the plane perpendicular to the free fall acceleration direction. From the result comparison we conclude that qualitatively the same configurations are obtained with both methods. Quantitatively there are small differences  due to approximations used. These differences will be discussed later in this section.  MD calculations for high fields give the same results with and without explicit gravity treatment. In the case without explicit gravity treatment it is assumed that particles sediment on a flat surface. This confirms that no extra error is introduced using this assumption. 

Also for most angles at strong magnetic fields the  theoretically calculated configurations using quasi two-dimensional and three-dimensional calculations without explicit gravity treatment are identical. For three-dimensional calculation there are no restriction apart form no overlapping of particles. In quasi two-dimensional calculation apart from no overlapping  it is  also required that particles touch the bottom surface of a capillary. In fact, the obtained configurations are the same whenever energetically favorable configuration is a straight chain without kinks in the no gravity case. For small magnetic fields in this case there are some differences, however, they are so small that it is not possible to  measure them experimentally or resolve in MD simulations. In MD calculations for high fields there are small differences, but they arise due to the superball approximation.

One has to note that simulated annealing simulations have to be repeated many times to find the global minimum as the system may get stuck in a local minimum if cooled too fast. In general, for longer chains it is harder to find the global minimum as there are more local minima.  In all cases of the moment orientation, when gravity is taken into account, the energetically favorable configurations are chains of cubes. This, however, is not the case when there is no gravity. When $\phi\approx 0^\circ$, i.e. magnetic moment is along a diagonal of a cube ([111] direction), we confirm the findings of \cite{Aoshima, Kantorovich2, Hematite_along_diogonal, Linse}. In this case without gravity cubes should form regions with closely packed 2D cubic structures. 

\begin{figure}[htbp]
\includegraphics[width=0.48\columnwidth]{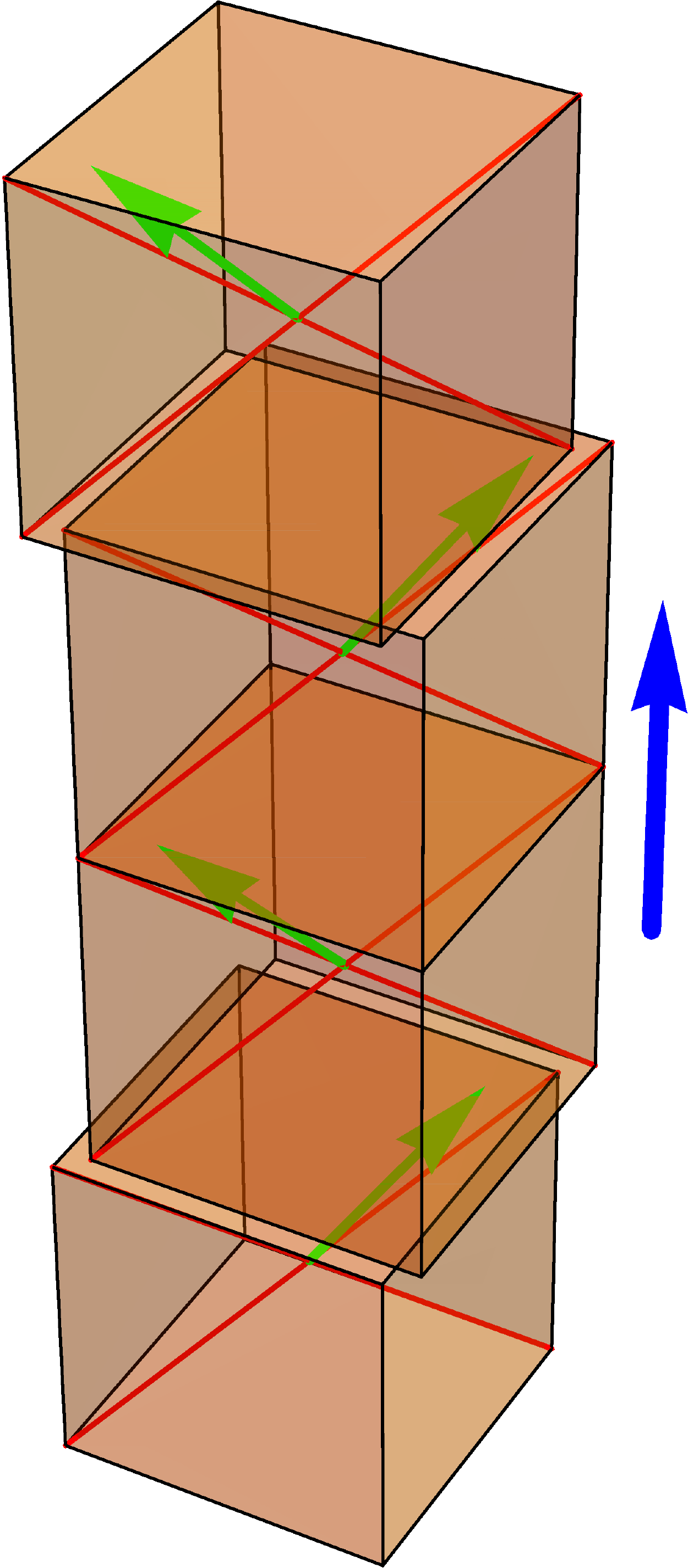}\hfill
\includegraphics[width=0.48\columnwidth]{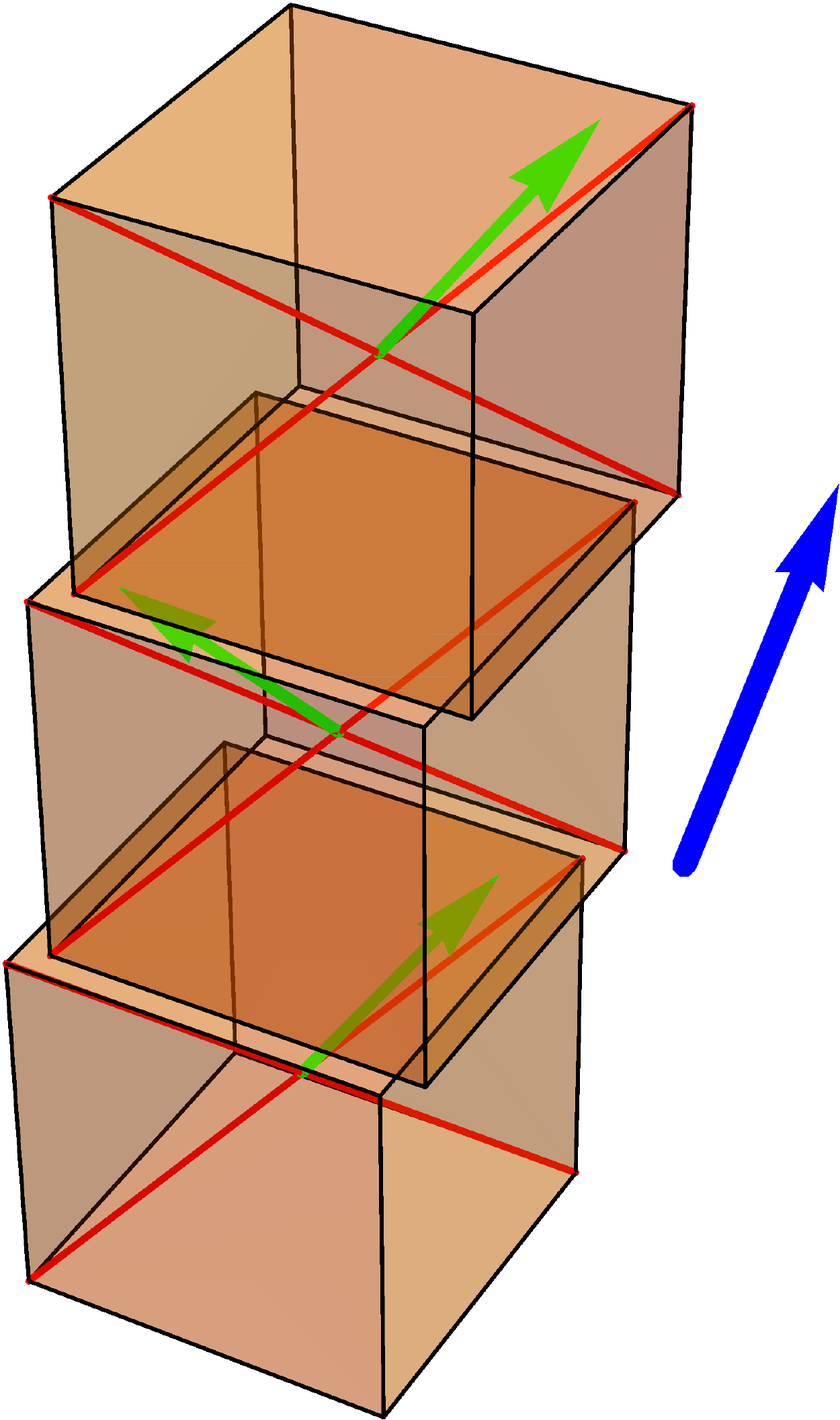}
\caption{Energetically favorable configurations for chains of four and three cubes in the case of small magnetic fields and absent gravity. The chains are straight and the shifts (how much the centers of neighboring cubes are shifted as compared to the case where cubes are on top of each other ) are \textbf{increased 100 times} to make them visible. The orientation of the magnetic moment is $\phi=$12$^{\circ}$.}
\label{fig:no_field}
\end{figure}

\begin{figure}[htbp]
\includegraphics[width=0.58\columnwidth]{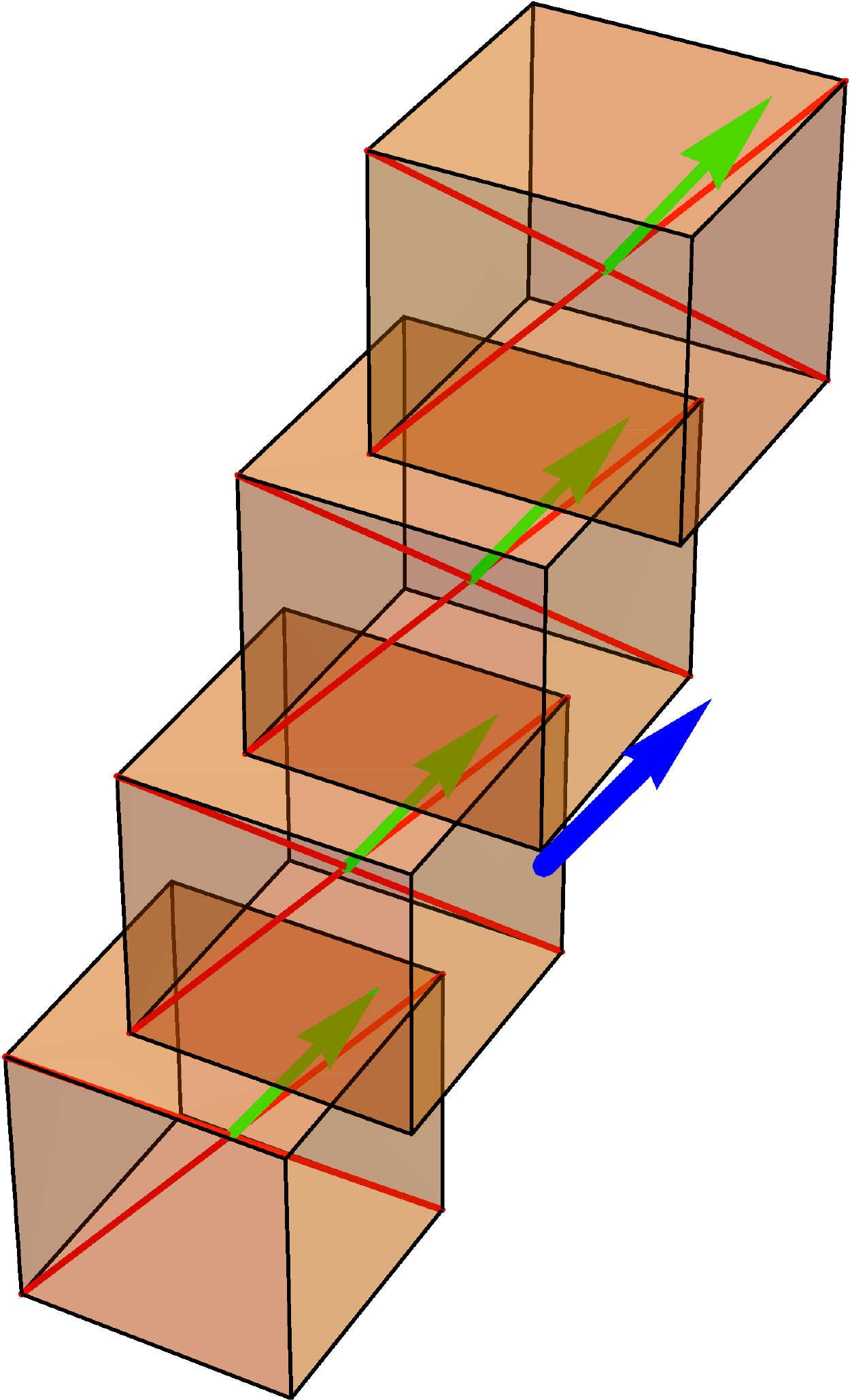}\hfill
\includegraphics[width=0.39\columnwidth]{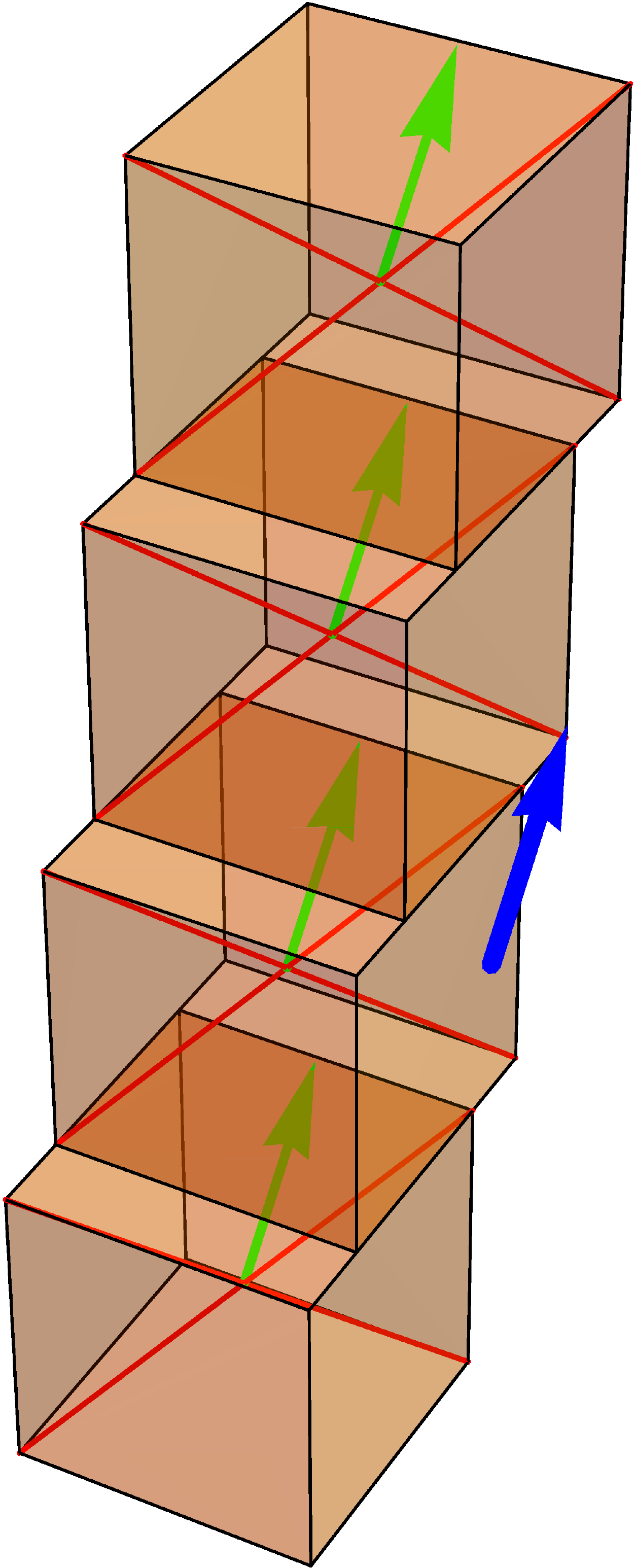}
\caption{Energetically favorable straight chains of four cubes in the case of strong magnetic fields and without gravity. Case 1 on the left ($\phi=12^\circ$). Case 2 on the right ($\Phi=0^\circ$).}
\label{fig:field}
\end{figure}

At small values of external field (or no external field) we observe  straight chains where each next cube is almost exactly on top  of the previous cube  (see Fig.~\ref{fig:no_field}) similar as in experiment \cite{Philipse2018, Rossi_phd}. Note that the phrase ``on top'' is applicable only to Fig.~\ref{fig:no_field}. In the case of gravity, which is the case in experiments, one has to rotate the chain such that the total energy is minimal. In practice, all cubes then touch the bottom surface with an edge or a face. For larger external magnetic field values the chains rearrange (see Fig.~\ref{fig:field}) and magnetic moments of each cube now are oriented parallel to the external magnetic field.

 Note that also values for $\Phi$ close to zero, i.e. in this case $\Phi\in[0^\circ, 2^\circ)$ and $\phi\in(-35^\circ, -32^\circ)$ (rounded to the nearest integer) the situation is slightly different. The energetically favorable chains for larger field values contain kinks (see Fig.~\ref{fig:kinks1} top). Thus, those cases are analyzed separately. Overall three different configurations are observed and are summarized in the table below:\vspace{0.2ex}
 
\begin{center}
 \begin{tabular}{@{}m{0.38\columnwidth}|m{0.6\columnwidth}@{}}\toprule
\textbf{Energetically favorable configuration} & \multicolumn{1}{c}{\textbf{When observed}}  \\ \midrule
Straight chain without shifts &  $B<B_{crit}$ \& $\forall\phi$, $\forall\Phi$; \linebreak  $\forall B$ \& $\phi\approx 55^\circ$  \\
\midrule
Straight chain with shifts & $B>B_{crit}$ \& $\phi\in(-32^\circ, 55^\circ)$; \linebreak $B>B_{crit}$ \& $\Phi\in[2^\circ, 30^\circ]$
\\
\midrule
Kinked chain &$B>B_{crit}$ \& $\phi\in[-35^\circ, -32^\circ)$; \linebreak $B>B_{crit}$ \& $\Phi\in[0^\circ, 2^\circ)$ \\
\bottomrule\end{tabular}
\end{center}

\subsection{Chains without kinks}

\begin{figure}[htbp]
\includegraphics[width=0.99\columnwidth]{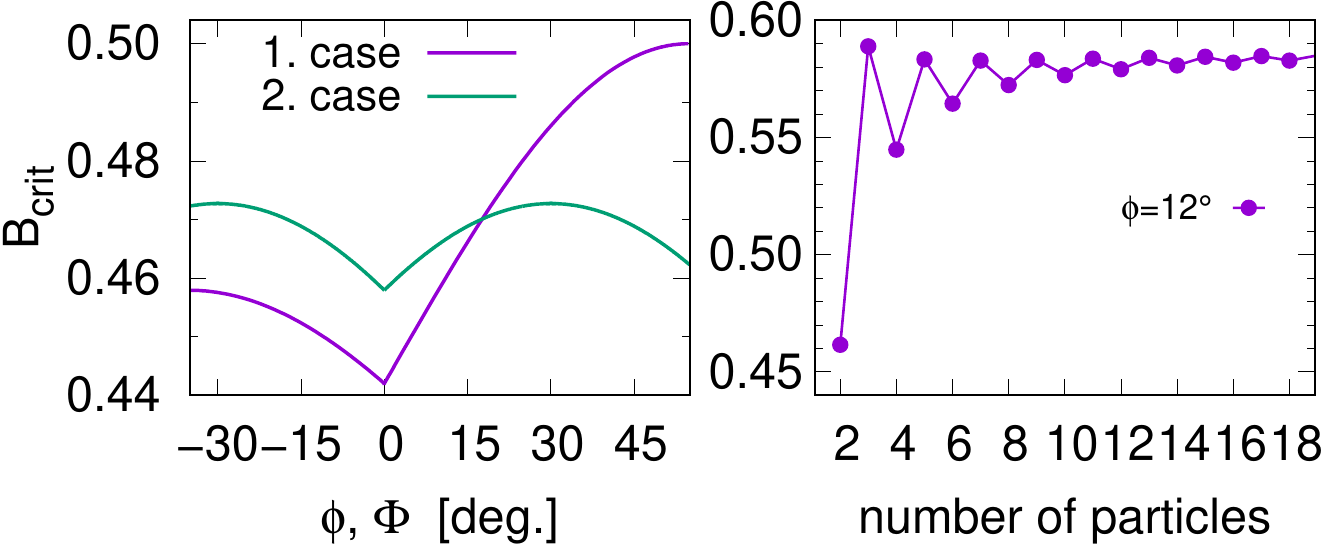}
\caption{The critical value of magnetic field  $\tilde B_{crit}$ for switching from the first to the second chain configuration dependence on the orientation of the magnetic moment and the length of the chain. On the right only results for $\phi=$12$^\circ$ are presented as all other orientations of moment show a similar behaviour.}
\label{fig:bcrit}
\end{figure}

From MD simulations and analytical/numerical calculations we obtain that in the case of $\phi\in(-32^\circ, 55^\circ)$ or $\Phi\in[2^\circ, 30^\circ]$  the energetically most favorable configuration is a straight chain and it does not depend on the magnitude of external magnetic field. If there is no gravity, then for $\phi\in(-32^\circ, 2^\circ)$ actually other configurations are energetically more favorable.  As already mentioned, two different chain configurations are observed. In one magnetic moments form zig-zag structure at low fields (below a critical value of the external magnetic field $B_{crit}$) as shown in Fig.~\ref{fig:no_field} and in the other magnetic moments are parallel to the external magnetic field at high field values (above critical value of the external magnetic field $B_{crit}$) as shown in Fig.~\ref{fig:field}. The critical magnetic field $B_{crit}$, as can be seen from Fig.~\ref{fig:bcrit}, depends on the orientation of the magnetic moment and the length of the chain. In dimensionless units it is in the range $\tilde B_{crit}\in (0.44; 0.7)$. For hematite $B_{crit}\in (97; 150)\,\mathrm{\mu T}$. This is larger, but comparable with the Earth's magnetic field. The critical value $B_{crit}$ itself only slightly varies with the angle of moment orientation. The dependence on the number of particles in chain is more pronounced. With an increased number of particles in the chain it converges to some value, although not monotonically. The critical value decreases by adding two particles to a chain with an odd number of particles,  but the critical value increases by adding two particles to a chain with even number particles.    

\begin{figure}[htbp]
\includegraphics[width=0.45\columnwidth]{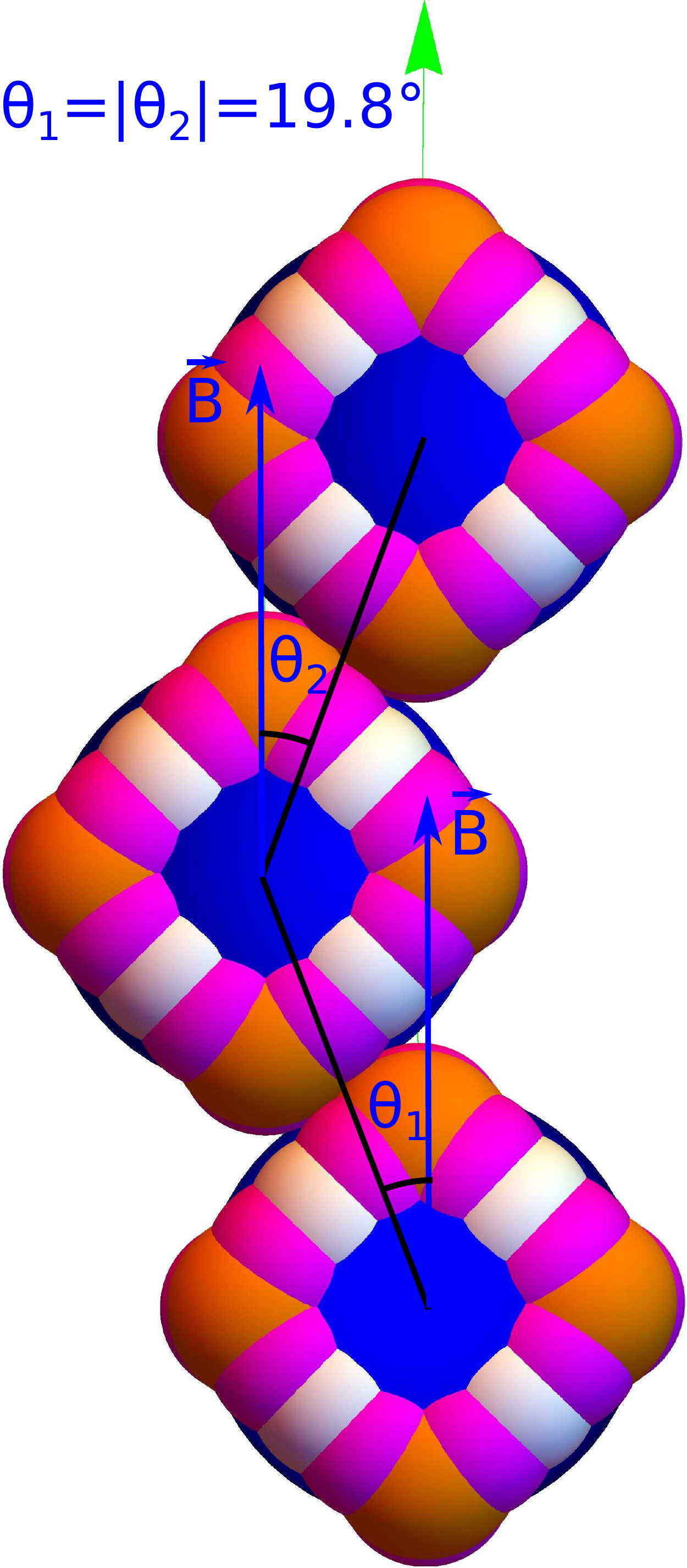}\hfill
\includegraphics[width=0.53\columnwidth]{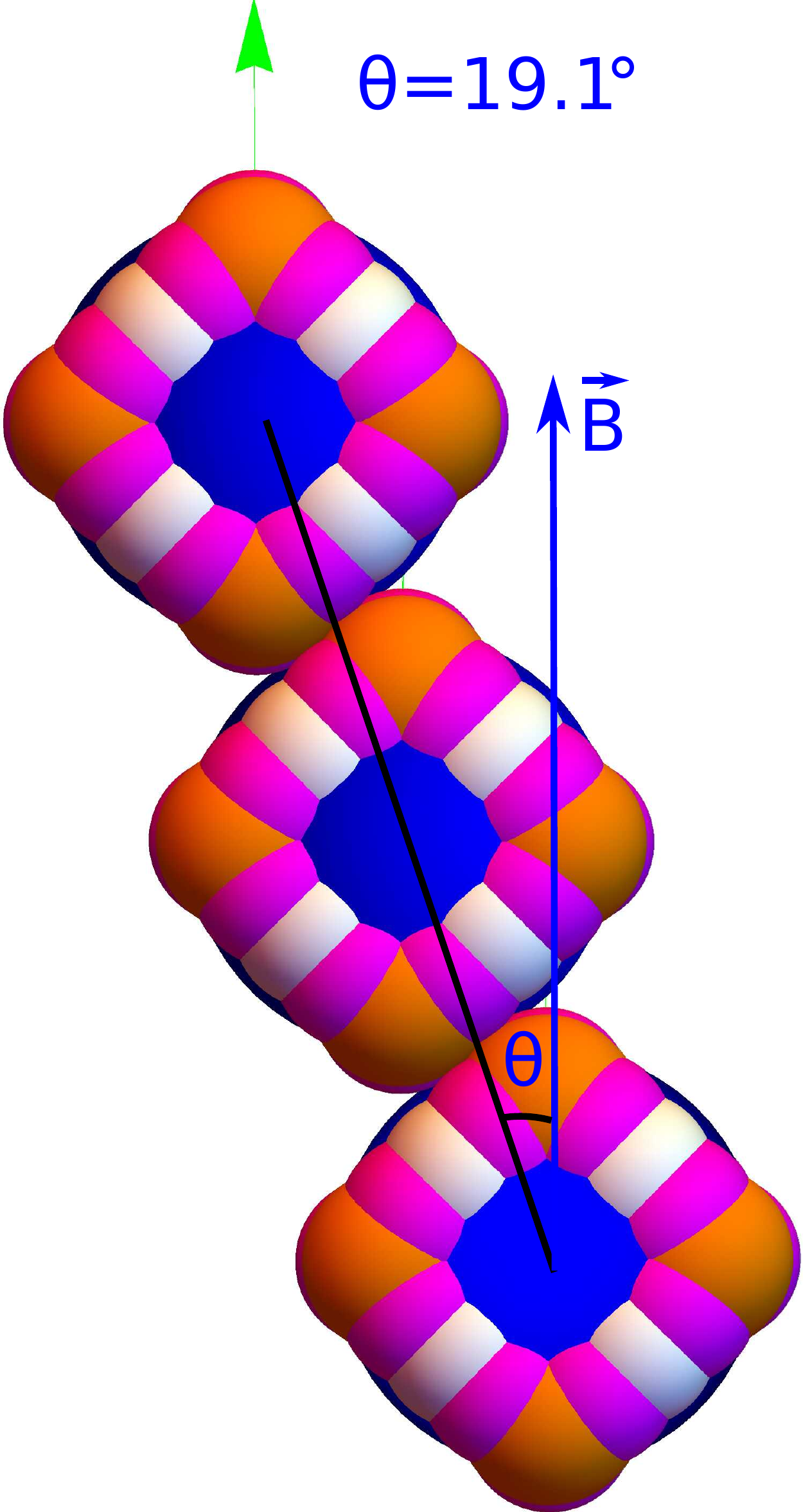}
\caption{Configurations obtained in the MD simulation with simulated annealing for $\Phi=$0$^{\circ}$ and fields above  $B_{crit}$. As we decrease the temperature with finite steps the global minimum not always is found. Sometimes only a local minimum is found as shown on the right.}
\label{fig:theta}
\end{figure}

For small field values without gravity the energetically favorable configurations are straight chains in both cases of moment orientation. Two touching cubes are almost on top of each other, but nevertheless the centers are slightly shifted. This difference we call shifts (also $b$ and $c$ parameters in Appendix \ref{sec:app_B}). Visually (not considering the orientation of magnetic moment) the chains consisting of even number of cubes are symmetrical against the midpoint of a chain and the chains containing odd number of cubes are anti-symmetric as one can see from Fig.~\ref{fig:no_field}. But these theoretically calculated shifts are tiny and can not be measured due to thermal fluctuation nor the resolution of a microscope. If one now adds gravity, these tiny shifts almost completely disappear.  The total magnetic moment of a chain is parallel to the external magnetic field $\sum_{i=1}^N \vect{m}_i \times \vect B=0$. The chains consisting of even number of particles are oriented along the magnetic field, but chains consisting of odd number of particles make some angle $\theta$  with the magnetic field (visually shown in Fig.~\ref{fig:theta}). This angle tends to zero with an increased number of particles. The individual magnetic moments form the so called zig-zag structures. Every next cube is rotated by 180$^\circ$ around the axis perpendicular to the face where cubes touche. This axis is parallel to the external magnetic field for chains with an even number of cubes. 

For high values of magnetic field the configuration is slightly different. In this case all individual moments are oriented along the external magnetic field. The shape (not considering the orientation of magnetic moment) of the configuration is anti-symmetric against the center of chain. In this case unlike at low magnetic fields it is independent of even or odd number of cubes in a chain. Unlike in the low field case the shifts for most cases are experimentally determinable. The shifts are also independent on the number of cubes in a chain, thus the angle $\theta$ which makes orientation of a chain with the external magnetic field is independent of the chain length. In the absence of gravity the energetically favorable configurations are shown in Fig.~\ref{fig:field}.
 \begin{figure}[htbp]
\includegraphics[width=0.99\columnwidth]{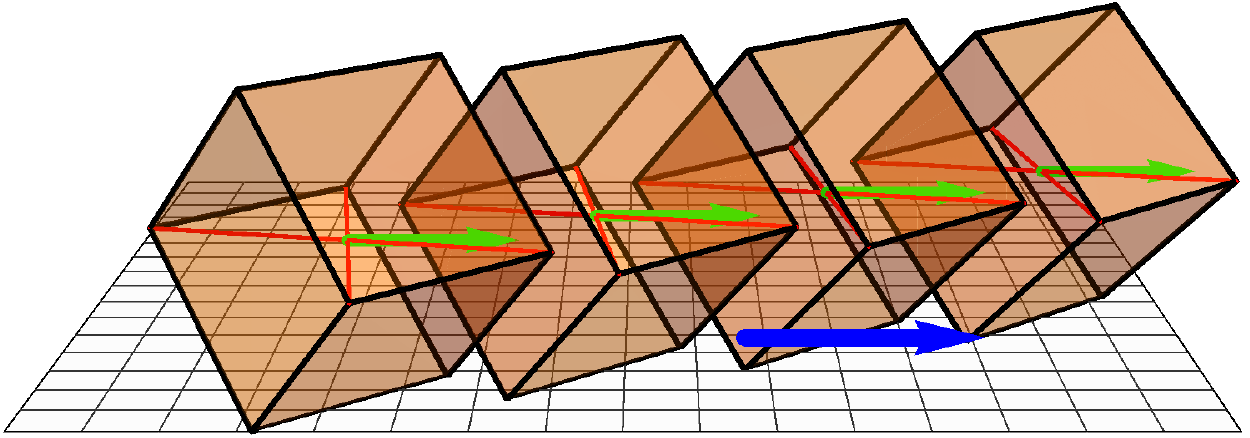}
\includegraphics[width=0.99\columnwidth]{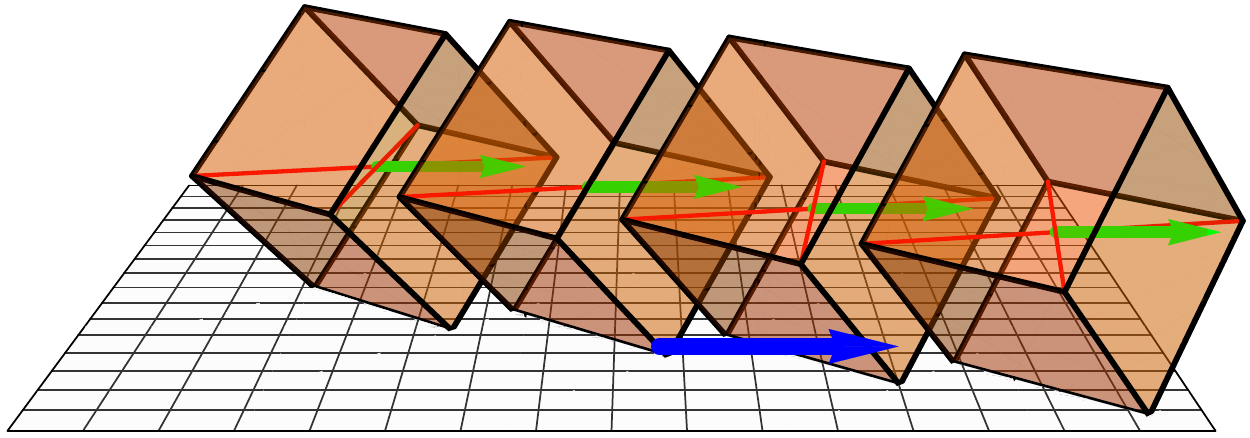}
\caption{Energetically favorable straight chain  configurations of four cubes with gravity  and $\phi=12^\circ$ in the case of strong magnetic fields. The chain can be aligned in two ways having the same energy.}
\label{fig:field_gravity}
\end{figure}

\begin{figure}[htbp]
\includegraphics[width=0.8\columnwidth]{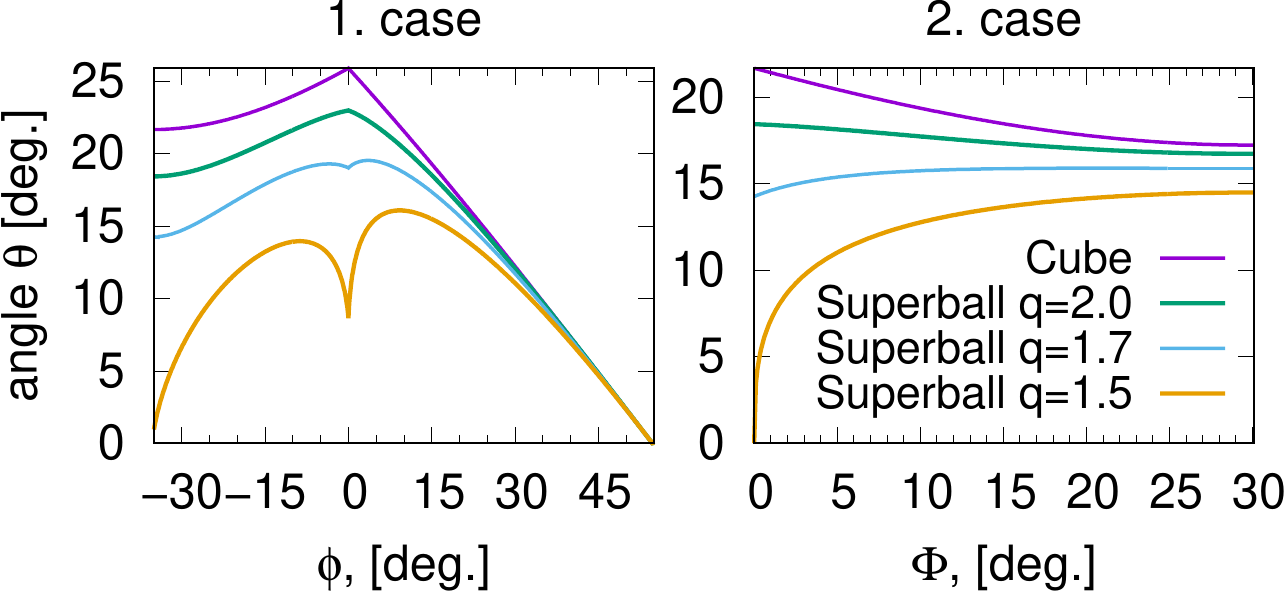}
\caption{Dependence of positive angle $\theta$ on different orientations of magnetic moment for superball particles with different shape factor $q$. The cube corresponds to the limit when $q\rightarrow\infty$. }
\label{fig:theta_}
\end{figure}

The obtained configurations in the case without gravity can be rotated around any axis parallel to the external magnetic field and the total energy does not change. In the case if magnetic field is in the plane perpendicular to the gravity it turns out that no additional calculations are  necessary in both cases of magnetic moment orientation. The most energetically favorable configuration of the straight chain if there is no gravity has to be found  and the obtained configuration has to be rotated  around this axis such that gravitational potential energy is minimal. The obtained configuration for $\phi=12^\circ$ is shown in Fig.~\ref{fig:field_gravity}. One has to note that energy does not change if one rotates the obtained configuration around the axis which goes through the center of mass of the chain and is parallel to the magnetic field by $180^\circ$ angle. Therefore, unless $\theta\neq0$ (this is in all cases except for $\phi=90^\circ-\arctan(\frac{\sqrt{2}}{2})\approx 55^\circ$ as visible from Fig.~\ref{fig:theta_}), we have two energetically favorable alignments of chains  with the same energy. Those configurations are mirror image of each other. Therefore, if for one configuration angle between the magnetic field and chain orientation is $\theta=\tilde\theta$ than for other $\theta=-\tilde\theta$. 

To compare  with different methods (theory and experiment) the obtained straight chain configurations, the angle $\theta$ is probably the best quantity for this task. The angle $\theta$ does not dependent on chain length but only on the shape of hematite particles and the orientation of magnetic moment. For examined configurations we see that cube is a good approximation to a superball with $q=2.0$ or vice versa, but nevertheless this approximation may lead to an error of up to $4^\circ$ for some orientations of magnetic moment as visible from Fig.~\ref{fig:theta_} and Tab.~\ref{tab:angles}. Thus, it is important to know the exact shape factor of particles and it should not differ too much in the experiment to obtain  good statics. In the case for $q=1.5$ already qualitatively different results may be obtained for some orientations of magnetic moment.

\subsection{Chains with kinks}

We already mentioned that in the case if $\Phi\in[0^\circ, 2^\circ)$ and $\phi\in[-35^\circ, -32^\circ)$ the energetically favorable solutions in strong external fields and gravity  are chains with kinks (see Fig.~\ref{fig:kinks1} top). Note that these are the structures which are observed in experiments \cite{Rossi_phd, Philipse2018}. Therefore, we examine this in higher detail. By performing simulated annealing simulations in Espresso 4.1.2 for three particle chains we qualitatively find only two configurations, as shown in Fig~\ref{fig:theta}. The large majority are configurations with a kink (Fig.~\ref{fig:theta} left), but there are some cases where also  straight chains (Fig.~\ref{fig:theta} right and its mirror image)  are observed. This is because the configuration with the kink has a lower energy and the simulated annealing when cooled too fast does not always find a global energy minimum. Angles $\theta$ in our analytical calculations, however, slightly differ from MD simulation results. As for chains with kinks this angle $\theta$ is not constant along the chain, it is better to measure angles $\theta_i$ between each two neighboring particles, as shown in Fig.~\ref{fig:theta}. For larger chains one can also define the most probable angle $\bar\theta$. As there is  quite a big variety of methods used, we introduce different superscripts to differentiate them:\vspace*{-1ex}
%
\begin{center}
 \begin{tabular}{@{}cl@{}}\toprule
\textbf{superscript} & \multicolumn{1}{c}{\textbf{meaning}}  \\ \midrule
c &  analytical calculation with cubes  \\
sb &  analytical calculation with superballs ($q=2.0$)  \\
exp &  
experiment with superballs ($q=2.0$)  \\
45 &  MD sim. with superballs ($q=2.0$, 45 spheres)  \\
93 &  MD sim. with superballs ($q=2.0$, 93 spheres)  \\
\bottomrule\end{tabular}
\end{center}

In three particle case two angles $\theta_1$ and $\theta_2$ as indicated in Fig.~\ref{fig:theta} left, are measured. Calculated results for this 3 cube system are summarised in Tab.~\ref{tab:angles}.
\begin{table}\centering
 \caption{Calculated angles for three particle chains.}
 \label{tab:angles}
 \begin{tabular}{cccc}\toprule
\textbf{angle straight chain} & \textbf{value} &\textbf{angle kinked chain} & \textbf{value} \\ \midrule
$\theta^{c}$ &  $\pm21.7^\circ$  &
$\theta^{c}_1=-\theta^{c}_2$ &  $\pm23.5^\circ$  \\
$\theta^{sb}$ &  $\pm18.4^\circ$  &
$\theta^{sb}_1=-\theta^{sb}_2$ &  $\pm20.9^\circ$  \\
$\theta^{45}$ &  $\pm19.1^\circ$  &
$\theta^{45}_1=-\theta^{45}_2$ &  $\pm19.8^\circ$  \\
\bottomrule\end{tabular}
\end{table}

%
%
In this particular case from Tab.~\ref{tab:angles} follows that the approximation of superball in MD calculations introduces an error of  less than $2^\circ$. This error is larger for the first case of moment orientation. In some cases it can lead to a difference of  up to $4^\circ$. The value of $\theta_i$ for longer chains  converges to a straight chain  orientation $\theta$ for cubes sufficiently far from the kink. Thus, the most probable value measured  in experiment should be in the range $|\bar \theta|\in(18.4^\circ; 20.9^\circ)$.  This in the error range of dipole approximation match the experimental results \cite{Philipse2018}. However,  in our experiments we observe that there are much more straight three-cube chains than kinked ones.  Also x-ray scattering experiment \cite{Philipse2018} suggests that these configurations are not possible as hematite cubes in magnetic field  should be tilted close to $45^\circ$ with respect to the horizontal plane and touch the bottom surface with an edge. But for those structures the tilt is close to $0^\circ$ and the cube touches the bottom surface with a face. Therefore, it is most plausible that  energy minimization can not explain kink formation. From this we have to conclude that in an experiment at room temperature we do not observe energetically favorable configurations and in our assumptions some effect is not included.

\subsection{Finite temperature MD simulations initialized  with energetically favorable configurations}

\begin{figure}[htbp]
\includegraphics[width=0.48\columnwidth]{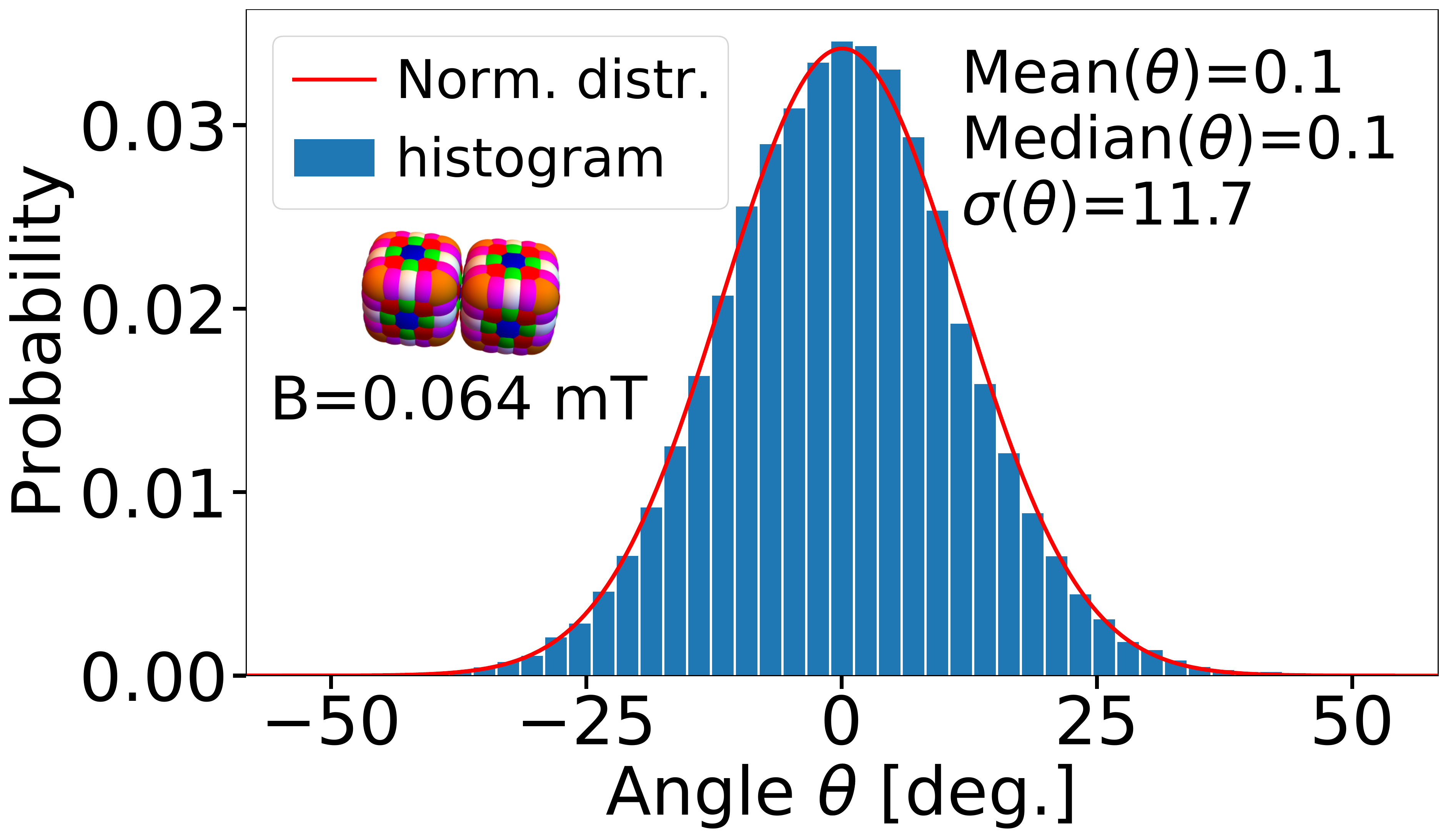}
\hfill
\includegraphics[width=0.48\columnwidth]{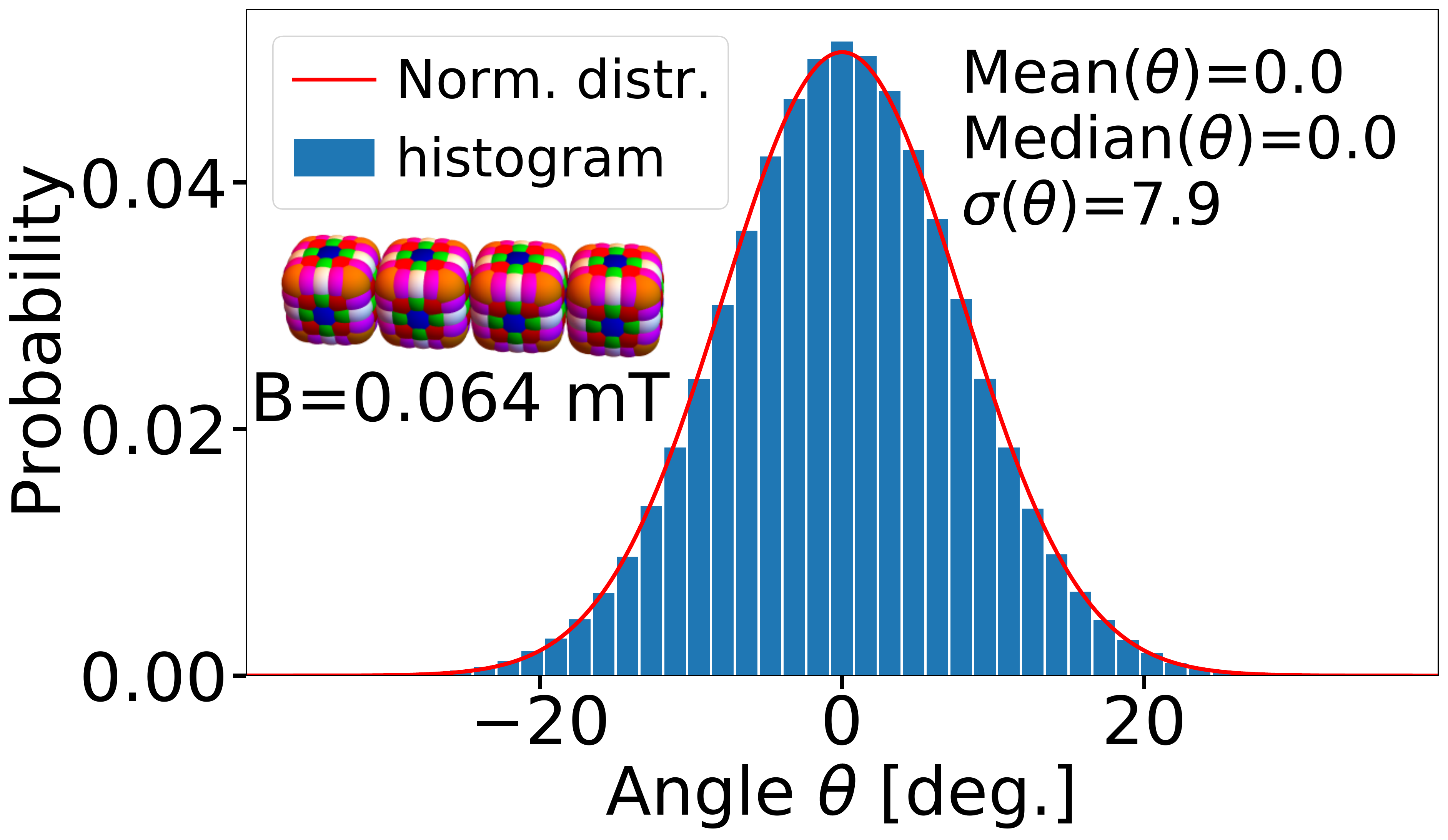}\\
\includegraphics[width=0.48\columnwidth]{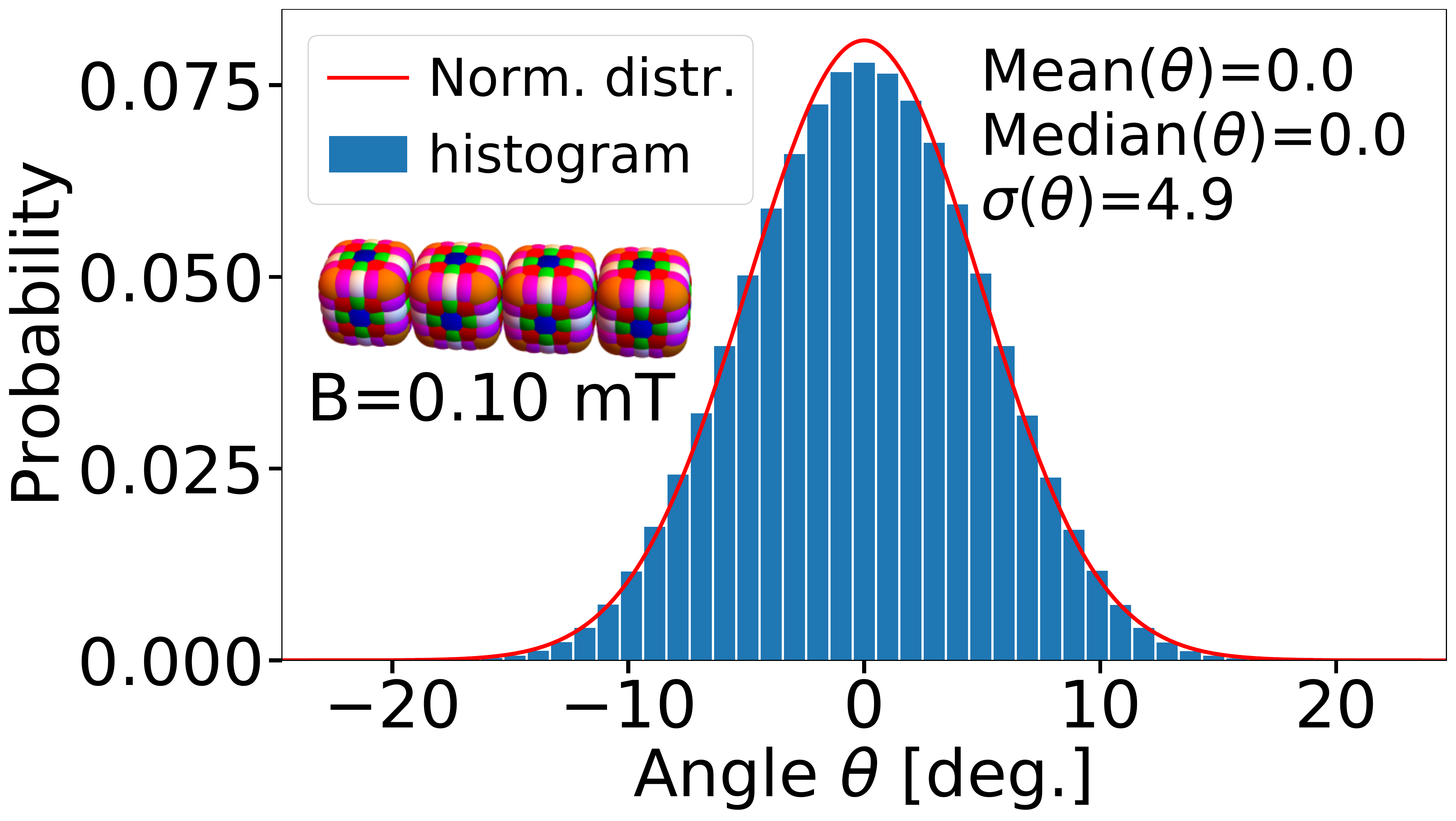}
\hfill
\includegraphics[width=0.48\columnwidth]{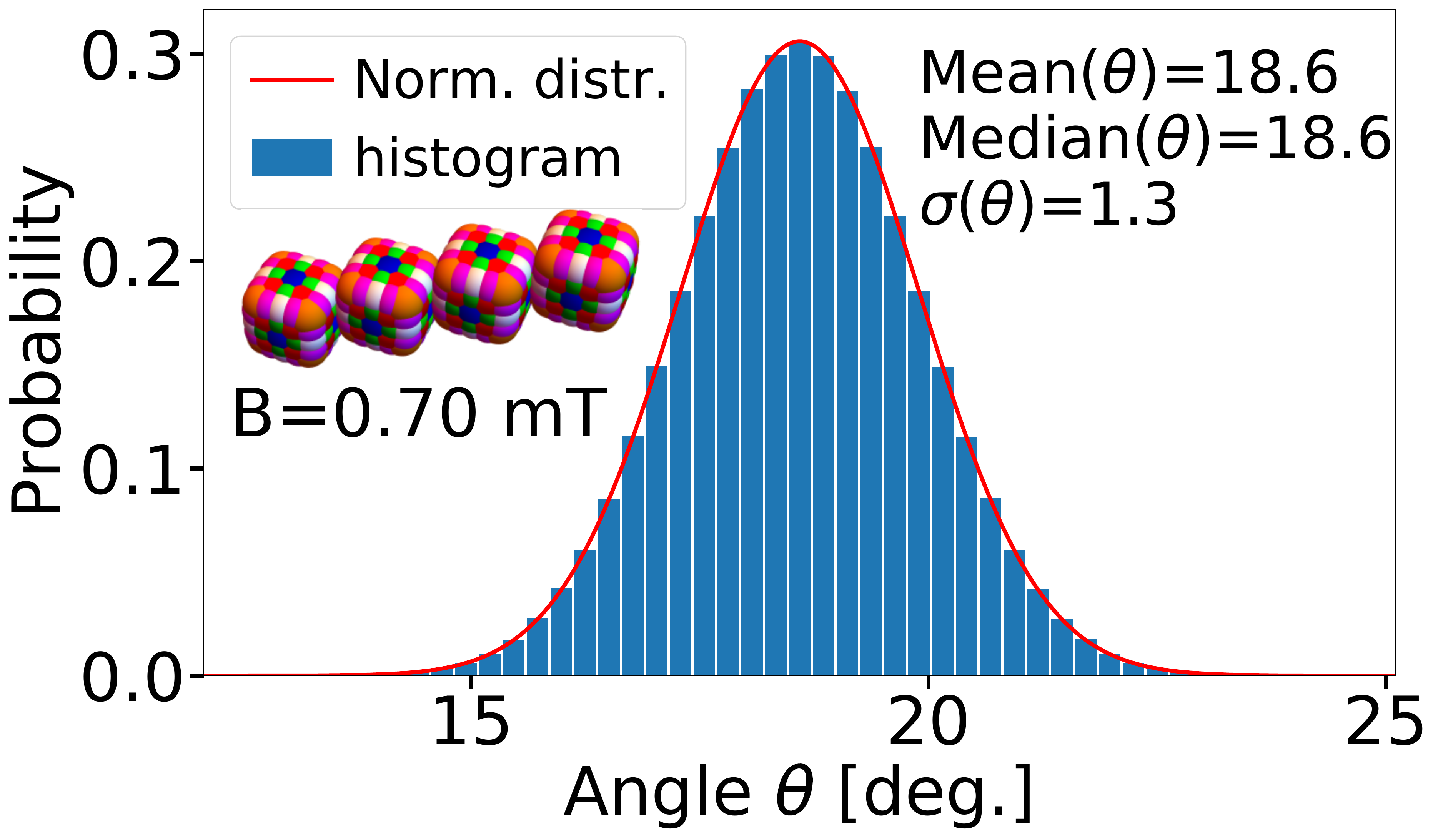}
\caption{Histograms of the angle $\theta$ for straight cube chains with $\phi=12^\circ$ at different strength of magnetic fields at room temperature calculated from MD simulations. The graphs in the top row are for the chain length of two (left) and four (right) at  $B=0.064\,\mathrm{mT}$, which is the magnetic field of the Earth at equator. The graphs in the bottom row are for four-cube chains at different values of the magnetic field. On the left the value of magnetic filed is slightly below $B_{crit}$, on the right it is significantly higher than $B_{crit}$ ($B=0.9B_{crit}\approx 0.10 \,\mathrm{mT}$ and $B=6B_{crit}\approx0.7 \,\mathrm{mT}$ respectively).}
\label{fig:MD_hist}
\end{figure}

In search for the missing effect we performed finite temperature MD simulations as in experiments at room temperature cubes and ellipsoids \cite{hem_ellipsoids,hem_ellipsoids2} significantly fluctuate. We initialized our calculations with energetically favorable configurations depending on the chain length and the moment orientation. To our surprise these configurations were still stable (no kinks where formed) at room temperature even for chains with more then 20 cubes which is an upper limit in our investigation.
At the room temperature, compared to calculations without thermal effects, the experiment and MD simulations show that cube chains notably fluctuate and thus chain orientation changes in time. This is more pronounced for smaller magnetic fields and shorter chains. The result of MD simulations for straight chains for $\phi=12^\circ$ at different strengths of magnetic field are shown in Fig.~\ref{fig:MD_hist}. It can be seen that the distribution of the angle $\theta$ is quite close to a normal distribution. For longer chains and stronger magnetic fields, which are below $B_{crit}$, deviations from normal distribution can be  observed. In this case clusters are formed consisting of two cubes with different magnetic moment orientation which fluctuate stronger than individual cubes in the chain. 

\subsection{Experimental measurements of configuration stability}

To confirm findings that energetically favorable configurations are stable at room temperature, we perform corresponding experimental measurements. The description of the experimental setup was given in \cite{Petrichenko_2020} (Riga setup). In short, we use an inverted microscope (Leica DMI3000B), equipped with a custom made coil system of three pairs of coils, and a video camera (Basler ac1920-155um). Power supplies (KEPCO) provide a current for field generation, which is controlled by a signal from DAQ card (NI). We use a self made LabView program to define the signal and synchronize it with the image   acquisition.

As the magnetic fields of interest are small (on the order of Earth magnetic field and smaller), we take special care of field calibration. This includes cancelling of unwanted parasitic fields sources, including Earth, equipment and other. To account for these fields, a magnetic sensor (HMC5883 GY-271 3V-5V Triple Axis Compass Magnetometer Sensor Module for Arduino) was placed in the position of the sample before each experiment, and then a current correction was added in the LabView code for the applied magnetic field to cancel the parasitic magnetic field. However, due to several experimental constraints, we were limited to a precision of $\Delta B\in(0.01;0.03)\,\mathrm{mT}$. We complement this approach with a further method of the field compensation, which relies on the fact that hematite cube chains orient even in very small external fields. That is, to improve the compensation, we try to broaden the angle distributions of hematite cube chain orientations, by slightly altering the current correction.   By searching for the point when the distribution become the broadest we were able to improve accuracy. Similarly, the direction of applied magnetic field, when its value is comparable or smaller than the field of the Earth, can be more precisely determined using angle $\theta$ which cube chains (consist of an even number of cubes) make with the external field. Therefore, for small magnetic fields the mean value of $\theta$ in the experimental graphs was set to zero.

For experiments we use hematite cubes, which were prepared and characterized as in \cite{Petrichenko_2020}, that is, cubes have the edge length $a\approx1.5\,\mathrm{\mu m}$ and shape factor $q\approx2.0$. To narrow the cube size distribution, we apply an extra step of gravitational sedimentation. Before every experiment, the sample solution is mixed and the pH level of the solution restored. 

\begin{figure}[htbp]
\includegraphics[width=0.24\columnwidth]{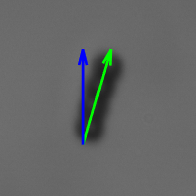}
\hfill
\includegraphics[width=0.24\columnwidth]{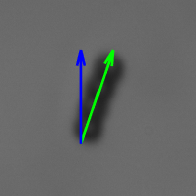}
\hfill
\includegraphics[width=0.24\columnwidth]{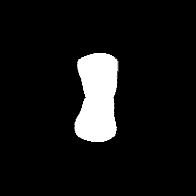}
\hfill
\includegraphics[width=0.24\columnwidth]{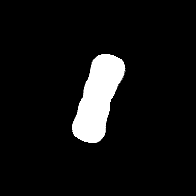}
\caption{Left to right: microscope images of a chain in $B = 0.70\,\mathrm{mT}$ field at furthest deviations from mean position (field aligned with vertical axis, indicated by blue arrow; chain direction as determined by algorithm used for field-chain angle calculation indicated by green arrow). Two rightmost images: shape described by all positions taken by the chain in a sequence of 400 images under an applied field $B = 0.10\,\mathrm{mT}$ (second from right) and $B = 0.70\,\mathrm{mT}$ (right).}
\label{fig:exp_images}
\end{figure}

The experimental procedure is as follows. We introduce a small volume of cube suspension in a glass capillary (Vitrocom, liquid thickness $100\,\mathrm{\mu m}$), seal it and place it in the microscope. To improve the resolution, we use an oil immersion $100\times$ objective. Then, a suitable hematite cube chain is centered in the field of view, selected external field is applied and an image series of a straight chain, fluctuating due to Brownian motion, is taken (see examples in Fig.\ref{fig:exp_images}(a)-(b)). This is repeated for a repeating staircase magnetic field ($B\approx 0.00\,\mathrm{mT}$, $B=0.064\,\mathrm{mT}$, $B=0.10\,\mathrm{mT}$, $B=0.70\,\mathrm{mT}$), with the field being rotated by $45^0$ at each field value change. This pattern was used to minimize the chance of any history being accumulated either by subjecting the chains to same orientation of field for prolonged periods of time or magnetizing elements of the experimental system. Thus, every full measurement of a sample consists of 24 combinations of field value and angle, provided that the chain survives without attracting other hematite cubes. Afterwards, the images of cube chains are processed using MatLab, to find their orientation with respect to the field direction and data is averaged over multiple samples to obtain the final histograms.

\begin{figure}[htbp]
\includegraphics[width=0.48\columnwidth]{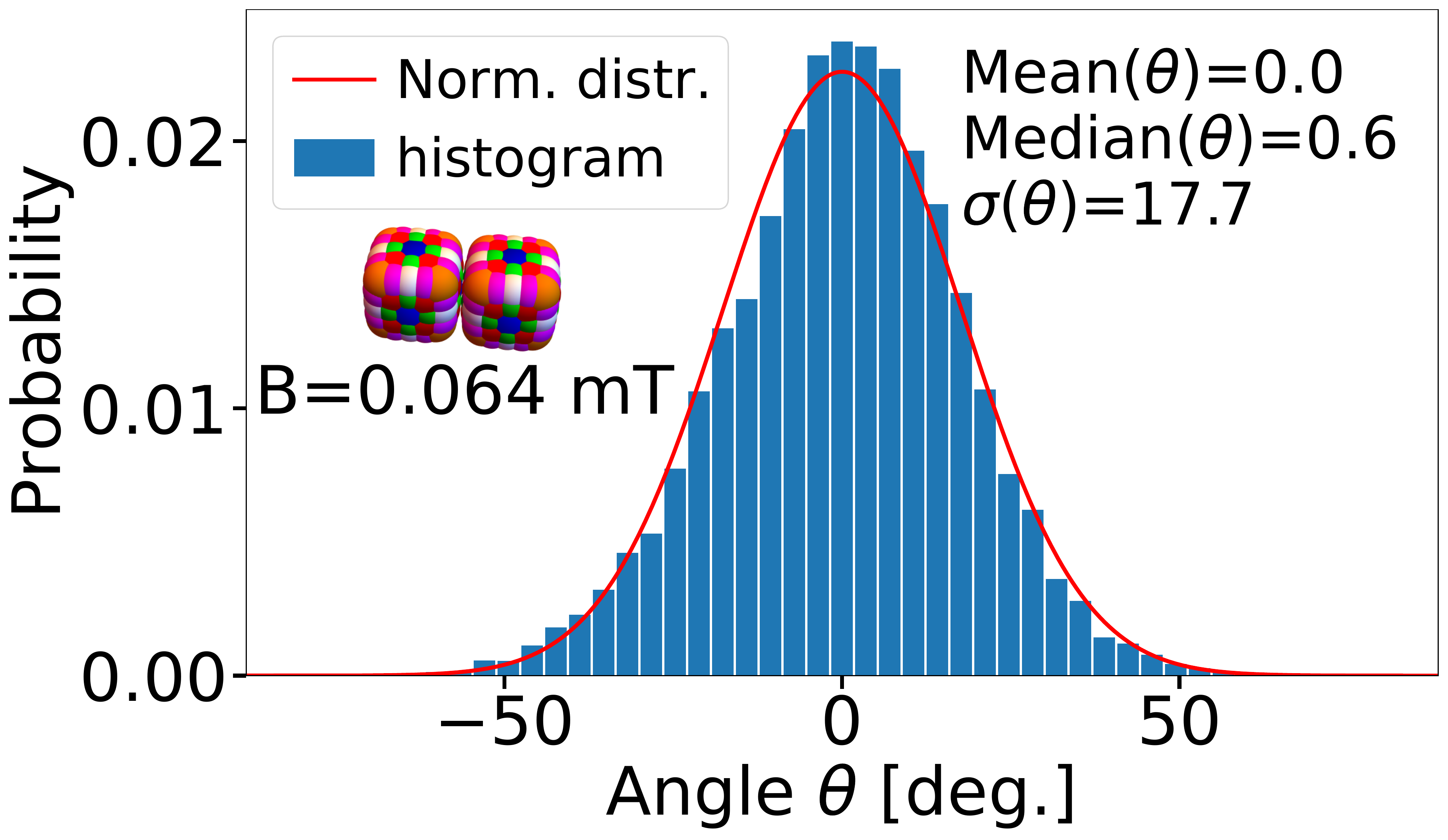}
\hfill
\includegraphics[width=0.48\columnwidth]{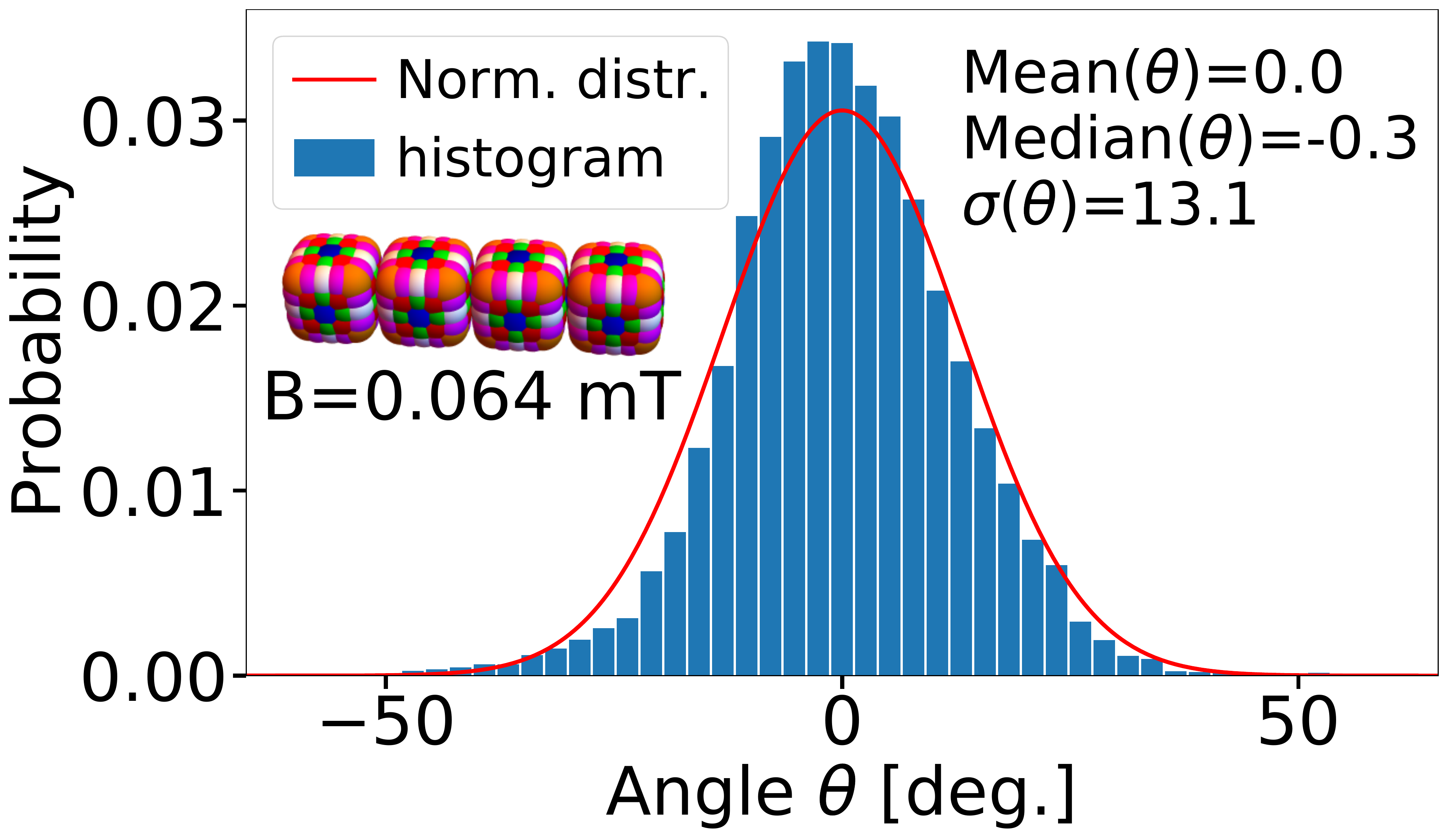}\\
\includegraphics[width=0.48\columnwidth]{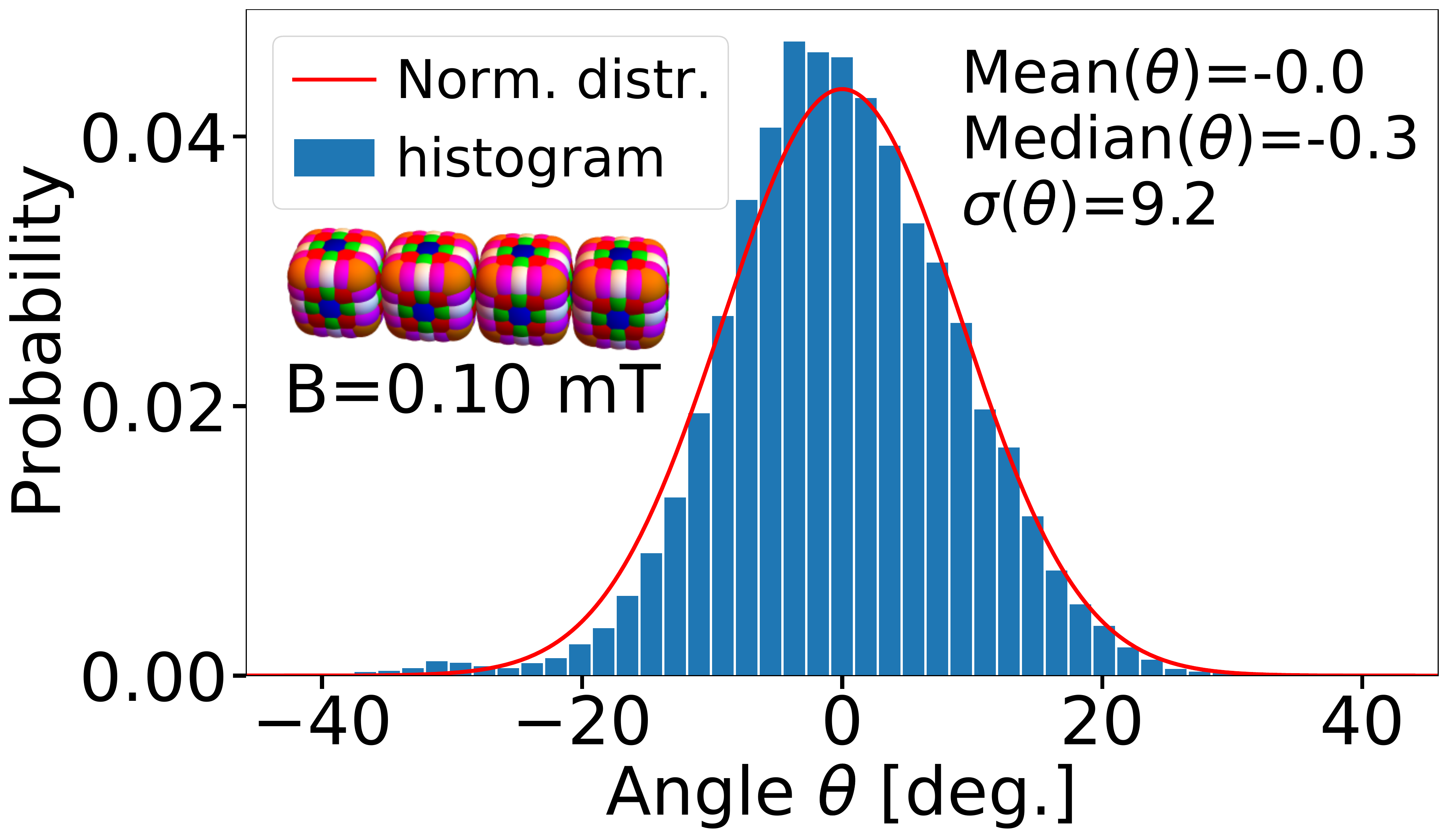}
\hfill
\includegraphics[width=0.48\columnwidth]{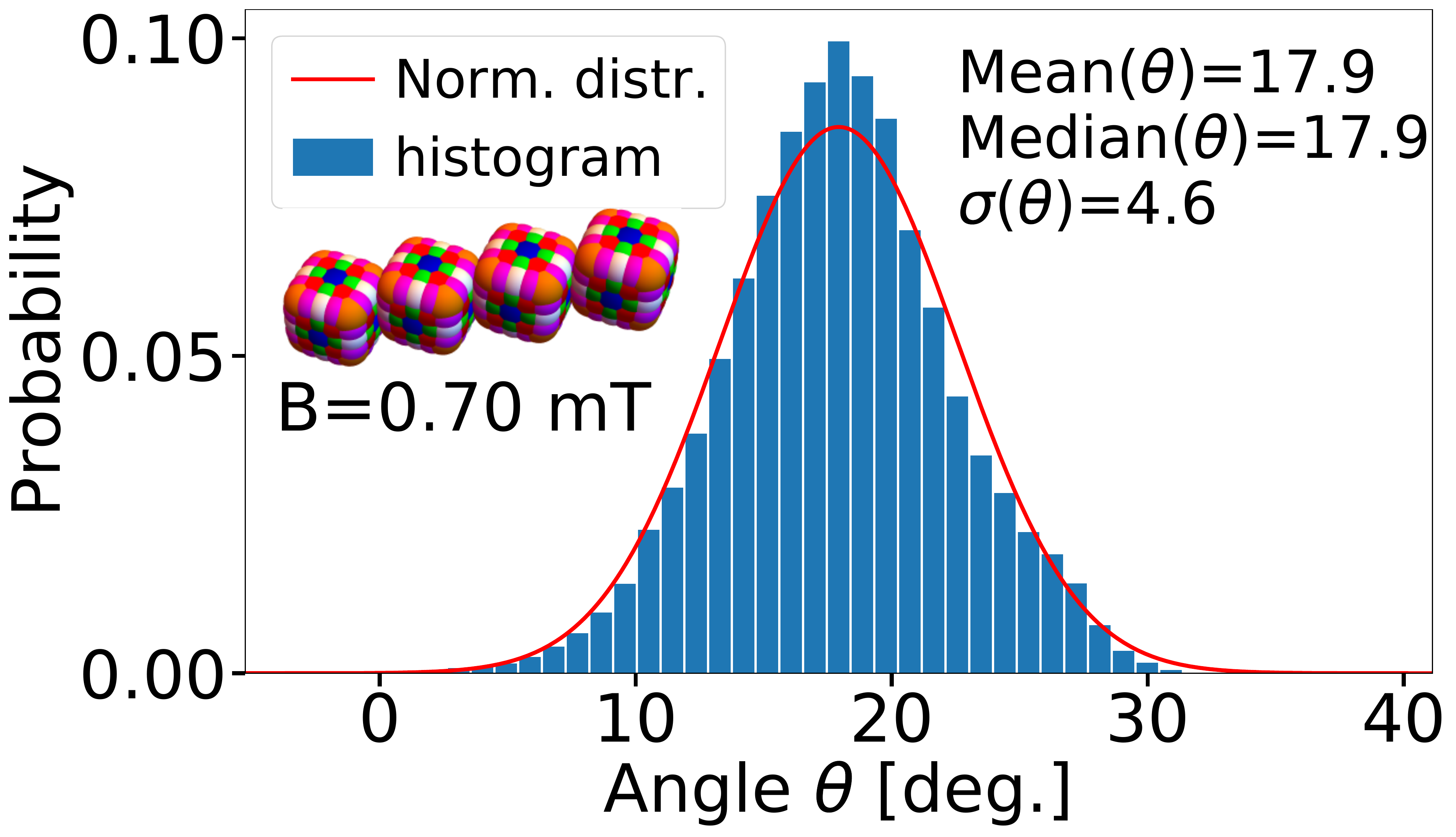}
\caption{Experimentally measured histograms of angle $\theta$ for straight cube chains at the same external magnetic fields as Fig.~\ref{fig:MD_hist}. For small external magnetic fields the mean value of angle $\theta$ is set to zero as cube chains allow  to determine the  direction of the magnetic field more precisely than the magnetic sensor used in the experiment.}
\label{fig:exp_hist}
\end{figure}

A view at the cube chains and their experimentally observed behavior under two magnetic field values is demonstrated in Fig.~\ref{fig:exp_images}. The experimentally measured histograms are shown in Fig.~\ref{fig:exp_hist}. When comparing theoretical and experimental results we observe similar trends. However, the experimental distributions are broader, which can be explained with the fact that we average over multiple different samples. An increase of chain length or external magnetic field narrows the distributions and above $B_{crit}$ we observe the rearrangement of chains. Quantitatively there is a small discrepancy ($\approx1^\circ$) for the mean angle in theory and experiment at $0.7 \,\mathrm{mT}$. This, however, can be explained with   the dipole approximation error, the approximation to superball in MD simulations, and the precision of angle determination, as well as cube shape and size variations in experiments.

\subsection{Randomly initialized finite temperature MD simulations}

\begin{figure}[htbp]
\includegraphics[height=0.80\columnwidth, angle=90]{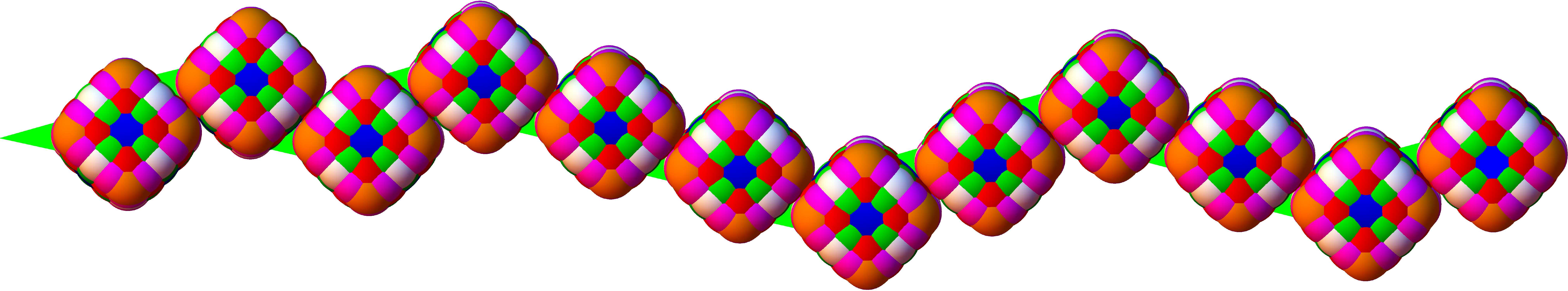}\\[2ex]
\includegraphics[height=0.78\columnwidth, angle=90]{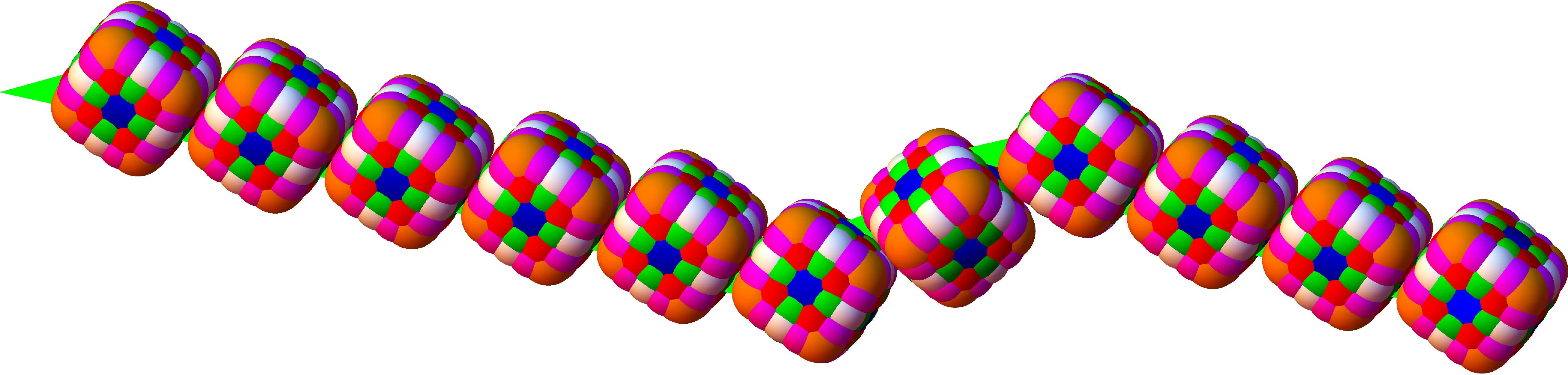}\\[2ex]
\includegraphics[height=0.80\columnwidth, angle=90]{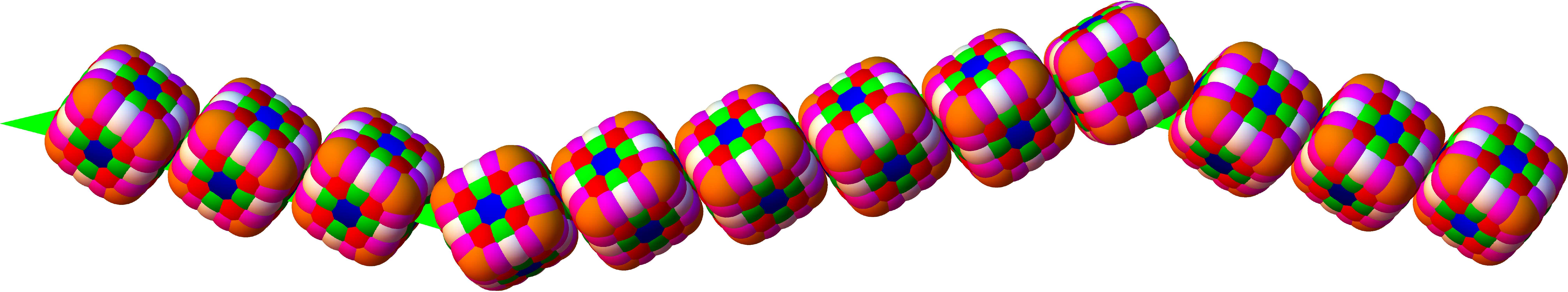}\\[2ex]
\includegraphics[height=0.80\columnwidth, angle=90]{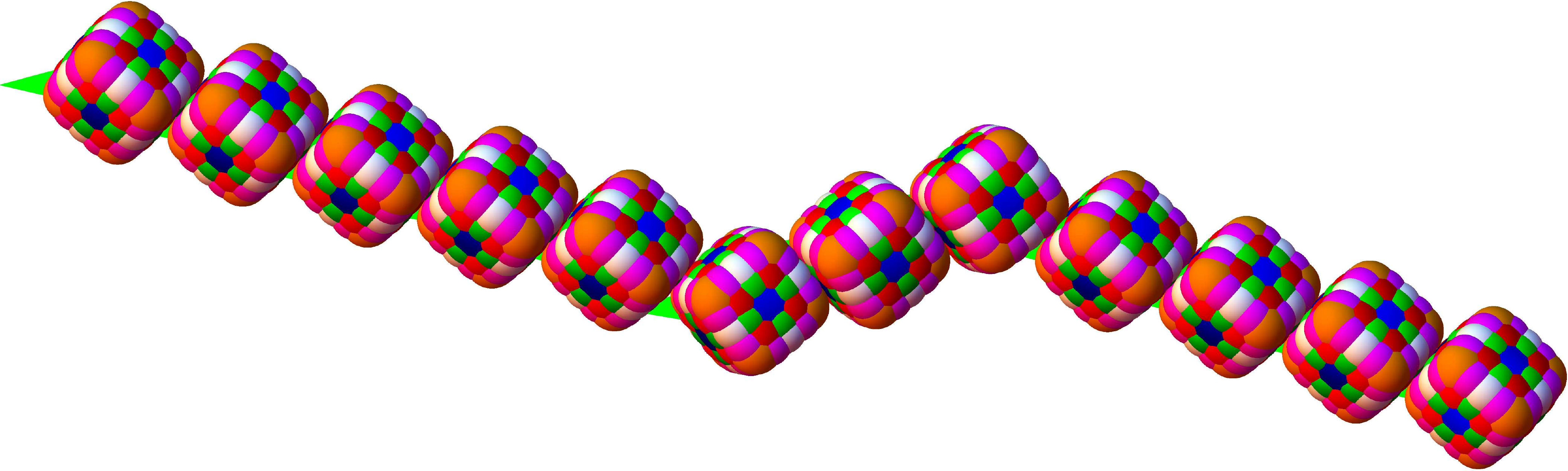}\\[2ex]
\includegraphics[height=0.80\columnwidth, angle=90]{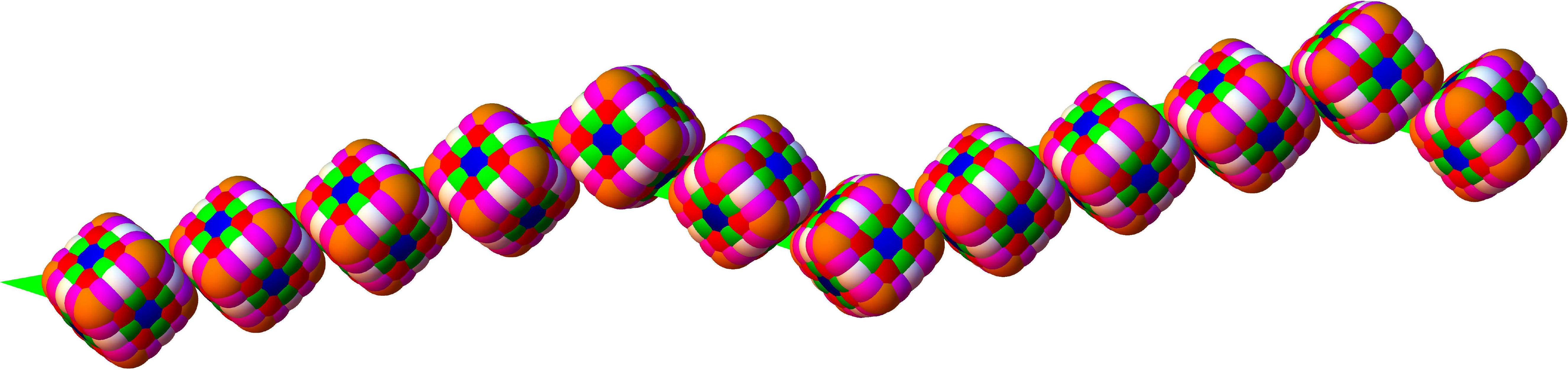}\\[2ex]
\caption{Chains with kinks obtained in room temperature ($300\,\mathrm{K}$) in MD simulations for the second case orientations: $\Phi=0^\circ$, $\Phi=20^\circ$, $\Phi=30^\circ$ and for the first case orientations: $\phi=12^\circ$, $\phi=6^\circ$ arranged from top to bottom. Note that $\Phi=30^\circ$ and $\phi=\arcsin(\frac{1}{3})\approx19^\circ$  correspond to the same orientation of the magnetic moment. A 93 sphere superball approximation is used to obtain these results.}
\label{fig:kinks}
\end{figure}

The situation, however, is quite different when MD simulations at room temperature are initialized from random configurations. Kinks are observed in all cases of the moment orientation except for $\phi\approx 55^\circ$. Some typical chain structures at high magnetic fields are shown in Fig.~\ref{fig:kinks}. This significantly differs to a situation when simulations are initialized with energetically favorable configuration, which is a chain without kinks. When simulations are started from energetically favorable configurations no kinks are formed even when $B>B_{crit}$. Chains remain straight. To obtain a chain with kinks in MD simulation from this initial condition, the magnetic field has to be reduced to $B\ll B_{crit}$ and then increased to $B\gg B_{crit}$. This suggests that kinks are formed during chain formation. The reduction of magnetic field to $B\ll B_{crit}$ from $B> B_{crit}$ as well as an increase to  $B\gg B_{crit}$ from $B< B_{crit}$ causes the rearrangement of chains. 

\begin{figure}[htbp]
\includegraphics[width=0.48\columnwidth]{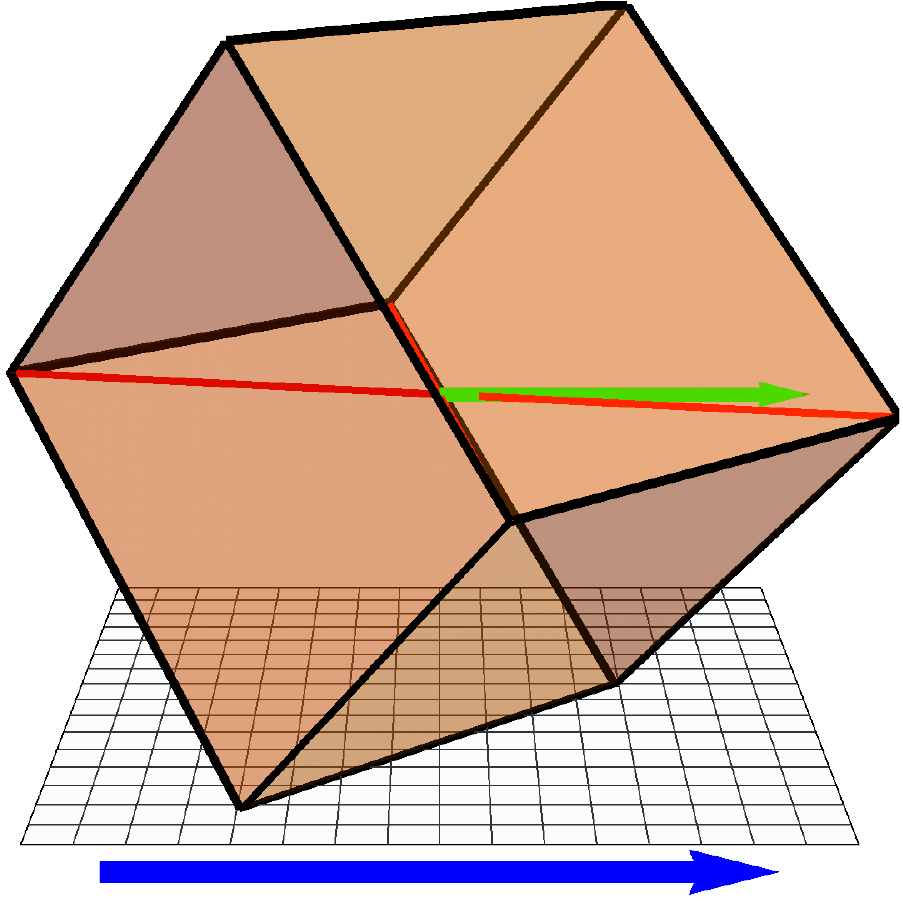}
\hfill
\includegraphics[width=0.48\columnwidth]{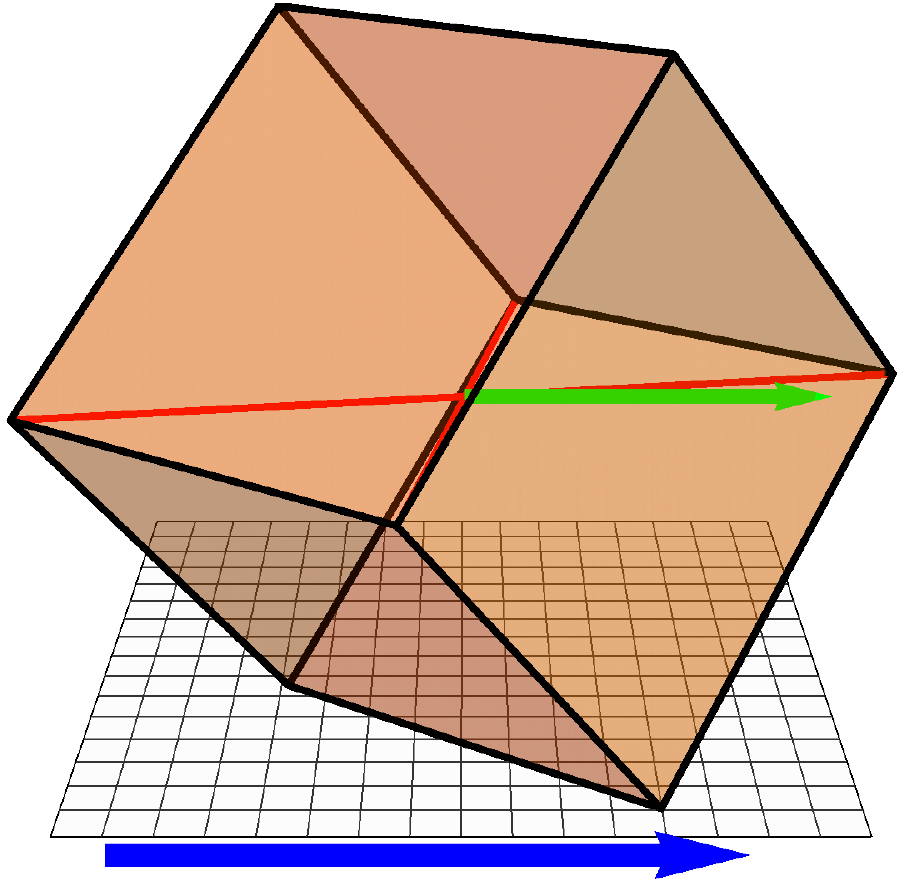}
\caption{Two alignments of cube which have the same minimal total energy in an external homogeneous magnetic field.}
\label{fig:alignments}
\end{figure}

\begin{figure}[htbp]
\includegraphics[width=0.48\columnwidth]{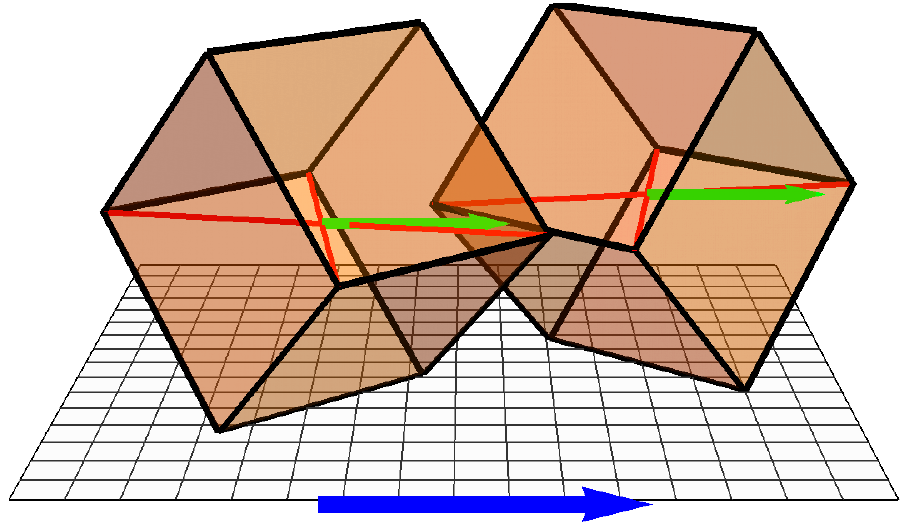}
\hfill
\includegraphics[width=0.48\columnwidth]{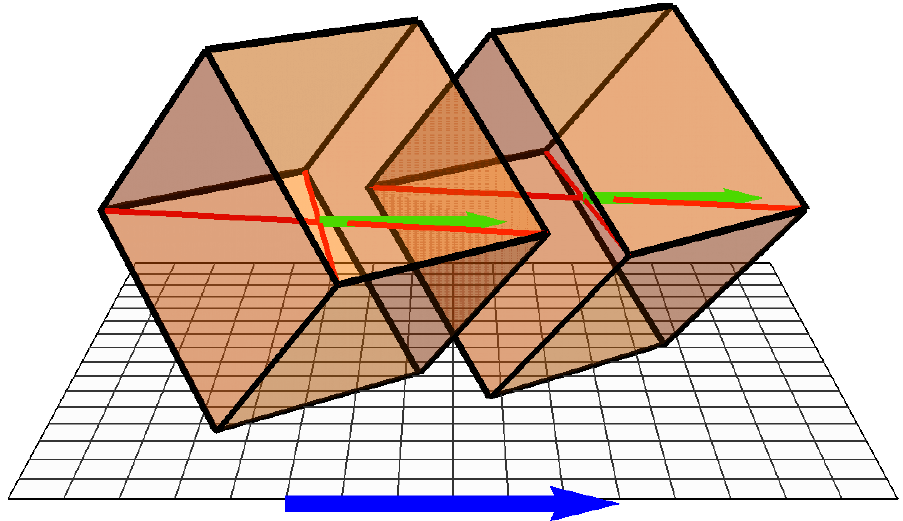}
\caption{Two cube chains. A cube chain with different alignments of cubes (left) and with the same alignment (right).}
\label{fig:2chain}
\end{figure}

In all cases of the moment orientation  except for $\phi\approx 55^\circ$ a cube in external magnetic field can be orientated in two alignments
(see Fig.~\ref{fig:alignments}) such that its total energy is minimal. The second alignment can be obtained by rotating the cube which is in the first alignment  by $180^\circ$  around an axis parallel to the magnetic field. If in the process of chain formation, two cubes or two cube chains with the same alignment attach, a straight chain is formed. Similarly, if a cube or a chain of cubes attaches to another cube with a different alignment (two cubes which are shown in Fig.~\ref{fig:alignments})  they form the chain shown in Fig.~\ref{fig:2chain} (right). As this configuration is not energetically favorable and the thermal energy is sufficient for one cube to change its alignment, the chain shown in Fig.~\ref{fig:2chain} (right) or its mirror image is formed. Thus, in experiments and room temperature MD simulations (even when simulation is start from the configuration Fig.~\ref{fig:2chain}) only two cube chains as shown in Fig.~\ref{fig:2chain} (right) are observed.

\begin{figure}[htbp]
\includegraphics[width=0.99\columnwidth]{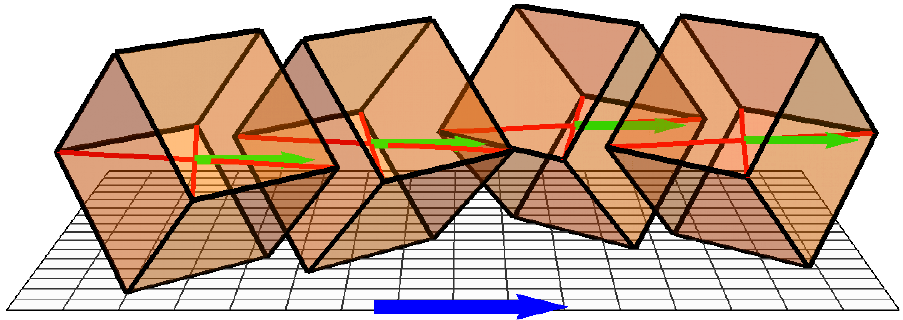}
\caption{Four cube  cube chain with a kink. It is formed by attaching two two-cube chains with different alignments.}
\label{fig:4chain}
\end{figure}

The situation, however, is different when two or more element chains with different alignments attach. Now, in the case of two two-cube chains, a four cube chain as shown in Fig.~\ref{fig:4chain} is formed. In this case the thermal energy is not sufficient anymore to rotate the whole two- or more- element chain around the axis of magnetic field and thus such kinked chains are observed in MD simulations and experiments. At an increased temperature chains which attach become shorter and therefore more kinks are formed. This is, in principle, an entropy effect, as stated in \cite{Philipse2018}.

For kinked chains the obtained values for the most probable angles $\bar\theta$ at room temperature are close to the results for angle $\theta$ for straight chains (difference is less than $1^\circ$) . This, however, is just a lucky cancellation of errors: the thermal fluctuations reduce the angle $\bar\theta$ but the magnetic interactions increase. Nevertheless, angle $\theta$ at zero temperature can be used to predict the value of $\bar\theta$ at room temperature. From MD simulations we can confirm the findings theoretical findings of \cite{Philipse2018} for the case of $\phi=12^\circ$. But the results of analytical calculation and MD simulation differ. With the simulated annealing one finds that for the  straight chains $\theta^{45}=\pm16,4^\circ$, however the analytically calculated value for $q=2$ is $\theta^{sb}=\pm19,0^\circ$. This suggests that with a better approximation used in MD simulations the actual most probable angle should be close to $\pm19,0^\circ$. Indeed, if we use a 93 sphere approximation in MD simulations, we obtain that for straight chains $\theta^{93}=\pm19,3^\circ$. This suggests that reason for good agreement between MD simulations and experiment in  \cite{Philipse2018}   is the cancellation of the errors -- dipole approximation and approximation to superball. Thus, a 45 sphere approximation is better choice when dipole approximation is used to describe experiment. 


\section{Conclusion}
\label{sec:concl}

In current work we examined equilibrium configurations for both cases of the possible magnetic moment orientation mentioned in the scientific literature but for all possible angles. Also, unlike in previous works \cite{Aoshima, Hematite_along_diogonal, Linse}, we do not restric that cubes have to touch the bottom surface with a face wich leads to different configurations and  allows to explain the reason for kink formation.

The first  in lieterature mentioned  case is when the moment is in the plane defined by two diagonals of a cube and the second is when the magnetic moment is perpendicular to the cube's diagonal. In both cases for any magnetic moment orientation in the presence of gravity the energetically favorable configurations are chains of particles. Depending on the strength of the external magnetic field (except for the case when magnetic moment is perpendicular to the face of the cube) two configurations are observed. The first configuration is observed for the values below the critical magnetic field value. The critical magnetic field $B_{crit}\approx 0.1\,\mathrm{mT}$ is typically larger, but comparable with the magnetic field of the Earth. In this case cubes arrange in straight chains and magnetic moments form zig-zag structures. The second configuration is obtained for magnetic fields larger than the critical value. In this case the magnetic moment direction is parallel to the external magnetic field and the cube centers are  shifted (in 2D), as compared to the previous case. In all cases  at high fields the structure reassembles staircases with the same step width. For the case of $\Phi\in[0^\circ, 2^\circ)$ and $\phi\in(-35^\circ, -32^\circ)$ the equilibrium structure is the staircase structure, but with kinks.

These kinked staircase structure are actually similar to structures that are observed in experiments \cite{Rossi_phd, Philipse2018}. However, these are not the structures that are observed in experiment, as in this case there is a different most probable angle between orientation of the chain and the magnetic field than measured in \cite{Rossi_phd, Philipse2018} and also the x-ray scattering experiment \cite{Philipse2018} suggests that such an orientation of the magnetic moment is not possible. Therefore, in experiments at room temperatures we plausibly do not observe energetically favorable structures but rather structures with  minimal free energy. 

In experiments \cite{Rossi_phd, Philipse2018} and from room temperature MD simulations (random initial conditions) typically observed structures are chains with kinks. Kink formation is not energetically favorable. The formation of kinks is an effect which arises due to an interplay of energy and thermal fluctuations. Kinks are formed during assembling process of chains. This is confirmed with room temperature MD simulations. If MD simulation is started from an energetically favorable chain without kinks even at high magnetic fields, no kinks are formed. Chains fluctuate more if the external field is smaller. To obtain a chain with kinks, the magnetic field has to be reduced to $B\ll B_{crit}$ and then increased to $B\gg B_{crit}$. The reduction of magnetic field to $B\ll B_{crit}$ form $B> B_{crit}$ as well as an increase to  $B\gg B_{crit}$ from $B< B_{crit}$ causes the rearrangement chains. 

For a single cube or a chain of cubes there are two alignments in magnetic field with the same minimal total energy. Therefore, during sedimentation process cubes (and short chains of cubes that are formed) are aligned in two alignments with approximately the same population. During the assembly process when the single cube  attaches to an already formed short chain of cubes or a single single cube with a differed alignment the thermal energy is sufficient for rotation of one cube and a straight chain is formed. However when already  a small chain of two or more cubes attaches to a longer chain with different alignment, the thermal energy is not sufficient for rotation and kinked structure is formed. This explains why there are kinks in the chains and thus the observations of \cite{Philipse2018}. 

In the current paper we also investigated for the first time the distribution of angle $\theta$  between short straight cube chain alignment direction and the static external  magnetic field in experiments and MD simulations  at room temperature. Such chains are building blocks of  swarms in rotating magnetic fields \cite{Petrichenko_2020}. In the magnetic field of the Earth the thermal fluctuations are very pronounced and therefore distributions are very broad. For two cube chains one observes that displacement can be more than $50^\circ$. By increasing chain length  or external magnetic field the distributions become narrower and above a critical value of the magnetic field we observe a rearrangement of chains. For $0.7 \,\mathrm{mT}$ the $\sigma(\theta)$ for individual four-cube chain is less than $2^\circ$. The distributions are quite close to the normal distributions both in the experiment and in the MD simulations, however, experimental distributions are broader as averaged over many different chains. There is a small discrepancy ($\approx1^\circ$) for the mean angle between theory and experiment at $0.7 \,\mathrm{mT}$, however, this can be explained with the approximations and the precision of angle measurements in experiment.


\section*{Acknowledgment}

Authors are very thankful to Andis Draguns for fruitful discussions and his bachelor thesis which inspired this study and  which resulted in the current paper, as well as Dr. Oksana Petrichenko for providing hematite cube samples.

M.B. acknowledges financial support from PostDocLatvia grant No. 1.1.1.2/VIAA/3/19/562 and G.K.   from PostDocLatvia Grant No. 1.1.1.2/VIAA/1/16/197. M.B., G.K. and A.C. acknowledge the support from M.era-net project FMF No.1.1.1.5./ERANET/18/04.

 \appendix
    \section{Superball out of spheres}
    \label{sec:app_A}
\begin{figure}[t]
\includegraphics[width=0.48\columnwidth]{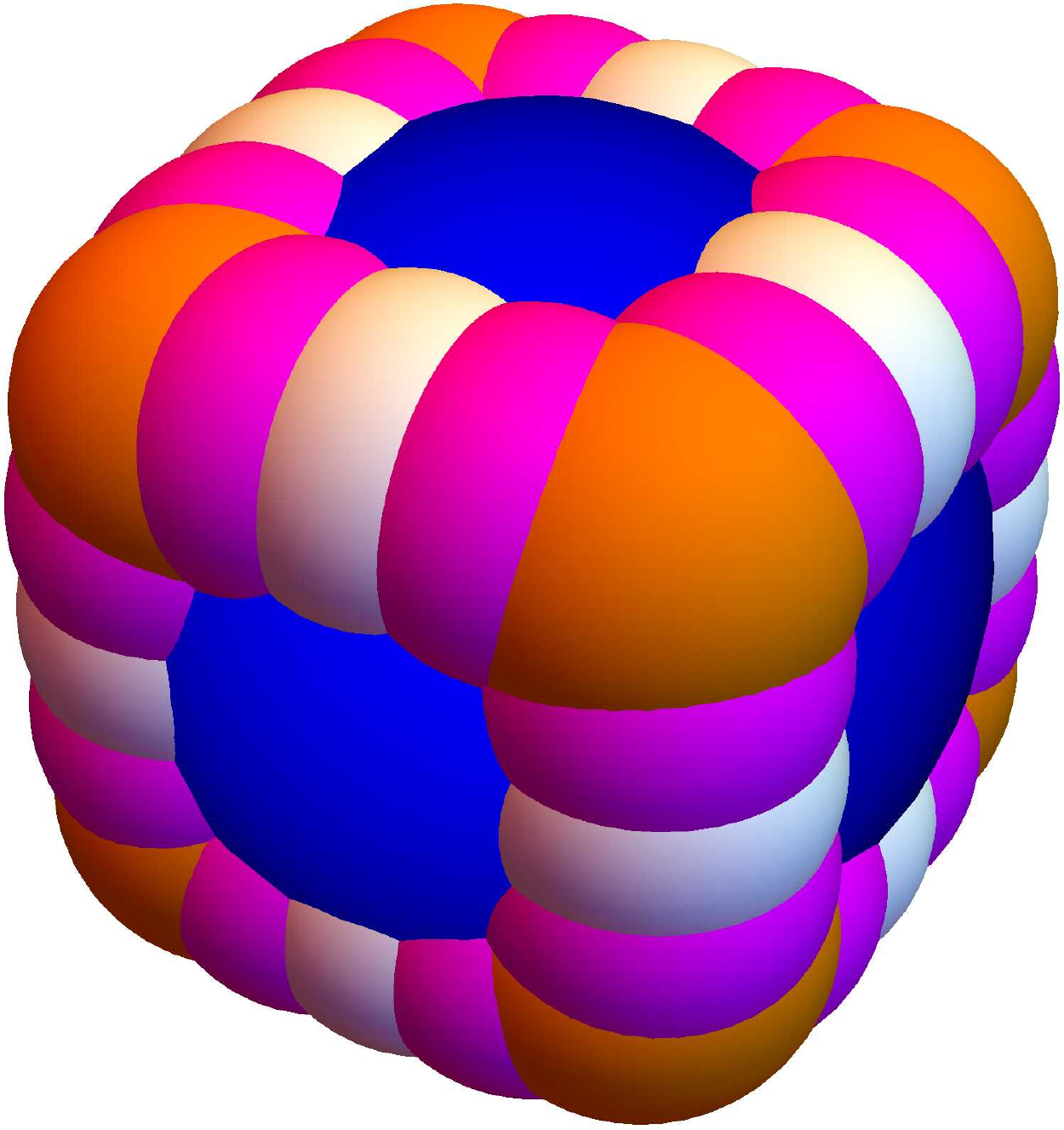}\hfill
\includegraphics[width=0.48\columnwidth]{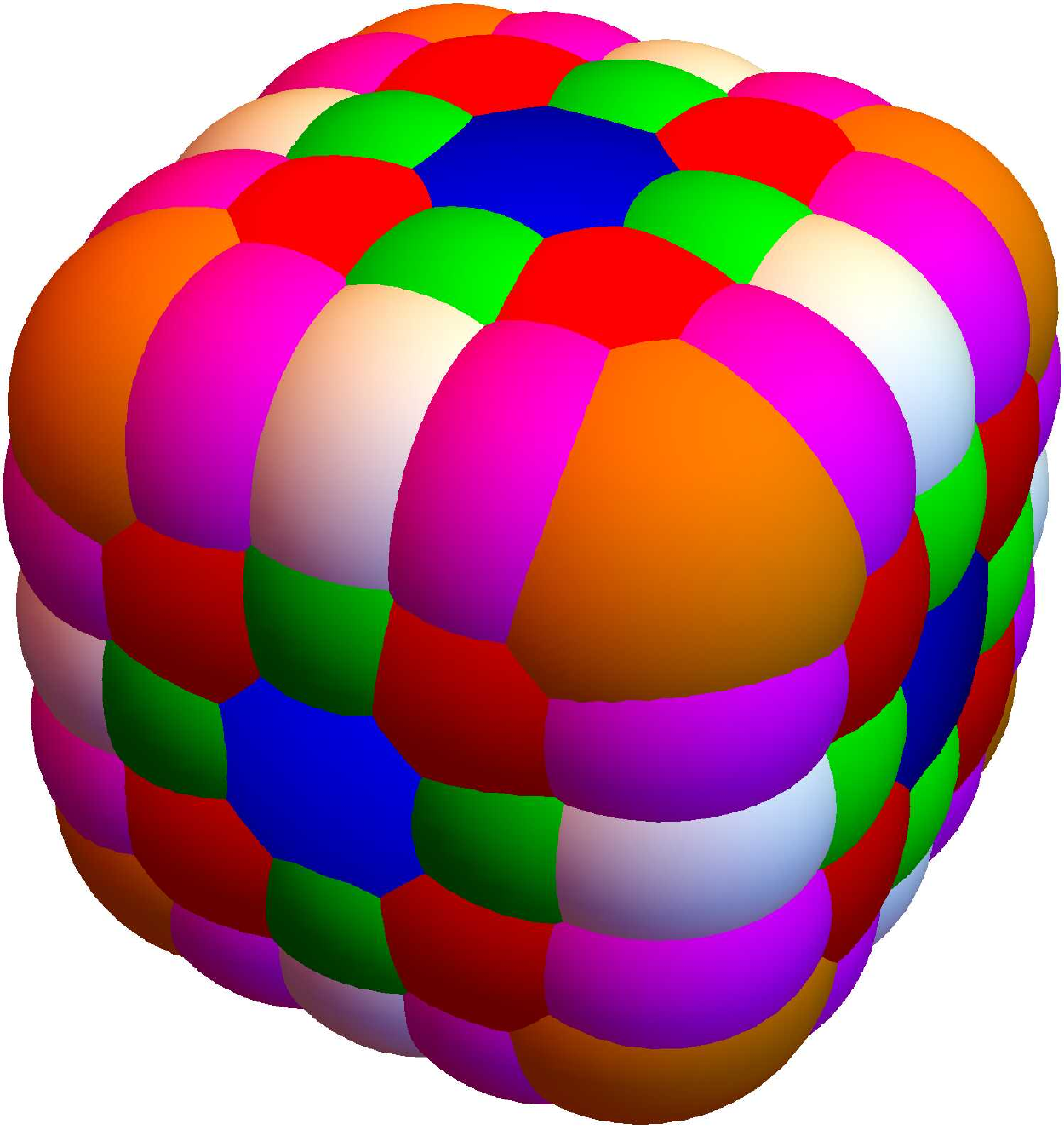}
\caption{Superball with $q=2.0$ approximated with 45 spheres(figure left) and with 93 spheres(figure right) .}
\label{fig:sball1}
\end{figure}

To construct a superball with $q=2.0$ out of spheres we are using approaches suggested in \cite{Kantorovich, Philipse2018}. Although we are using 45 sphere approximation instead 46 (it is not clear to authors of this paper where this extra sphere is placed in \cite{Philipse2018}).

One sphere with radius $R_c=\frac{a}{2}$ is placed at the center followed by 8 spheres with radius $R_v$ placed at the vertices of superball and 12 spheres with radius $R_e$ placed at the midpoints of edges. These spheres in Figs.~\ref{fig:sball} and \ref{fig:sball1} are depicted with blue, orange and white color respectively. The radius and position of the sphere is chosen such that at vertices and edge midpoints the shape and curvature matches with superball's.

\begin{figure}[h]
\includegraphics[width=0.48\columnwidth]{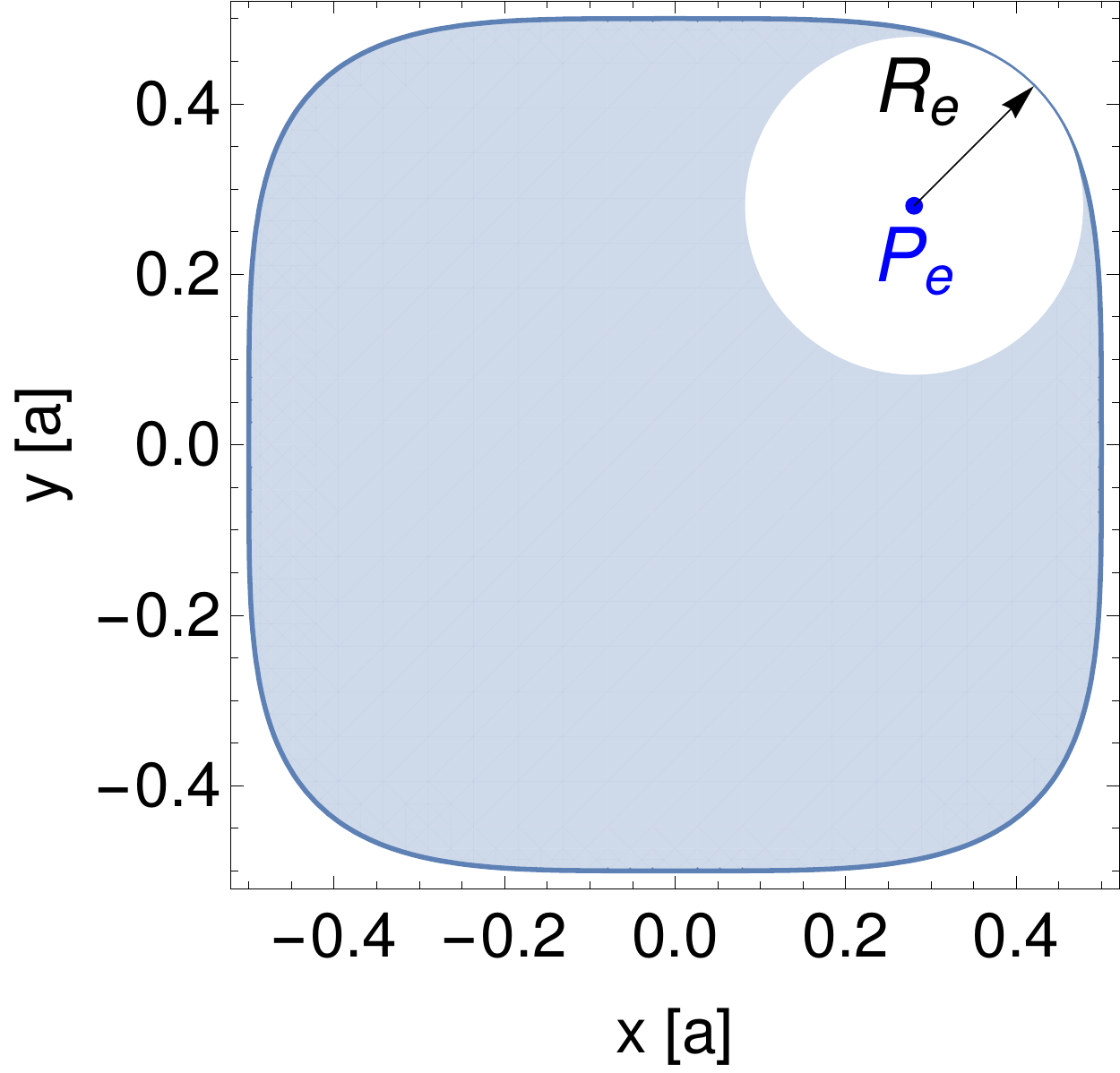}\hfill
\includegraphics[width=0.48\columnwidth]{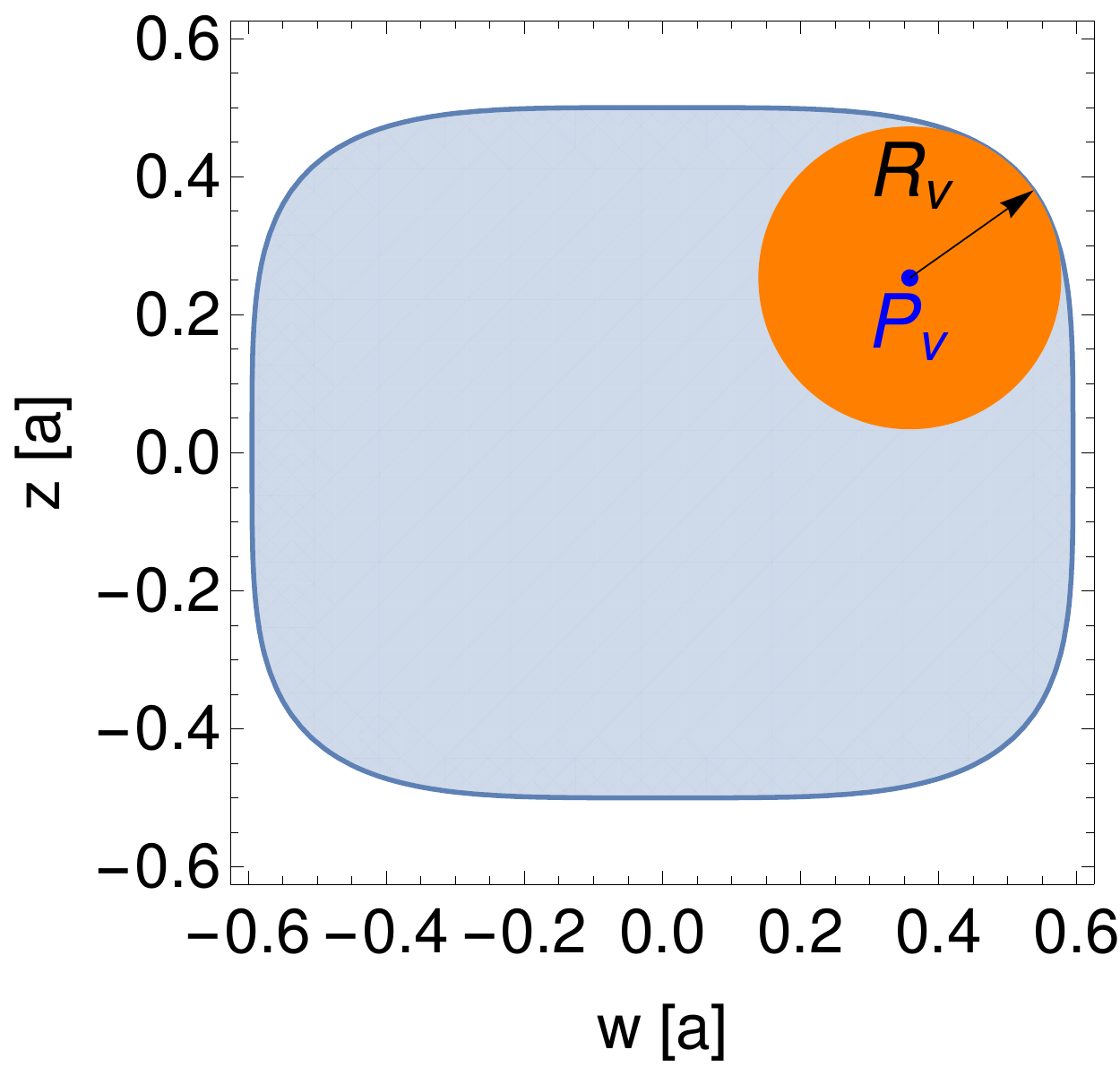}
\caption{Superballs with $q=2.0$ cross-section view in the plane $z=0$ (figure left) and with the plane defined by two diagonals of superball (figure right). Here $w$ is diagonal in $xy$ plane and formally can be written $\vect{w}=x(\vect{\hat x}+\vect{\hat y})$.}
\label{fig:re_rv}
\end{figure}

The radii $R_v(q)$, $R_e(q)$ and corresponding radius vectors of centers of those spheres $P_v(q)$, $P_e(q)$ can be calculated analytically. This can be done by parameterizing superballs cross-section borders for $x>0$, $y>0$, $y>0$ (see Fig.~\ref{fig:re_rv}): 
\begin{align}
 &x=\frac{a}{2}\cos(t)^{\frac{1}{q}}; \quad &w=\frac{a}{2}2^{\frac{q-1}{2q}}\cos(t)^{\frac{1}{q}};\\
 &y=\frac{a}{2}\sin(t)^{\frac{1}{q}}; \quad &z=\frac{a}{2}\sin(t)^{\frac{1}{q}},
 \label{eq:par}
\end{align}
and calculating curvature at $t=\frac{\pi}{4}$ or $t=\arctan[2^{-q/4}]$ for edge midpoint and vertices respectively as in\cite{Kantorovich}. The general expression especially for radius vectors $P_v$ is rather lengthy therefor we will not present here. For $q=2$ one finds that $R_e(2)=\frac{\sqrt[4]{2}}{4}a$, $R_v(2)=\frac{\sqrt[4]{3}}{6}a$ and $P_e(2)=\fiek{\frac{1}{3\sqrt[4]{2}}a, \frac{1}{3\sqrt[4]{2}}a, 0}$ and  $P_v(2)=\fiek{\frac{1}{3\sqrt[4]{3}}a, \frac{1}{3\sqrt[4]{3}}a, \frac{1}{3\sqrt[4]{3}}a}$. 

To make this approximation suitable for larger values of $q$ we add two extra spheres to  every edge (magenta spheres in Fig.~\ref{fig:sball1}) as was done \cite{Philipse2018}. We chose to place those spheres at $t\approx\frac{\pi}{30}$ in $wz$ plane. The radius of those spheres can be obtain from  principal curvatures of superball at given point which can be obtained  by parametrisizing the surface of superball similar to \ref{eq:par} and using expressions from \cite{Curvature} for calculations. This then gives 45 spheres in total and one option is shown in Fig.~\ref{fig:sball1} left. To explain the differences with our analytical calculations  we repeat MD simulations also with an approximation of 93 spheres which shown in Fig.~\ref{fig:sball1} right. There we added symmetrically 8 spheres to each face. The radius of newly added spheres has to be smaller than calculated from curvature at this point. It turns out that in this case the  radius is limited by the fact that centers of spheres should not be to close. Otherwise the potential energy of steric repulsion for superball becomes so large that the round-off errors start to play crucial role.

\section{Energy minimization}
    \label{sec:app_B}
In this section we demonstrate how the total energy of two hematite cube system can be minimized analytically in the absence of gravity for the first case of moment orientation. This is the simplest case, but similarly it can be done for all other cases, although number of variational parameters increases and equations becomes more complicated for more particles. 

The coordinates of center for one cube can be chosen arbitrary, so we put it at the origin of coordinate system and assign index zero to it. The radius vector of the first cube's center is $\tilde{\vect{r}}_0=\fiek{0,0,0}$. Instead of rotating the first cube, we fix the orientation of the first cube and rotate magnetic field such that energy is minimal (this can not be done in the case of gravity, but it leads to the same results as stated before). The orientation of the first cube is chosen such that all edges are parallel to either $x$ or $y$, or $z$ and the orientation of magnetic moment is $\tilde{\vect{\mu}}_0=\fiek{s,s,\sqrt{1-2s^2}}$ with $0\leq s\leq\sqrt{3}/3$ for non-negative angles $\phi$ and $\tilde{\vect{\mu}}_0=\fiek{s,\sqrt{1-2s^2},s}$ with $\sqrt{3}/3< s\leq\sqrt{2}/2$ for negative angles $\phi$. The value of $s$ for given angle $\phi$  can be easily determined as $\tilde{\vect{\mu}}_0\sim\fiek{1,1,\sqrt{2}\tan(\arctan(\sqrt{2}/2)+\phi)}$ which leads for $s=\sqrt{2}/2\cos(\arctan(\sqrt{2}/2)+\phi)$. If we do so then magnetic field has two free parameters. Quite convenient it is to write in form $\tilde{\vect B}= \tilde B \fiek{\sin(\eta)\cos(\zeta),\sin(\eta)\sin(\zeta),\cos(\eta)}$, where $0\leq\eta\leq\pi$ and $0\leq\zeta<2\pi$. Note that angle $\zeta$ in not defined if $\sin(\eta)=0$.

Two cubes which touch with faces can be arranged in 36 different ways (each cube has six faces). The number of combination can be reduced by factor of two as the ordering which is the first and which is the second cube is not important, however, there are still 18 cases. At this point one can argue or check numerically that energy is minimal if the first cubes top face touches the bottom face of the second cube and second cube can be rotated by arbitrary angle  $\xi\in[0;2\pi)$ around $z$ axis. Thus the magnetic moment and radius vector of the second cube can be written as 
{\small
\begin{align}
 \tilde{\vect{\mu}}_1&=\{s[\cos(\xi)-\sin(\xi)],s[\cos(\xi)+\sin(\xi)],\sqrt{1-2s^2}\},\\
 \tilde{\vect{r}}_1&=\fiek{b,c,1},\quad \tilde{\vect{r}}_{1\,0}= \tilde{\vect{r}}_{1}- \tilde{\vect{r}}_{0}=\fiek{b,c,1},
\end{align}
}
with $b,c\in[-1,1]$. For the positive angle $\phi$ due to symmetry actually $c=b$. For negative angles $\phi$ the parameter ratio is $c/b=\sqrt{1-2s^2}/s$ which is the ratio of corresponding components of $\tilde{ \vect{\mu}}_0$.

As positive and negative angles $\phi$ requires slightly different treatment, lets look at positive angles $\phi$. Note that expressions for energy bellow are given also for negative $\phi$. To find the energetically favorable configuration we have to find the global energy minimum for every value of $s\in[0^\circ,\sqrt{3}/3]$ by varying parameters $\eta$, $\zeta$, $\xi$, $b$. The total energy in this case reads:
{\small
\begin{equation}
\begin{split}
\tilde E_{tot}&=\frac{\tilde{\vect \mu}_0\cdot\tilde {\vect  \mu}_1}{\tilde r_{1\,0}^3}-\frac{3(\tilde{\vect  \mu}_0\cdot\tilde{\vect  r}_{1\,0})(\tilde{\vect\mu}_1\cdot\tilde{\vect r}_{1\,0})}{\tilde r_{1\,0}^5}-(\tilde{\vect{\mu}}_0+\tilde{\vect{\mu}}_1)\cdot\tilde{\vect{B}}. 
\end{split}
\label{eq:eos1}
\end{equation}
}

Analytically this is hard task in general case, however, doable in this case as function $\tilde E_{tot}(\eta, \zeta, \xi, b)$ for fixed $s$ and $\tilde B$ has only up to 12 extremes in given parameter range. To find global minimum one can search extreme points, determine which of them is local minima from  Hessian matrix  and choose the one with the lowest energy. Extremes $\vect P=\fiek{\eta, \zeta, \xi, b} $ can be found by setting partial derivatives of $\tilde E_{tot}(\eta, \zeta, \xi, b)$ equal to zero and solve the resulting equations simultaneously. In this case it is particularly simple, as partial derivative to $\zeta$ leads to
\begin{equation}
 s\cos(\xi/2)\sin(\eta)[\sin(\zeta - \xi/2)-\cos(\zeta - \xi/2)] =0\,,
 \label{eq:2app1}
\end{equation}
which means that one have to solve only system of three coupled equations. By choosing either $\cos(\xi/2)=0$ or $\sin(\eta)$ (one implies other) leads to first local minimum $\vect P^1_{min}=\fiek{0, 
\forall\zeta, \pi, 0 }$ and up to five other extremes. Note that as $\sin(\eta)=0$ parameter $\zeta$ is arbitrary and thus term ``point'' used in broader sense. The energy of this configuration is
\begin{align}
 \tilde E_{tot}^1(s, \tilde B)&= F_1(s)+G_1(s)\tilde B,\\
 F_1(s)&=
\begin{cases}
-2+2s^2, \text{ for } 0\leq s\leq\sqrt{3}/3,\\
-1-s^2, \text{ for } \sqrt{3}/3\leq s\leq\sqrt{2}/2,\\
\end{cases}\\
G_1(s)&=
\begin{cases}
-2\sqrt{1-2s^2}, \text{ for } 0\leq s\leq\sqrt{3}/3,\\
-2s, \text{ for } \sqrt{3}/3\leq s\leq\sqrt{2}/2.
\end{cases}
\end{align}

\begin{figure}[h]
\includegraphics[width=0.48\columnwidth]{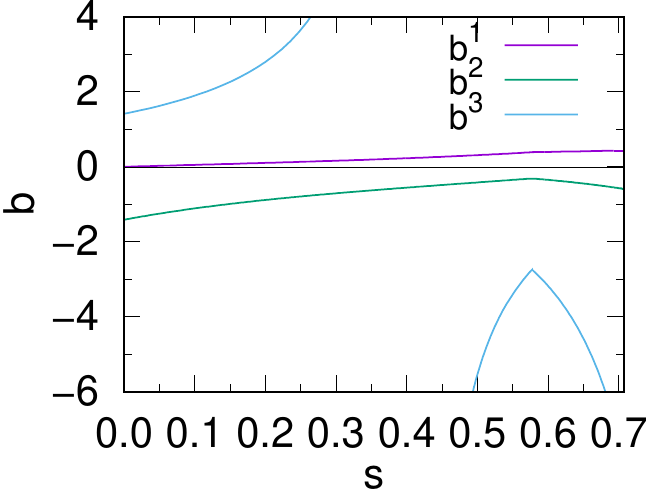}\hfill
\includegraphics[width=0.48\columnwidth]{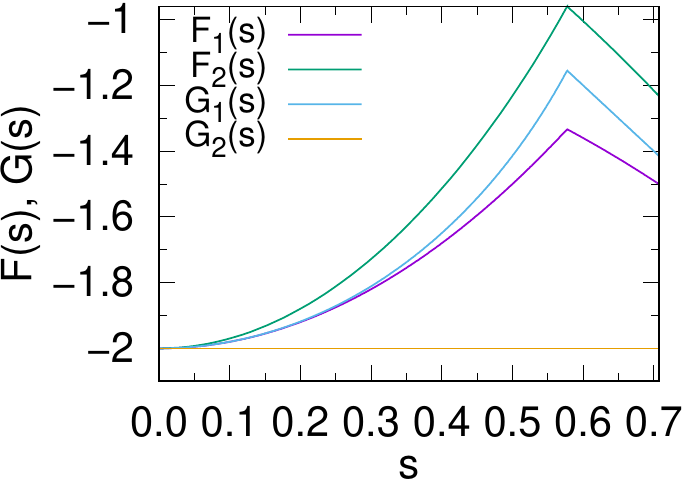}
\caption{\textbf{Left figure:} Roots of Eq. \eqref{eq:2app1} and analog equation for negatives values of $\phi$. \textbf{Left figure:} Functions $F(s)$ and $G(s)$ for first and the second minima.}
\label{fig:root}
\end{figure}

By choosing $sin(\zeta - \xi/2)=\cos(\zeta - \xi/2)$ one is able to find 
the second local minimum $\vect P^2_{min}=\fiek{\arcsin(\sqrt(2)s), \pi/4, 0, b^{+}}$ and up to five other extremes. Here $b^{+}$ is positive solution of
{\small
\begin{equation}
 b (2 - 7 s^2) - s \sqrt{1 - 2 s^2} + 8 b^2 s \sqrt{1 - 2 s^2} + 
 b^3 (-1 + 6 s^2)=0
\end{equation}
} in the range $b\in[0^\circ,1]$. As one can see from left graph of Fig.~\ref{fig:root} for all value of $s$ in given range there is exactly one solution.

The energy of the second minimum can be written formally in the form
\begin{equation}
\tilde E_{tot}^2(s, \tilde B)= F_2(s)+G_2(s)\tilde B, G_2(s)=-2.
\end{equation}
The function $F_2(s)$ is rather complicated complicated expression therefore it is shown in right graph of Fig.~\ref{fig:root}. From Fig.~\ref{fig:root} one can clearly see that $F_2(s)\geq F_1(s)$ and $G_1(s)> G_2(s)$, therefore for all $s$ except $s=0$ at small fields $\tilde B$ first minimum is the global minimum of the system, but at stronger field the second minimum is the global minimum. The critical value of $\tilde B$ one can find by setting  
$E_{tot}^1(s, \tilde B)=E_{tot}^2(s, \tilde B)$. The determined value matches exactly with the one determined numerically and shown in Fig. \ref{fig:bcrit}. For $s=0$ the parameter $\xi$ is not defined therefore both minima are the same point and  there is only one energetically favorable configuration independent of $\tilde B$. 

For more particles and in case of gravity  the analytical equation become impractical to solve therefore we search for global minimum numerically using Simplex algorithm of Nelder and Mead \cite{minimization} implemented in GNU Scientific Library \cite{GSL}. For longer chains without kinks one finds that there are still two configurations - one which is favorable at low fields and the one which is at field strength above critical value.

\section{Estimation of error caused by dipole approximation}
    \label{sec:app_C}
Hematite is a weak ferromagnetic material and has a permanent spontaneous  magnetization $M_s=2.2\times10^3\,\mathrm{A/m}$ \cite{Philipse2018,lowrie_2007}. Micron-sized hematite particles can be synthesized in different shapes: cubes, disks, ellipsoids, peanuts, and others \cite{Rossi_phd,hematite_col_syntheis,hematite_col_syntheis2,Rossi2021}. These particles have high coercivity (up to 2T) \cite{LIU2010387,lowrie_2007} and those particles can be single domain up to 15 $\mu m$ size \cite{lowrie_2007}. Let's estimate whether our 1.5 $\mu m$ are single-domain particle or not. From \cite{LIU2010387} we estimate that coercivity of our sample is $B_c=550\, \mathrm{mT}$, (we obtain similar value calculated from measured magnetization curve of a our samples). Thus, by applying external magnetic field of $0.7\, \mathrm{mT}$ in experiment, the orientation of magnetic moment in cube does not changes.

Assuming uniaxial anisotropy we can estimate that uniaxial anisotropy constant is  
\begin{equation}
 K_u=\frac{M_sB_c}{2} =6.05 \cdot 10^2 \,\mathrm{J/m^3}. 
\end{equation}
Now using open-source micromagnetic software OOMMF \cite{oommf} we can check whether our particles are really single-domane ones. To do this we assume exchange stiffness of $A=10 \,\mathrm{pJ/m}$. 

From calculations we see that independent of orientation of easy axis of magnetization our 1.5 $\mu m$ cubes are still single-domain.  Assuming the same values for $K_u$ and $A$ we can check that particles remain single-domane even up to $15\,\mathrm{ \mu m}$ as stated in \cite{lowrie_2007}.

For $1.5\, \mathrm{\mu m}$ and larger hematite particles with cubic shape we do not observe that energetically favorable state is a pronounced flower state \cite{BJORK2021168057,BONILLA2017394} as for magnetite. In our case the magnetization is almost constant. In the case of the constant magnetization it is convenient to introduce the magnetostatic potential $\phi_i(r)$ due to the magnetostatic surface charges $\sigma_{il}$ \cite{Mag_pot,chikazumi2009physics}. For hematite it is possible as it has relatively high anisotropy and weak magnetization, therefore the volume distributions of magnetostatic  charges for hematite are negligible even when magnetization is not perpendicular to the face of a cube.

The surface charge is constant on each cubes face and can be calculated using $\sigma_{il}=\vect{M}_i\cdot \vect{n}_{il}$, where $\vect{n}_{il}$ is $i$-th cube, $l$-th face outer normal. The analogy to electrostatics allow us to calculate the exact interaction energy of two cubes
\begin{equation}
 W_{ij}=\sum_{l=0}^6 \iint \sigma_{jl} \phi_i\, \mathrm{d}S,
\end{equation}
with
\begin{equation}
 \phi_i(r)=\sum_{l=0}^6 \iint \frac{\sigma_{il}}{\sqrt{(r-r')^2}}\, \mathrm{d^2}r'  .
\end{equation}

Unfortunately,  in general case it is hard numerically to evaluate  those integrals and analytical calculations  are very lengthy and complicated. However, for the case when cube faces and edges are parallel (this corresponds to the restriction that angle $\xi\in{0; \pi/2;\pi; 3\pi/2}$ in the Appendix~\ref{sec:app_B} ), in paper \cite{Mag_pot} there is analytical expression available. The exact energy expression is quite lengthy (more than hundred terms), therefore, we put just reference to \cite{Mag_pot} and mention that in Eq.~18 of \cite{Mag_pot} there is small typo. The term $\mathrm{ln}(R-1)$ have to be replaced with $\mathrm{ln}(R-Y)$. But this allows us to estimate what error dipole approximation introduces. 

\begin{figure}[h]
\includegraphics[width=0.99\columnwidth]{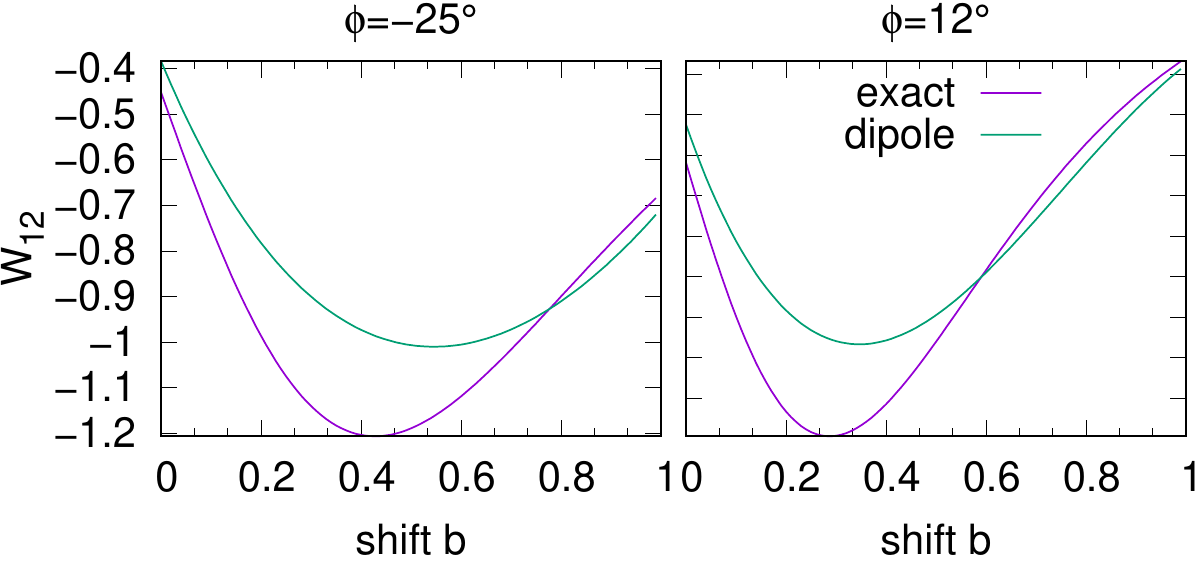}
\caption{ Two cube magnetic interaction energy (exact and using dipole approximation) vs. shift b (see Appendix~\ref{sec:app_B} for definition) for magnetic moment orientation $\phi=-25^\circ$ and $\phi=12^\circ$. }
\label{fig:int_pot}
\end{figure}

\begin{figure}[htbp]
\includegraphics[width=0.99\columnwidth]{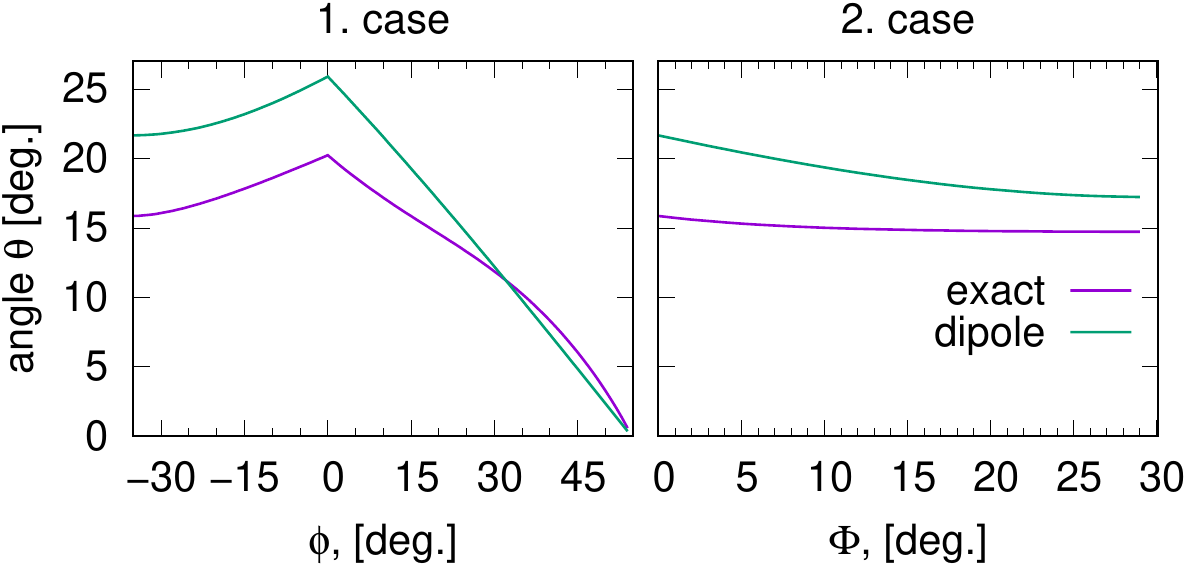}
\caption{Estimation of error that introduces dipole approximation for angle $\theta$ in the case of cubic hematite particles by recalculating Fig~\ref{fig:theta_} using exact magnetic interaction.}
\label{fig:theta_error}
\end{figure}

Comparing the exact two cube magnetic interaction energy with the results calculated using dipole approximation, we confirm findings of \cite{Mag_pot}. The dipole approximation introduces a relative error which for some cube configurations is as large as 18\%. The error is the biggest in the case when magnetic moment is along $z$-axis ($\phi \approx 55^\circ$) and second cube is on top of the first one. 18\% is significant, but, concerning configurations, dipole approximation leads  only to quantitative changes but not qualitative ones. This can be understood from Fig.~\ref{fig:int_pot} where is shown how changes the two cube magnetic interaction energy when one cube is shifted for two different magnetization orientations. Qualitatively the potentials are similar. Dipole approximation for small shifts overestimate interaction energy as the result interaction potentials in dipole approximation are deeper and narrower. As the result also the position of potential minimums shifts. However, the minima count is the same therefore dipole approximation do not change count of local minima in  Appendix~\ref{sec:app_B} and moreover the shift ration $c/b$ is still exactly the same, therefore qualitatively there are no changes. Quantitatively dipole approximation, when shift $b$ in not very small, overestimate angles $\theta$ as can be seen from Fig.~\ref{fig:theta_error}. The overestimation can be as large as $5^\circ$ for $\phi\approx 35^\circ$ or $\Phi\approx 0^\circ$. For  $\phi\approx 12^\circ$, which should be actual orientation of magnetic moment in a hematite cube, the dipole  approximation overestimates the angle $\theta$ by $\approx3^\circ$. This is not negligible, however, knowing this, it is possible to account for this error.

It is also clear that dipole approximation causes distributions for angle $\theta$ at finite temperatures to be more narrow as potentials (Fig.~\ref{fig:int_pot}) in the dipole approximation case are deeper narrower. However, dipole approximation error in this case is order smaller than the error which is in experiment due to cube size distribution.

\bibliography{biblography}

\begin{thebibliography}{40}%
\makeatletter
\providecommand \@ifxundefined [1]{%
 \@ifx{#1\undefined}
}%
\providecommand \@ifnum [1]{%
 \ifnum #1\expandafter \@firstoftwo
 \else \expandafter \@secondoftwo
 \fi
}%
\providecommand \@ifx [1]{%
 \ifx #1\expandafter \@firstoftwo
 \else \expandafter \@secondoftwo
 \fi
}%
\providecommand \natexlab [1]{#1}%
\providecommand \enquote  [1]{``#1''}%
\providecommand \bibnamefont  [1]{#1}%
\providecommand \bibfnamefont [1]{#1}%
\providecommand \citenamefont [1]{#1}%
\providecommand \href@noop [0]{\@secondoftwo}%
\providecommand \href [0]{\begingroup \@sanitize@url \@href}%
\providecommand \@href[1]{\@@startlink{#1}\@@href}%
\providecommand \@@href[1]{\endgroup#1\@@endlink}%
\providecommand \@sanitize@url [0]{\catcode `\\12\catcode `\$12\catcode
  `\&12\catcode `\#12\catcode `\^12\catcode `\_12\catcode `\%12\relax}%
\providecommand \@@startlink[1]{}%
\providecommand \@@endlink[0]{}%
\providecommand \url  [0]{\begingroup\@sanitize@url \@url }%
\providecommand \@url [1]{\endgroup\@href {#1}{\urlprefix }}%
\providecommand \urlprefix  [0]{URL }%
\providecommand \Eprint [0]{\href }%
\providecommand \doibase [0]{https://doi.org/}%
\providecommand \selectlanguage [0]{\@gobble}%
\providecommand \bibinfo  [0]{\@secondoftwo}%
\providecommand \bibfield  [0]{\@secondoftwo}%
\providecommand \translation [1]{[#1]}%
\providecommand \BibitemOpen [0]{}%
\providecommand \bibitemStop [0]{}%
\providecommand \bibitemNoStop [0]{.\EOS\space}%
\providecommand \EOS [0]{\spacefactor3000\relax}%
\providecommand \BibitemShut  [1]{\csname bibitem#1\endcsname}%
\let\auto@bib@innerbib\@empty
\bibitem [{\citenamefont {Rossi}(2012)}]{Rossi_phd}%
  \BibitemOpen
  \bibfield  {author} {\bibinfo {author} {\bibfnamefont {L.}~\bibnamefont
  {Rossi}},\ }\emph {\bibinfo {title} {Colloidal Superballs}},\ \href@noop {}
  {Ph.D. thesis},\ \bibinfo  {school} {Utrecht University}, \bibinfo {address}
  {Utrecht, Nederlands} (\bibinfo {year} {2012})\BibitemShut {NoStop}%
\bibitem [{\citenamefont {Kusior}\ \emph {et~al.}(2019)\citenamefont {Kusior},
  \citenamefont {Michalec}, \citenamefont {Jelen},\ and\ \citenamefont
  {Radecka}}]{hematite_col_syntheis}%
  \BibitemOpen
  \bibfield  {author} {\bibinfo {author} {\bibfnamefont {A.}~\bibnamefont
  {Kusior}}, \bibinfo {author} {\bibfnamefont {K.}~\bibnamefont {Michalec}},
  \bibinfo {author} {\bibfnamefont {P.}~\bibnamefont {Jelen}},\ and\ \bibinfo
  {author} {\bibfnamefont {M.}~\bibnamefont {Radecka}},\ }\bibfield  {title}
  {\bibinfo {title} {Shaped fe2o3 nanoparticles – synthesis and enhanced
  photocatalytic degradation towards rhb},\ }\href
  {https://doi.org/https://doi.org/10.1016/j.apsusc.2018.12.113} {\bibfield
  {journal} {\bibinfo  {journal} {Applied Surface Science}\ }\textbf {\bibinfo
  {volume} {476}},\ \bibinfo {pages} {342 } (\bibinfo {year}
  {2019})}\BibitemShut {NoStop}%
\bibitem [{\citenamefont {Das}\ \emph {et~al.}(2014)\citenamefont {Das},
  \citenamefont {Mondal},\ and\ \citenamefont
  {Mukherjee}}]{hematite_col_syntheis2}%
  \BibitemOpen
  \bibfield  {author} {\bibinfo {author} {\bibfnamefont {P.}~\bibnamefont
  {Das}}, \bibinfo {author} {\bibfnamefont {B.}~\bibnamefont {Mondal}},\ and\
  \bibinfo {author} {\bibfnamefont {K.}~\bibnamefont {Mukherjee}},\ }\bibfield
  {title} {\bibinfo {title} {Facile synthesis of pseudo-peanut shaped hematite
  iron oxide nano-particles and their promising ethanol and formaldehyde
  sensing characteristics},\ }\href {https://doi.org/10.1039/C4RA03098B}
  {\bibfield  {journal} {\bibinfo  {journal} {RSC Adv.}\ }\textbf {\bibinfo
  {volume} {4}},\ \bibinfo {pages} {31879} (\bibinfo {year}
  {2014})}\BibitemShut {NoStop}%
\bibitem [{\citenamefont {Meijer}\ and\ \citenamefont
  {Rossi}(2021)}]{Rossi2021}%
  \BibitemOpen
  \bibfield  {author} {\bibinfo {author} {\bibfnamefont {J.~M.}\ \bibnamefont
  {Meijer}}\ and\ \bibinfo {author} {\bibfnamefont {L.}~\bibnamefont {Rossi}},\
  }\bibfield  {title} {\bibinfo {title} {Preparation{,} properties{,} and
  applications of magnetic hematite microparticles},\ }\href
  {https://doi.org/10.1039/D0SM01977A} {\bibfield  {journal} {\bibinfo
  {journal} {Soft Matter}\ ,\ } (\bibinfo {year} {2021})}\BibitemShut {NoStop}%
\bibitem [{\citenamefont {Rossi}\ \emph {et~al.}(2018)\citenamefont {Rossi},
  \citenamefont {Donaldson}, \citenamefont {Meijer}, \citenamefont {Petukhov},
  \citenamefont {Kleckner}, \citenamefont {Kantorovich}, \citenamefont
  {Irvine}, \citenamefont {Philipse},\ and\ \citenamefont
  {Sacanna}}]{Philipse2018}%
  \BibitemOpen
  \bibfield  {author} {\bibinfo {author} {\bibfnamefont {L.}~\bibnamefont
  {Rossi}}, \bibinfo {author} {\bibfnamefont {J.~G.}\ \bibnamefont
  {Donaldson}}, \bibinfo {author} {\bibfnamefont {J.-M.}\ \bibnamefont
  {Meijer}}, \bibinfo {author} {\bibfnamefont {A.~V.}\ \bibnamefont
  {Petukhov}}, \bibinfo {author} {\bibfnamefont {D.}~\bibnamefont {Kleckner}},
  \bibinfo {author} {\bibfnamefont {S.~S.}\ \bibnamefont {Kantorovich}},
  \bibinfo {author} {\bibfnamefont {W.~T.~M.}\ \bibnamefont {Irvine}}, \bibinfo
  {author} {\bibfnamefont {A.~P.}\ \bibnamefont {Philipse}},\ and\ \bibinfo
  {author} {\bibfnamefont {S.}~\bibnamefont {Sacanna}},\ }\bibfield  {title}
  {\bibinfo {title} {Self-organization in dipolar cube fluids constrained by
  competing anisotropies},\ }\href {https://doi.org/10.1039/C7SM02174G}
  {\bibfield  {journal} {\bibinfo  {journal} {Soft Matter}\ }\textbf {\bibinfo
  {volume} {14}},\ \bibinfo {pages} {1080} (\bibinfo {year}
  {2018})}\BibitemShut {NoStop}%
\bibitem [{\citenamefont {Lowrie}(2007)}]{lowrie_2007}%
  \BibitemOpen
  \bibfield  {author} {\bibinfo {author} {\bibfnamefont {W.}~\bibnamefont
  {Lowrie}},\ }\href {https://doi.org/10.1017/CBO9780511807107} {\emph
  {\bibinfo {title} {Fundamentals of Geophysics}}},\ \bibinfo {edition} {2nd}\
  ed.\ (\bibinfo  {publisher} {Cambridge University Press},\ \bibinfo {year}
  {2007})\BibitemShut {NoStop}%
\bibitem [{\citenamefont {Massana-Cid}\ \emph {et~al.}(2017)\citenamefont
  {Massana-Cid}, \citenamefont {Martinez-Pedrero}, \citenamefont {Cebers},\
  and\ \citenamefont {Tierno}}]{hem_ellipsoids}%
  \BibitemOpen
  \bibfield  {author} {\bibinfo {author} {\bibfnamefont {H.}~\bibnamefont
  {Massana-Cid}}, \bibinfo {author} {\bibfnamefont {F.}~\bibnamefont
  {Martinez-Pedrero}}, \bibinfo {author} {\bibfnamefont {A.}~\bibnamefont
  {Cebers}},\ and\ \bibinfo {author} {\bibfnamefont {P.}~\bibnamefont
  {Tierno}},\ }\bibfield  {title} {\bibinfo {title} {Orientational dynamics of
  fluctuating dipolar particles assembled in a mesoscopic colloidal ribbon},\
  }\href {https://doi.org/10.1103/PhysRevE.96.012607} {\bibfield  {journal}
  {\bibinfo  {journal} {Phys. Rev. E}\ }\textbf {\bibinfo {volume} {96}},\
  \bibinfo {pages} {012607} (\bibinfo {year} {2017})}\BibitemShut {NoStop}%
\bibitem [{\citenamefont {Martinez-Pedrero}\ \emph
  {et~al.}(2016{\natexlab{a}})\citenamefont {Martinez-Pedrero}, \citenamefont
  {Cebers},\ and\ \citenamefont {Tierno}}]{hem_ellipsoids3}%
  \BibitemOpen
  \bibfield  {author} {\bibinfo {author} {\bibfnamefont {F.}~\bibnamefont
  {Martinez-Pedrero}}, \bibinfo {author} {\bibfnamefont {A.}~\bibnamefont
  {Cebers}},\ and\ \bibinfo {author} {\bibfnamefont {P.}~\bibnamefont
  {Tierno}},\ }\bibfield  {title} {\bibinfo {title} {Orientational dynamics of
  colloidal ribbons self-assembled from microscopic magnetic ellipsoids},\
  }\href {https://doi.org/10.1039/C5SM02823J} {\bibfield  {journal} {\bibinfo
  {journal} {Soft Matter}\ }\textbf {\bibinfo {volume} {12}},\ \bibinfo {pages}
  {3688} (\bibinfo {year} {2016}{\natexlab{a}})}\BibitemShut {NoStop}%
\bibitem [{\citenamefont {Martinez-Pedrero}\ \emph
  {et~al.}(2016{\natexlab{b}})\citenamefont {Martinez-Pedrero}, \citenamefont
  {Cebers},\ and\ \citenamefont {Tierno}}]{hem_ellipsoids2}%
  \BibitemOpen
  \bibfield  {author} {\bibinfo {author} {\bibfnamefont {F.}~\bibnamefont
  {Martinez-Pedrero}}, \bibinfo {author} {\bibfnamefont {A.}~\bibnamefont
  {Cebers}},\ and\ \bibinfo {author} {\bibfnamefont {P.}~\bibnamefont
  {Tierno}},\ }\bibfield  {title} {\bibinfo {title} {Dipolar rings of
  microscopic ellipsoids: Magnetic manipulation and cell entrapment},\ }\href
  {https://doi.org/10.1103/PhysRevApplied.6.034002} {\bibfield  {journal}
  {\bibinfo  {journal} {Phys. Rev. Applied}\ }\textbf {\bibinfo {volume} {6}},\
  \bibinfo {pages} {034002} (\bibinfo {year} {2016}{\natexlab{b}})}\BibitemShut
  {NoStop}%
\bibitem [{\citenamefont {Soni}\ \emph {et~al.}(2019)\citenamefont {Soni},
  \citenamefont {Bililign}, \citenamefont {Magkiriadou}, \citenamefont
  {Sacanna}, \citenamefont {Bartolo}, \citenamefont {Shelley},\ and\
  \citenamefont {Irvine}}]{Soni}%
  \BibitemOpen
  \bibfield  {author} {\bibinfo {author} {\bibfnamefont {V.}~\bibnamefont
  {Soni}}, \bibinfo {author} {\bibfnamefont {E.~S.}\ \bibnamefont {Bililign}},
  \bibinfo {author} {\bibfnamefont {S.}~\bibnamefont {Magkiriadou}}, \bibinfo
  {author} {\bibfnamefont {S.}~\bibnamefont {Sacanna}}, \bibinfo {author}
  {\bibfnamefont {D.}~\bibnamefont {Bartolo}}, \bibinfo {author} {\bibfnamefont
  {M.~J.}\ \bibnamefont {Shelley}},\ and\ \bibinfo {author} {\bibfnamefont
  {W.~T.~M.}\ \bibnamefont {Irvine}},\ }\bibfield  {title} {\bibinfo {title}
  {The odd free surface flows of a colloidal chiral fluid},\ }\href
  {https://doi.org/10.1038/s41567-019-0603-8} {\bibfield  {journal} {\bibinfo
  {journal} {Nature Physics}\ }\textbf {\bibinfo {volume} {15}},\ \bibinfo
  {pages} {1188} (\bibinfo {year} {2019})}\BibitemShut {NoStop}%
\bibitem [{\citenamefont {Petrichenko}\ \emph {et~al.}(2020)\citenamefont
  {Petrichenko}, \citenamefont {Kitenbergs}, \citenamefont {Brics},
  \citenamefont {Dubois}, \citenamefont {Perzynski},\ and\ \citenamefont
  {C{\={e}}bers}}]{Petrichenko_2020}%
  \BibitemOpen
  \bibfield  {author} {\bibinfo {author} {\bibfnamefont {O.}~\bibnamefont
  {Petrichenko}}, \bibinfo {author} {\bibfnamefont {G.}~\bibnamefont
  {Kitenbergs}}, \bibinfo {author} {\bibfnamefont {M.}~\bibnamefont {Brics}},
  \bibinfo {author} {\bibfnamefont {E.}~\bibnamefont {Dubois}}, \bibinfo
  {author} {\bibfnamefont {R.}~\bibnamefont {Perzynski}},\ and\ \bibinfo
  {author} {\bibfnamefont {A.}~\bibnamefont {C{\={e}}bers}},\ }\bibfield
  {title} {\bibinfo {title} {Swarming of micron-sized hematite cubes in a
  rotating magnetic field {\textendash} experiments},\ }\href
  {https://doi.org/10.1016/j.jmmm.2020.166404} {\bibfield  {journal} {\bibinfo
  {journal} {Journal of Magnetism and Magnetic Materials}\ }\textbf {\bibinfo
  {volume} {500}},\ \bibinfo {pages} {166404} (\bibinfo {year}
  {2020})}\BibitemShut {NoStop}%
\bibitem [{\citenamefont {Aubret}\ \emph {et~al.}(2018)\citenamefont {Aubret},
  \citenamefont {Youssef}, \citenamefont {Sacanna},\ and\ \citenamefont
  {Palacci}}]{Aubret_2018}%
  \BibitemOpen
  \bibfield  {author} {\bibinfo {author} {\bibfnamefont {A.}~\bibnamefont
  {Aubret}}, \bibinfo {author} {\bibfnamefont {M.}~\bibnamefont {Youssef}},
  \bibinfo {author} {\bibfnamefont {S.}~\bibnamefont {Sacanna}},\ and\ \bibinfo
  {author} {\bibfnamefont {J.}~\bibnamefont {Palacci}},\ }\bibfield  {title}
  {\bibinfo {title} {Targeted assembly and synchronization of self-spinning
  microgears},\ }\href {https://doi.org/10.1038/s41567-018-0227-4} {\bibfield
  {journal} {\bibinfo  {journal} {Nature Physics}\ }\textbf {\bibinfo {volume}
  {14}},\ \bibinfo {pages} {1114} (\bibinfo {year} {2018})}\BibitemShut
  {NoStop}%
\bibitem [{\citenamefont {Driscoll}\ \emph {et~al.}(2016)\citenamefont
  {Driscoll}, \citenamefont {Delmotte}, \citenamefont {Youssef}, \citenamefont
  {Sacanna}, \citenamefont {Donev},\ and\ \citenamefont
  {Chaikin}}]{Driscoll_2016}%
  \BibitemOpen
  \bibfield  {author} {\bibinfo {author} {\bibfnamefont {M.}~\bibnamefont
  {Driscoll}}, \bibinfo {author} {\bibfnamefont {B.}~\bibnamefont {Delmotte}},
  \bibinfo {author} {\bibfnamefont {M.}~\bibnamefont {Youssef}}, \bibinfo
  {author} {\bibfnamefont {S.}~\bibnamefont {Sacanna}}, \bibinfo {author}
  {\bibfnamefont {A.}~\bibnamefont {Donev}},\ and\ \bibinfo {author}
  {\bibfnamefont {P.}~\bibnamefont {Chaikin}},\ }\bibfield  {title} {\bibinfo
  {title} {Unstable fronts and motile structures formed by~microrollers},\
  }\href {https://doi.org/10.1038/nphys3970} {\bibfield  {journal} {\bibinfo
  {journal} {Nature Physics}\ }\textbf {\bibinfo {volume} {13}},\ \bibinfo
  {pages} {375} (\bibinfo {year} {2016})}\BibitemShut {NoStop}%
\bibitem [{\citenamefont {Castillo}\ \emph {et~al.}(2014)\citenamefont
  {Castillo}, \citenamefont {Pompe}, \citenamefont {van Mourik}, \citenamefont
  {Verbart}, \citenamefont {Thies-Weesie}, \citenamefont {de~Jongh},\ and\
  \citenamefont {Philipse}}]{Castillo_2014}%
  \BibitemOpen
  \bibfield  {author} {\bibinfo {author} {\bibfnamefont {S.~I.~R.}\
  \bibnamefont {Castillo}}, \bibinfo {author} {\bibfnamefont {C.~E.}\
  \bibnamefont {Pompe}}, \bibinfo {author} {\bibfnamefont {J.}~\bibnamefont
  {van Mourik}}, \bibinfo {author} {\bibfnamefont {D.~M.~A.}\ \bibnamefont
  {Verbart}}, \bibinfo {author} {\bibfnamefont {D.~M.~E.}\ \bibnamefont
  {Thies-Weesie}}, \bibinfo {author} {\bibfnamefont {P.~E.}\ \bibnamefont
  {de~Jongh}},\ and\ \bibinfo {author} {\bibfnamefont {A.~P.}\ \bibnamefont
  {Philipse}},\ }\bibfield  {title} {\bibinfo {title} {Colloidal cubes for the
  enhanced degradation of organic dyes},\ }\href
  {https://doi.org/10.1039/c4ta01373e} {\bibfield  {journal} {\bibinfo
  {journal} {Journal of Materials Chemistry A}\ }\textbf {\bibinfo {volume}
  {2}},\ \bibinfo {pages} {10193} (\bibinfo {year} {2014})}\BibitemShut
  {NoStop}%
\bibitem [{\citenamefont {Meijer}\ \emph {et~al.}(2013)\citenamefont {Meijer},
  \citenamefont {Byelov}, \citenamefont {Rossi}, \citenamefont {Snigirev},
  \citenamefont {Snigireva}, \citenamefont {Philipse},\ and\ \citenamefont
  {Petukhov}}]{Meijer_2013}%
  \BibitemOpen
  \bibfield  {author} {\bibinfo {author} {\bibfnamefont {J.-M.}\ \bibnamefont
  {Meijer}}, \bibinfo {author} {\bibfnamefont {D.~V.}\ \bibnamefont {Byelov}},
  \bibinfo {author} {\bibfnamefont {L.}~\bibnamefont {Rossi}}, \bibinfo
  {author} {\bibfnamefont {A.}~\bibnamefont {Snigirev}}, \bibinfo {author}
  {\bibfnamefont {I.}~\bibnamefont {Snigireva}}, \bibinfo {author}
  {\bibfnamefont {A.~P.}\ \bibnamefont {Philipse}},\ and\ \bibinfo {author}
  {\bibfnamefont {A.~V.}\ \bibnamefont {Petukhov}},\ }\bibfield  {title}
  {\bibinfo {title} {Self-assembly of colloidal hematite cubes: a microradian
  x-ray diffraction exploration of sedimentary crystals},\ }\href
  {https://doi.org/10.1039/c3sm51553b} {\bibfield  {journal} {\bibinfo
  {journal} {Soft Matter}\ }\textbf {\bibinfo {volume} {9}},\ \bibinfo {pages}
  {10729} (\bibinfo {year} {2013})}\BibitemShut {NoStop}%
\bibitem [{\citenamefont {Rossi}\ \emph {et~al.}(2015)\citenamefont {Rossi},
  \citenamefont {Soni}, \citenamefont {Ashton}, \citenamefont {Pine},
  \citenamefont {Philipse}, \citenamefont {Chaikin}, \citenamefont {Dijkstra},
  \citenamefont {Sacanna},\ and\ \citenamefont {Irvine}}]{Rossi2}%
  \BibitemOpen
  \bibfield  {author} {\bibinfo {author} {\bibfnamefont {L.}~\bibnamefont
  {Rossi}}, \bibinfo {author} {\bibfnamefont {V.}~\bibnamefont {Soni}},
  \bibinfo {author} {\bibfnamefont {D.~J.}\ \bibnamefont {Ashton}}, \bibinfo
  {author} {\bibfnamefont {D.~J.}\ \bibnamefont {Pine}}, \bibinfo {author}
  {\bibfnamefont {A.~P.}\ \bibnamefont {Philipse}}, \bibinfo {author}
  {\bibfnamefont {P.~M.}\ \bibnamefont {Chaikin}}, \bibinfo {author}
  {\bibfnamefont {M.}~\bibnamefont {Dijkstra}}, \bibinfo {author}
  {\bibfnamefont {S.}~\bibnamefont {Sacanna}},\ and\ \bibinfo {author}
  {\bibfnamefont {W.~T.~M.}\ \bibnamefont {Irvine}},\ }\bibfield  {title}
  {\bibinfo {title} {Shape-sensitive crystallization in colloidal superball
  fluids},\ }\href {https://doi.org/10.1073/pnas.1415467112} {\bibfield
  {journal} {\bibinfo  {journal} {Proceedings of the National Academy of
  Sciences}\ }\textbf {\bibinfo {volume} {112}},\ \bibinfo {pages} {5286}
  (\bibinfo {year} {2015})},\ \Eprint
  {https://arxiv.org/abs/https://www.pnas.org/content/112/17/5286.full.pdf}
  {https://www.pnas.org/content/112/17/5286.full.pdf} \BibitemShut {NoStop}%
\bibitem [{\citenamefont {Meijer}(2015)}]{Meijer2015}%
  \BibitemOpen
  \bibfield  {author} {\bibinfo {author} {\bibfnamefont {J.-M.}\ \bibnamefont
  {Meijer}},\ }\bibinfo {title} {Preparation and characterization of colloidal
  cubes},\ in\ \href {https://doi.org/10.1007/978-3-319-14809-0_5} {\emph
  {\bibinfo {booktitle} {Colloidal Crystals of Spheres and Cubes in Real and
  Reciprocal Space}}}\ (\bibinfo  {publisher} {Springer International
  Publishing},\ \bibinfo {address} {Cham},\ \bibinfo {year} {2015})\ pp.\
  \bibinfo {pages} {73--87}\BibitemShut {NoStop}%
\bibitem [{\citenamefont {Palacci}\ \emph {et~al.}(2013)\citenamefont
  {Palacci}, \citenamefont {Sacanna}, \citenamefont {Steinberg}, \citenamefont
  {Pine},\ and\ \citenamefont {Chaikin}}]{Palacci_2013}%
  \BibitemOpen
  \bibfield  {author} {\bibinfo {author} {\bibfnamefont {J.}~\bibnamefont
  {Palacci}}, \bibinfo {author} {\bibfnamefont {S.}~\bibnamefont {Sacanna}},
  \bibinfo {author} {\bibfnamefont {A.~P.}\ \bibnamefont {Steinberg}}, \bibinfo
  {author} {\bibfnamefont {D.~J.}\ \bibnamefont {Pine}},\ and\ \bibinfo
  {author} {\bibfnamefont {P.~M.}\ \bibnamefont {Chaikin}},\ }\bibfield
  {title} {\bibinfo {title} {Living crystals of light-activated colloidal
  surfers},\ }\href {https://doi.org/10.1126/science.1230020} {\bibfield
  {journal} {\bibinfo  {journal} {Science}\ }\textbf {\bibinfo {volume}
  {339}},\ \bibinfo {pages} {936} (\bibinfo {year} {2013})}\BibitemShut
  {NoStop}%
\bibitem [{\citenamefont {Ozaki}\ \emph {et~al.}(1986)\citenamefont {Ozaki},
  \citenamefont {Suzuki}, \citenamefont {Takahashi},\ and\ \citenamefont
  {Matijević}}]{hematite1}%
  \BibitemOpen
  \bibfield  {author} {\bibinfo {author} {\bibfnamefont {M.}~\bibnamefont
  {Ozaki}}, \bibinfo {author} {\bibfnamefont {H.}~\bibnamefont {Suzuki}},
  \bibinfo {author} {\bibfnamefont {K.}~\bibnamefont {Takahashi}},\ and\
  \bibinfo {author} {\bibfnamefont {E.}~\bibnamefont {Matijević}},\ }\bibfield
   {title} {\bibinfo {title} {Reversible ordered agglomeration of hematite
  particles due to weak magnetic interactions},\ }\href
  {https://doi.org/https://doi.org/10.1016/0021-9797(86)90207-9} {\bibfield
  {journal} {\bibinfo  {journal} {Journal of Colloid and Interface Science}\
  }\textbf {\bibinfo {volume} {113}},\ \bibinfo {pages} {76 } (\bibinfo {year}
  {1986})}\BibitemShut {NoStop}%
\bibitem [{\citenamefont {Ozaki}\ \emph {et~al.}(1988)\citenamefont {Ozaki},
  \citenamefont {Egami}, \citenamefont {Sugiyama},\ and\ \citenamefont
  {Matijević}}]{hematite2}%
  \BibitemOpen
  \bibfield  {author} {\bibinfo {author} {\bibfnamefont {M.}~\bibnamefont
  {Ozaki}}, \bibinfo {author} {\bibfnamefont {T.}~\bibnamefont {Egami}},
  \bibinfo {author} {\bibfnamefont {N.}~\bibnamefont {Sugiyama}},\ and\
  \bibinfo {author} {\bibfnamefont {E.}~\bibnamefont {Matijević}},\ }\bibfield
   {title} {\bibinfo {title} {Agglomeration in colloidal hematite dispersions
  due to weak magnetic interactions: Ii. the effects of particle size and
  shape},\ }\href
  {https://doi.org/https://doi.org/10.1016/0021-9797(88)90114-2} {\bibfield
  {journal} {\bibinfo  {journal} {Journal of Colloid and Interface Science}\
  }\textbf {\bibinfo {volume} {126}},\ \bibinfo {pages} {212 } (\bibinfo {year}
  {1988})}\BibitemShut {NoStop}%
\bibitem [{\citenamefont {Park}\ \emph {et~al.}(1996)\citenamefont {Park},
  \citenamefont {Shindo}, \citenamefont {Waseda},\ and\ \citenamefont
  {Sugimoto}}]{hematite3}%
  \BibitemOpen
  \bibfield  {author} {\bibinfo {author} {\bibfnamefont {G.-S.}\ \bibnamefont
  {Park}}, \bibinfo {author} {\bibfnamefont {D.}~\bibnamefont {Shindo}},
  \bibinfo {author} {\bibfnamefont {Y.}~\bibnamefont {Waseda}},\ and\ \bibinfo
  {author} {\bibfnamefont {T.}~\bibnamefont {Sugimoto}},\ }\bibfield  {title}
  {\bibinfo {title} {Internal structure analysis of monodispersed pseudocubic
  hematite particles by electron microscopy},\ }\href
  {https://doi.org/https://doi.org/10.1006/jcis.1996.0021} {\bibfield
  {journal} {\bibinfo  {journal} {Journal of Colloid and Interface Science}\
  }\textbf {\bibinfo {volume} {177}},\ \bibinfo {pages} {198 } (\bibinfo {year}
  {1996})}\BibitemShut {NoStop}%
\bibitem [{\citenamefont {Buzzaccaro}\ \emph {et~al.}(2008)\citenamefont
  {Buzzaccaro}, \citenamefont {Tripodi}, \citenamefont {Rusconi}, \citenamefont
  {Vigolo},\ and\ \citenamefont {Piazza}}]{grav_len}%
  \BibitemOpen
  \bibfield  {author} {\bibinfo {author} {\bibfnamefont {S.}~\bibnamefont
  {Buzzaccaro}}, \bibinfo {author} {\bibfnamefont {A.}~\bibnamefont {Tripodi}},
  \bibinfo {author} {\bibfnamefont {R.}~\bibnamefont {Rusconi}}, \bibinfo
  {author} {\bibfnamefont {D.}~\bibnamefont {Vigolo}},\ and\ \bibinfo {author}
  {\bibfnamefont {R.}~\bibnamefont {Piazza}},\ }\bibfield  {title} {\bibinfo
  {title} {Kinetics of sedimentation in colloidal suspensions},\ }\href
  {https://doi.org/10.1088/0953-8984/20/49/494219} {\bibfield  {journal}
  {\bibinfo  {journal} {Journal of Physics: Condensed Matter}\ }\textbf
  {\bibinfo {volume} {20}},\ \bibinfo {pages} {494219} (\bibinfo {year}
  {2008})}\BibitemShut {NoStop}%
\bibitem [{\citenamefont {Donaldson}\ \emph {et~al.}(2017)\citenamefont
  {Donaldson}, \citenamefont {Linse},\ and\ \citenamefont
  {Kantorovich}}]{Kantorovich}%
  \BibitemOpen
  \bibfield  {author} {\bibinfo {author} {\bibfnamefont {J.~G.}\ \bibnamefont
  {Donaldson}}, \bibinfo {author} {\bibfnamefont {P.}~\bibnamefont {Linse}},\
  and\ \bibinfo {author} {\bibfnamefont {S.~S.}\ \bibnamefont {Kantorovich}},\
  }\bibfield  {title} {\bibinfo {title} {How cube-like must magnetic
  nanoparticles be to modify their self-assembly?},\ }\href
  {https://doi.org/10.1039/C7NR01245D} {\bibfield  {journal} {\bibinfo
  {journal} {Nanoscale}\ }\textbf {\bibinfo {volume} {9}},\ \bibinfo {pages}
  {6448} (\bibinfo {year} {2017})}\BibitemShut {NoStop}%
\bibitem [{\citenamefont {Weik}\ \emph {et~al.}(2019)\citenamefont {Weik},
  \citenamefont {Weeber}, \citenamefont {Szuttor}, \citenamefont
  {Breitsprecher}, \citenamefont {de~Graaf}, \citenamefont {Kuron},
  \citenamefont {Landsgesell}, \citenamefont {Menke}, \citenamefont {Sean},\
  and\ \citenamefont {Holm}}]{Espresso}%
  \BibitemOpen
  \bibfield  {author} {\bibinfo {author} {\bibfnamefont {F.}~\bibnamefont
  {Weik}}, \bibinfo {author} {\bibfnamefont {R.}~\bibnamefont {Weeber}},
  \bibinfo {author} {\bibfnamefont {K.}~\bibnamefont {Szuttor}}, \bibinfo
  {author} {\bibfnamefont {K.}~\bibnamefont {Breitsprecher}}, \bibinfo {author}
  {\bibfnamefont {J.}~\bibnamefont {de~Graaf}}, \bibinfo {author}
  {\bibfnamefont {M.}~\bibnamefont {Kuron}}, \bibinfo {author} {\bibfnamefont
  {J.}~\bibnamefont {Landsgesell}}, \bibinfo {author} {\bibfnamefont
  {H.}~\bibnamefont {Menke}}, \bibinfo {author} {\bibfnamefont
  {D.}~\bibnamefont {Sean}},\ and\ \bibinfo {author} {\bibfnamefont
  {C.}~\bibnamefont {Holm}},\ }\bibfield  {title} {\bibinfo {title} {Espresso
  4.0 -- an extensible software package for simulating soft matter systems},\
  }\href {https://doi.org/10.1140/epjst/e2019-800186-9} {\bibfield  {journal}
  {\bibinfo  {journal} {The European Physical Journal Special Topics}\ }\textbf
  {\bibinfo {volume} {227}},\ \bibinfo {pages} {1789} (\bibinfo {year}
  {2019})}\BibitemShut {NoStop}%
\bibitem [{\citenamefont {Arnold}\ \emph {et~al.}(2013)\citenamefont {Arnold},
  \citenamefont {Lenz}, \citenamefont {Kesselheim}, \citenamefont {Weeber},
  \citenamefont {Fahrenberger}, \citenamefont {Roehm}, \citenamefont
  {Ko{\v{s}}ovan},\ and\ \citenamefont {Holm}}]{Espresso2}%
  \BibitemOpen
  \bibfield  {author} {\bibinfo {author} {\bibfnamefont {A.}~\bibnamefont
  {Arnold}}, \bibinfo {author} {\bibfnamefont {O.}~\bibnamefont {Lenz}},
  \bibinfo {author} {\bibfnamefont {S.}~\bibnamefont {Kesselheim}}, \bibinfo
  {author} {\bibfnamefont {R.}~\bibnamefont {Weeber}}, \bibinfo {author}
  {\bibfnamefont {F.}~\bibnamefont {Fahrenberger}}, \bibinfo {author}
  {\bibfnamefont {D.}~\bibnamefont {Roehm}}, \bibinfo {author} {\bibfnamefont
  {P.}~\bibnamefont {Ko{\v{s}}ovan}},\ and\ \bibinfo {author} {\bibfnamefont
  {C.}~\bibnamefont {Holm}},\ }\bibfield  {title} {\bibinfo {title} {Espresso
  3.1: Molecular dynamics software for coarse-grained models},\ }in\ \href@noop
  {} {\emph {\bibinfo {booktitle} {Meshfree Methods for Partial Differential
  Equations VI}}},\ \bibinfo {editor} {edited by\ \bibinfo {editor}
  {\bibfnamefont {M.}~\bibnamefont {Griebel}}\ and\ \bibinfo {editor}
  {\bibfnamefont {M.~A.}\ \bibnamefont {Schweitzer}}}\ (\bibinfo  {publisher}
  {Springer Berlin Heidelberg},\ \bibinfo {address} {Berlin, Heidelberg},\
  \bibinfo {year} {2013})\ pp.\ \bibinfo {pages} {1--23}\BibitemShut {NoStop}%
\bibitem [{\citenamefont {Limbach}\ \emph {et~al.}(2006)\citenamefont
  {Limbach}, \citenamefont {Arnold}, \citenamefont {Mann},\ and\ \citenamefont
  {Holm}}]{Espresso3}%
  \BibitemOpen
  \bibfield  {author} {\bibinfo {author} {\bibfnamefont {H.}~\bibnamefont
  {Limbach}}, \bibinfo {author} {\bibfnamefont {A.}~\bibnamefont {Arnold}},
  \bibinfo {author} {\bibfnamefont {B.}~\bibnamefont {Mann}},\ and\ \bibinfo
  {author} {\bibfnamefont {C.}~\bibnamefont {Holm}},\ }\bibfield  {title}
  {\bibinfo {title} {Espresso—an extensible simulation package for research
  on soft matter systems},\ }\href
  {https://doi.org/https://doi.org/10.1016/j.cpc.2005.10.005} {\bibfield
  {journal} {\bibinfo  {journal} {Computer Physics Communications}\ }\textbf
  {\bibinfo {volume} {174}},\ \bibinfo {pages} {704 } (\bibinfo {year}
  {2006})}\BibitemShut {NoStop}%
\bibitem [{\citenamefont {Donaldson}\ and\ \citenamefont
  {Kantorovich}(2015)}]{Kantorovich2}%
  \BibitemOpen
  \bibfield  {author} {\bibinfo {author} {\bibfnamefont {J.~G.}\ \bibnamefont
  {Donaldson}}\ and\ \bibinfo {author} {\bibfnamefont {S.~S.}\ \bibnamefont
  {Kantorovich}},\ }\bibfield  {title} {\bibinfo {title} {Directional
  self-assembly of permanently magnetised nanocubes in quasi two dimensional
  layers},\ }\href {https://doi.org/10.1039/C4NR07101H} {\bibfield  {journal}
  {\bibinfo  {journal} {Nanoscale}\ }\textbf {\bibinfo {volume} {7}},\ \bibinfo
  {pages} {3217} (\bibinfo {year} {2015})}\BibitemShut {NoStop}%
\bibitem [{\citenamefont {Weeks}\ \emph {et~al.}(1971)\citenamefont {Weeks},
  \citenamefont {Chandler},\ and\ \citenamefont {Andersen}}]{WCA}%
  \BibitemOpen
  \bibfield  {author} {\bibinfo {author} {\bibfnamefont {J.~D.}\ \bibnamefont
  {Weeks}}, \bibinfo {author} {\bibfnamefont {D.}~\bibnamefont {Chandler}},\
  and\ \bibinfo {author} {\bibfnamefont {H.~C.}\ \bibnamefont {Andersen}},\
  }\bibfield  {title} {\bibinfo {title} {Role of repulsive forces in
  determining the equilibrium structure of simple liquids},\ }\href
  {https://doi.org/10.1063/1.1674820} {\bibfield  {journal} {\bibinfo
  {journal} {The Journal of Chemical Physics}\ }\textbf {\bibinfo {volume}
  {54}},\ \bibinfo {pages} {5237} (\bibinfo {year} {1971})},\ \Eprint
  {https://arxiv.org/abs/https://doi.org/10.1063/1.1674820}
  {https://doi.org/10.1063/1.1674820} \BibitemShut {NoStop}%
\bibitem [{\citenamefont {Aoshima}\ \emph {et~al.}(2012)\citenamefont
  {Aoshima}, \citenamefont {Ozaki},\ and\ \citenamefont {Satoh}}]{Aoshima}%
  \BibitemOpen
  \bibfield  {author} {\bibinfo {author} {\bibfnamefont {M.}~\bibnamefont
  {Aoshima}}, \bibinfo {author} {\bibfnamefont {M.}~\bibnamefont {Ozaki}},\
  and\ \bibinfo {author} {\bibfnamefont {A.}~\bibnamefont {Satoh}},\ }\bibfield
   {title} {\bibinfo {title} {Structural analysis of self-assembled lattice
  structures composed of cubic hematite particles},\ }\href
  {https://doi.org/10.1021/jp301645x} {\bibfield  {journal} {\bibinfo
  {journal} {The Journal of Physical Chemistry C}\ }\textbf {\bibinfo {volume}
  {116}},\ \bibinfo {pages} {17862} (\bibinfo {year} {2012})},\ \Eprint
  {https://arxiv.org/abs/https://doi.org/10.1021/jp301645x}
  {https://doi.org/10.1021/jp301645x} \BibitemShut {NoStop}%
\bibitem [{\citenamefont {Okada}\ and\ \citenamefont
  {Satoh}(2018)}]{Hematite_along_diogonal}%
  \BibitemOpen
  \bibfield  {author} {\bibinfo {author} {\bibfnamefont {K.}~\bibnamefont
  {Okada}}\ and\ \bibinfo {author} {\bibfnamefont {A.}~\bibnamefont {Satoh}},\
  }\bibfield  {title} {\bibinfo {title} {Dependence of the regime change in
  particle aggregates on the composition ratio of magnetic cubic particles with
  different magnetic moment directions},\ }\href
  {https://doi.org/10.1016/j.colsurfa.2017.07.078} {\bibfield  {journal}
  {\bibinfo  {journal} {Colloids and Surfaces A: Physicochemical and
  Engineering Aspects}\ }\textbf {\bibinfo {volume} {557}},\ \bibinfo {pages}
  {146 } (\bibinfo {year} {2018})},\ \bibinfo {note} {“A Collection of Papers
  Presented at the 31st ECIS Meeting, Madrid, Spain, 3-8 September,
  2017”}\BibitemShut {NoStop}%
\bibitem [{\citenamefont {Linse}(2015)}]{Linse}%
  \BibitemOpen
  \bibfield  {author} {\bibinfo {author} {\bibfnamefont {P.}~\bibnamefont
  {Linse}},\ }\bibfield  {title} {\bibinfo {title} {Quasi-2d fluids of dipolar
  superballs in an external field},\ }\href
  {https://doi.org/10.1039/C5SM00338E} {\bibfield  {journal} {\bibinfo
  {journal} {Soft Matter}\ }\textbf {\bibinfo {volume} {11}},\ \bibinfo {pages}
  {3900} (\bibinfo {year} {2015})}\BibitemShut {NoStop}%
\bibitem [{\citenamefont {Goldman}(2005)}]{Curvature}%
  \BibitemOpen
  \bibfield  {author} {\bibinfo {author} {\bibfnamefont {R.}~\bibnamefont
  {Goldman}},\ }\bibfield  {title} {\bibinfo {title} {Curvature formulas for
  implicit curves and surfaces},\ }\href
  {https://doi.org/https://doi.org/10.1016/j.cagd.2005.06.005} {\bibfield
  {journal} {\bibinfo  {journal} {Computer Aided Geometric Design}\ }\textbf
  {\bibinfo {volume} {22}},\ \bibinfo {pages} {632 } (\bibinfo {year}
  {2005})},\ \bibinfo {note} {geometric Modelling and Differential
  Geometry}\BibitemShut {NoStop}%
\bibitem [{\citenamefont {Nelder}\ and\ \citenamefont
  {Mead}(1965)}]{minimization}%
  \BibitemOpen
  \bibfield  {author} {\bibinfo {author} {\bibfnamefont {J.~A.}\ \bibnamefont
  {Nelder}}\ and\ \bibinfo {author} {\bibfnamefont {R.}~\bibnamefont {Mead}},\
  }\bibfield  {title} {\bibinfo {title} {{A Simplex Method for Function
  Minimization}},\ }\href {https://doi.org/10.1093/comjnl/7.4.308} {\bibfield
  {journal} {\bibinfo  {journal} {The Computer Journal}\ }\textbf {\bibinfo
  {volume} {7}},\ \bibinfo {pages} {308} (\bibinfo {year} {1965})},\ \Eprint
  {https://arxiv.org/abs/https://academic.oup.com/comjnl/article-pdf/7/4/308/1013182/7-4-308.pdf}
  {https://academic.oup.com/comjnl/article-pdf/7/4/308/1013182/7-4-308.pdf}
  \BibitemShut {NoStop}%
\bibitem [{\citenamefont {Galassi}\ \emph {et~al.}(2018)\citenamefont {Galassi}
  \emph {et~al.}}]{GSL}%
  \BibitemOpen
  \bibfield  {author} {\bibinfo {author} {\bibfnamefont {M.}~\bibnamefont
  {Galassi}} \emph {et~al.},\ }\href {https://www.gnu.org/software/gsl/}
  {\bibinfo {title} {Gnu scientific library reference manual}} (\bibinfo {year}
  {2018})\BibitemShut {NoStop}%
\bibitem [{\citenamefont {Liu}\ \emph {et~al.}(2010)\citenamefont {Liu},
  \citenamefont {Barrón}, \citenamefont {Torrent}, \citenamefont {Qin},\ and\
  \citenamefont {Yu}}]{LIU2010387}%
  \BibitemOpen
  \bibfield  {author} {\bibinfo {author} {\bibfnamefont {Q.}~\bibnamefont
  {Liu}}, \bibinfo {author} {\bibfnamefont {V.}~\bibnamefont {Barrón}},
  \bibinfo {author} {\bibfnamefont {J.}~\bibnamefont {Torrent}}, \bibinfo
  {author} {\bibfnamefont {H.}~\bibnamefont {Qin}},\ and\ \bibinfo {author}
  {\bibfnamefont {Y.}~\bibnamefont {Yu}},\ }\bibfield  {title} {\bibinfo
  {title} {The magnetism of micro-sized hematite explained},\ }\href
  {https://doi.org/https://doi.org/10.1016/j.pepi.2010.08.008} {\bibfield
  {journal} {\bibinfo  {journal} {Physics of the Earth and Planetary
  Interiors}\ }\textbf {\bibinfo {volume} {183}},\ \bibinfo {pages} {387}
  (\bibinfo {year} {2010})}\BibitemShut {NoStop}%
\bibitem [{\citenamefont {Donahue}\ and\ \citenamefont {Porter}(2016)}]{oommf}%
  \BibitemOpen
  \bibfield  {author} {\bibinfo {author} {\bibfnamefont {M.~J.}\ \bibnamefont
  {Donahue}}\ and\ \bibinfo {author} {\bibfnamefont {D.~G.}\ \bibnamefont
  {Porter}},\ }\href {https://doi.org/doi:10.21981/8RRA-5656} {\bibinfo {title}
  {Oommf: Object oriented micromagnetic framework}} (\bibinfo {year}
  {2016})\BibitemShut {NoStop}%
\bibitem [{\citenamefont {Bjørk}\ \emph {et~al.}(2021)\citenamefont {Bjørk},
  \citenamefont {Poulsen}, \citenamefont {Nielsen},\ and\ \citenamefont
  {Insinga}}]{BJORK2021168057}%
  \BibitemOpen
  \bibfield  {author} {\bibinfo {author} {\bibfnamefont {R.}~\bibnamefont
  {Bjørk}}, \bibinfo {author} {\bibfnamefont {E.}~\bibnamefont {Poulsen}},
  \bibinfo {author} {\bibfnamefont {K.}~\bibnamefont {Nielsen}},\ and\ \bibinfo
  {author} {\bibfnamefont {A.}~\bibnamefont {Insinga}},\ }\bibfield  {title}
  {\bibinfo {title} {Magtense: A micromagnetic framework using the analytical
  demagnetization tensor},\ }\href
  {https://doi.org/https://doi.org/10.1016/j.jmmm.2021.168057} {\bibfield
  {journal} {\bibinfo  {journal} {Journal of Magnetism and Magnetic Materials}\
  }\textbf {\bibinfo {volume} {535}},\ \bibinfo {pages} {168057} (\bibinfo
  {year} {2021})}\BibitemShut {NoStop}%
\bibitem [{\citenamefont {Bonilla}\ \emph {et~al.}(2017)\citenamefont
  {Bonilla}, \citenamefont {Lacroix},\ and\ \citenamefont
  {Blon}}]{BONILLA2017394}%
  \BibitemOpen
  \bibfield  {author} {\bibinfo {author} {\bibfnamefont {F.}~\bibnamefont
  {Bonilla}}, \bibinfo {author} {\bibfnamefont {L.-M.}\ \bibnamefont
  {Lacroix}},\ and\ \bibinfo {author} {\bibfnamefont {T.}~\bibnamefont
  {Blon}},\ }\bibfield  {title} {\bibinfo {title} {Magnetic ground states in
  nanocuboids of cubic magnetocrystalline anisotropy},\ }\href
  {https://doi.org/https://doi.org/10.1016/j.jmmm.2016.12.069} {\bibfield
  {journal} {\bibinfo  {journal} {Journal of Magnetism and Magnetic Materials}\
  }\textbf {\bibinfo {volume} {428}},\ \bibinfo {pages} {394} (\bibinfo {year}
  {2017})}\BibitemShut {NoStop}%
\bibitem [{\citenamefont {Schabes}\ and\ \citenamefont
  {Aharoni}(1987)}]{Mag_pot}%
  \BibitemOpen
  \bibfield  {author} {\bibinfo {author} {\bibfnamefont {M.}~\bibnamefont
  {Schabes}}\ and\ \bibinfo {author} {\bibfnamefont {A.}~\bibnamefont
  {Aharoni}},\ }\bibfield  {title} {\bibinfo {title} {Magnetostatic interaction
  fields for a three-dimensional array of ferromagnetic cubes},\ }\href
  {https://doi.org/10.1109/TMAG.1987.1065775} {\bibfield  {journal} {\bibinfo
  {journal} {IEEE Transactions on Magnetics}\ }\textbf {\bibinfo {volume}
  {23}},\ \bibinfo {pages} {3882} (\bibinfo {year} {1987})}\BibitemShut
  {NoStop}%
\bibitem [{\citenamefont {Chikazumi}(2009)}]{chikazumi2009physics}%
  \BibitemOpen
  \bibfield  {author} {\bibinfo {author} {\bibfnamefont {S.}~\bibnamefont
  {Chikazumi}},\ }\href {https://books.google.lv/books?id=AZVfuxXF2GsC} {\emph
  {\bibinfo {title} {Physics of Ferromagnetism}}},\ International Series of
  Monographs on Physics\ (\bibinfo  {publisher} {OUP Oxford},\ \bibinfo {year}
  {2009})\BibitemShut {NoStop}%
\end{thebibliography}%


%

\end{document}